\documentclass{emulateapj}
\usepackage{amsfonts,lscape,epsfig,amsmath,enumerate,color}
\begin{document}

\title{An Evolutionary Paradigm for Dusty Active Galaxies at Low Redshift}

\author{D.~Farrah\altaffilmark{1,2},
B.~Connolly\altaffilmark{3},
N.~Connolly\altaffilmark{4},
H.~W.~W.~Spoon\altaffilmark{2}
S.~Oliver\altaffilmark{1},
H.~B.~Prosper\altaffilmark{5},
L.~Armus\altaffilmark{6},
J.~R.~Houck\altaffilmark{2},
A.~R.~Liddle\altaffilmark{1},
V.~Desai\altaffilmark{6},
}
\altaffiltext{1}{Astronomy Centre, University of Sussex, Brighton, United Kingdom}
\altaffiltext{2}{Astronomy Department, Cornell University, Ithaca, NY 14853, USA}
\altaffiltext{3}{Department of Physics and Astronomy, University of Pennsylvania, Philadelphia, PA 19104-6396, USA}
\altaffiltext{4}{Physics Department, Hamilton College, Clinton, NY 13323, USA}
\altaffiltext{5}{Department of Physics, Florida State University, Tallahassee, FL 32306, USA}
\altaffiltext{6}{Spitzer Science Center, California Institute of Technology, 1200 East California Boulevard, Pasadena, CA 91125, USA}
\email{D.Farrah@sussex.ac.uk}
\date{\today}

\begin{abstract}
We apply methods from Bayesian inferencing and graph theory to a dataset of 102 mid-infrared spectra, and archival data from the optical to the millimeter, to construct an evolutionary paradigm for $z<0.4$ infrared-luminous galaxies (ULIRGs). We propose that the ULIRG lifecycle consists of three phases. The first phase lasts from the initial encounter until approximately coalescence. It is characterized by homogeneous mid-IR spectral shapes, and IR emission mainly from star formation, with a contribution from an AGN in some cases. At the end of this phase, a ULIRG enters one of two evolutionary paths depending on the dynamics of the merger, the available quantities of gas, and the masses of the black holes in the progenitors. On one branch, the contributions from the starburst and the AGN to the total IR luminosity decline and increase respectively. The IR spectral shapes are heterogeneous, likely due to feedback from AGN-driven winds. Some objects go through a brief QSO phase at the end. On the other branch, the decline of the starburst relative to the AGN is less pronounced, and few or no objects go through a QSO phase. We show that the 11.2$\mu$m PAH feature is a remarkably good diagnostic of evolutionary phase, and identify six ULIRGs that may be archetypes of key stages in this lifecycle. 
\end{abstract}

\keywords{methods: data analysis - methods: statistical - galaxies: evolution - galaxies: active - infrared: galaxies - galaxies: interactions}

\section{Introduction}
Ultraluminous Infrared Galaxies (ULIRGS, objects with rest-frame 1-1000$\mu$m luminosities of $>10^{12}$L$_{\odot}$) play a fundamental role in the cosmological evolution of galaxies and large-scale structures. First discovered in significant numbers by the Infrared Astronomical Satellite \citep{soi84}, they are almost invariably mergers \citep{far01,bus02,vei02,vei06}, powered by star formation and AGN activity, with the star formation usually dominating \citep{gen98,vei99,rig99,ima07,veg08}. ULIRGs are rare at low redshift, with less than fifty at $z\lesssim0.1$, but become much more numerous at high redshifts, reaching a density on the sky of several hundred per square degree at $z\gtrsim1$ \citep{rr97,dol,hug,bar98,bor,mor}. The high redshift ULIRGs appear similar in some ways to those in the local Universe, in that many of them are starburst dominated mergers \citep{far02b,cha03,sma04,tak06,bor06,val07,ber07,bri07,lon09,hua09}, though there are also signs of differences, {\it e.g.} a higher fraction of systems with no signs of interaction \citep{mel08}, systematically different mid-IR spectral shapes \citep{far08}, and overdense local environments \citep{bla04,far06a,mag07}. Reviews of their properties can be found in \citet{san96} and \citet{lfs06}.

The cosmological significance of ULIRGs makes a solid understanding of them at low redshifts important, but there remain several uncertainties over the life-cycle of low redshift ULIRGs. We know that the merger activity triggers star formation and AGN activity, but we do not know how long the starburst lasts, whether or not there exist distinct `AGN plus Starburst' or `AGN dominated' phases, the fraction of ULIRGs that become Quasars, or if there are multiple evolutionary paths that a ULIRG can take.

In this paper, we explore a new approach to study the evolution of the low redshift ULIRG population. Since many of the best diagnostics of star formation and AGN activity lie in the mid-infrared, we start with a mid-infrared spectroscopic dataset of $z<0.4$ ULIRGs, taken with the Infrared Spectrograph (IRS, \citealt{hou04}) on board the Spitzer Space Telescope \citep{wer04,soi08}. To this dataset we apply two novel analysis methods to determine trends in mid-IR spectral shape across the sample. First, we present a Bayesian based estimator of the degree of similarity between a pair of ULIRG spectra. This estimator uses data from every resolution element, marginalizing over measurement error, luminosity, and foreground obscuration, to produce Bayes factors that describe the degree of resemblance between every possible pair of spectra. Second, we use methods developed using graph theory to study interconnected groups of entities to produce a `network' diagram that visualizes these Bayes factors across the whole sample simultaneously. We combine these results with archival data to propose an evolutionary description of ULIRGs in the low-redshift Universe. 

This paper is structured as follows. In \S\ref{methodstuff} we describe the sample selection and data, outline the method by which we calculate the Bayes factors, and describe the network construction. In \S\ref{resultsstuff} we present the Bayes factors and network diagram. In \S\ref{sectrelia} we assess the robustness of the Bayes factors and the network diagram, and possible sources of error. Discussion can be found in \S\ref{discussstuff}. We summarize our conclusions in \S\ref{concstuff}. We assume a spatially flat cosmology with $H_{0}=70$ km s$^{-1}$ Mpc$^{-1}$, $\Omega=1$, and $\Omega_{m}=0.3$.

\begin{figure*}[tbhp]
\begin{center}
\includegraphics[width=0.35\textwidth,angle=90.0]{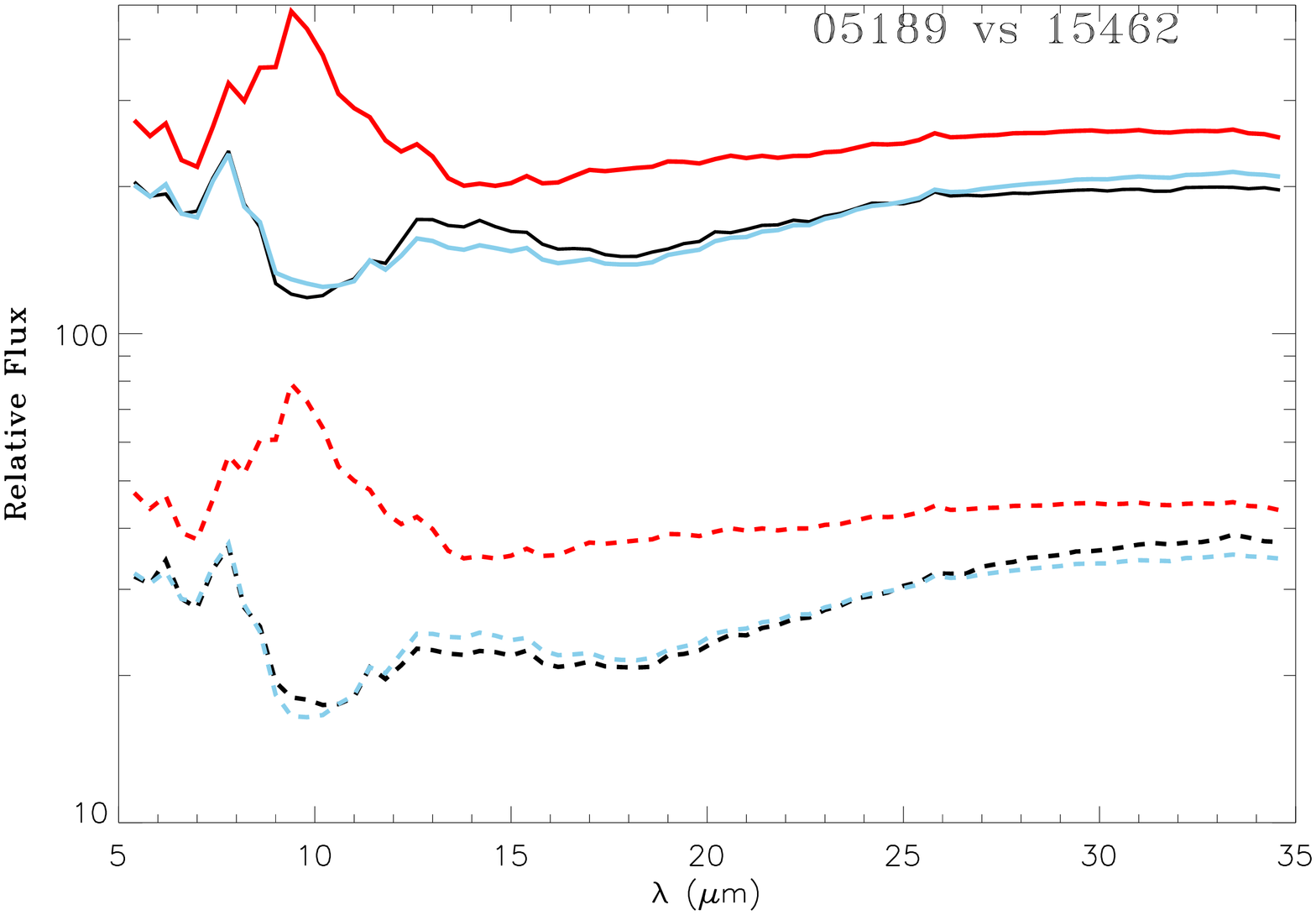}
\includegraphics[width=0.35\textwidth,angle=90.0]{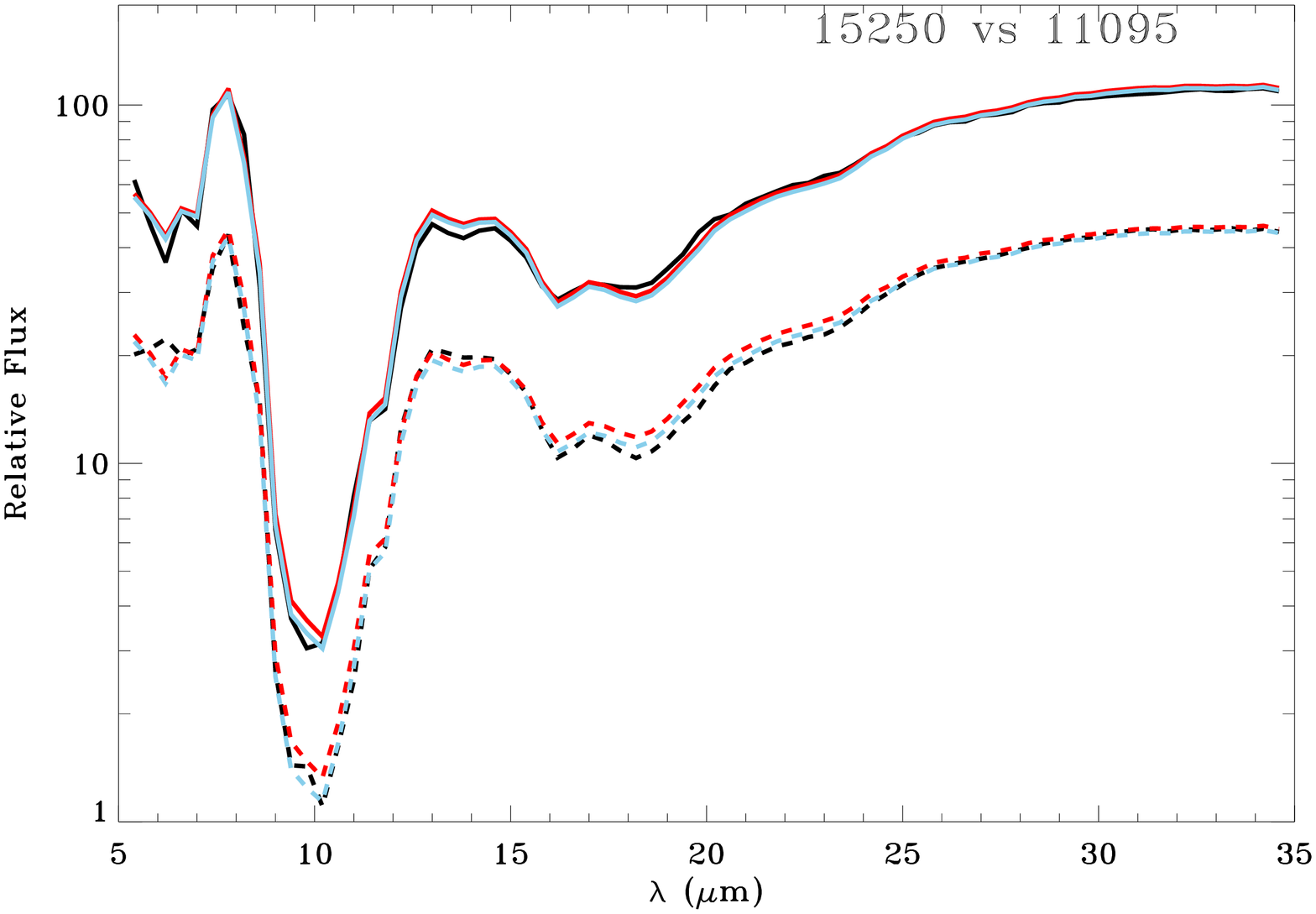}
\includegraphics[width=0.35\textwidth,angle=90.0]{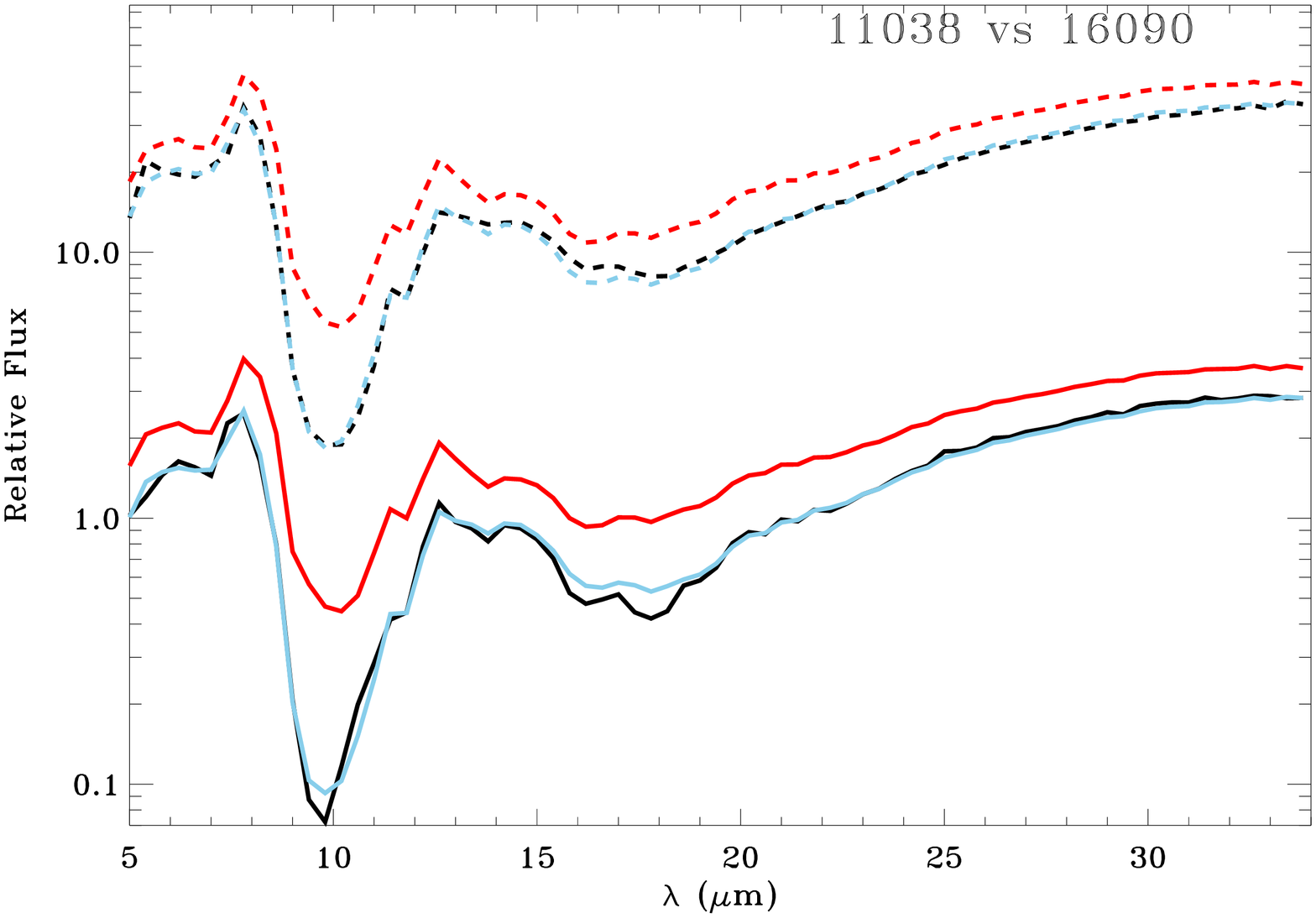}
\includegraphics[width=0.35\textwidth,angle=90.0]{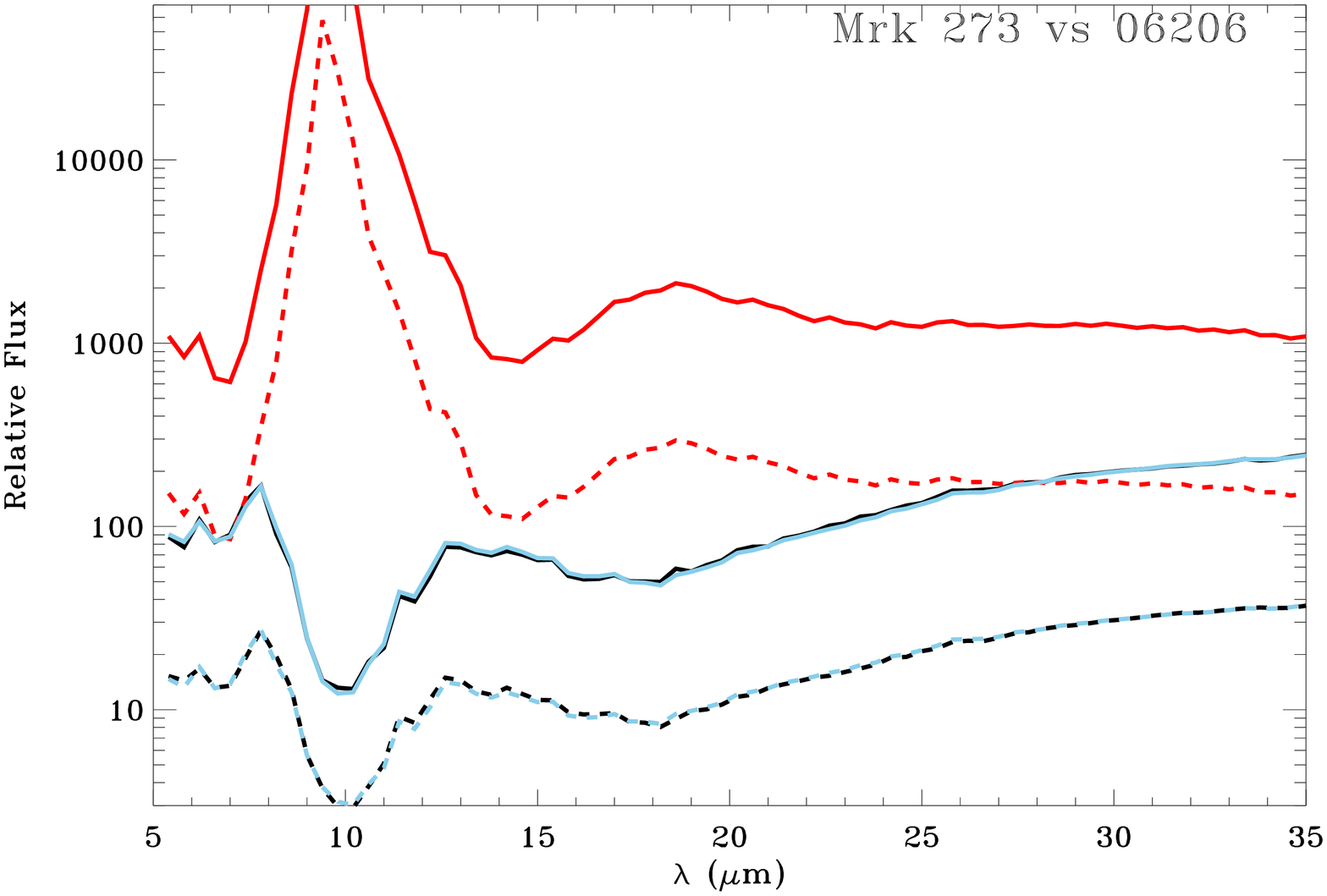}
\includegraphics[width=0.35\textwidth,angle=90.0]{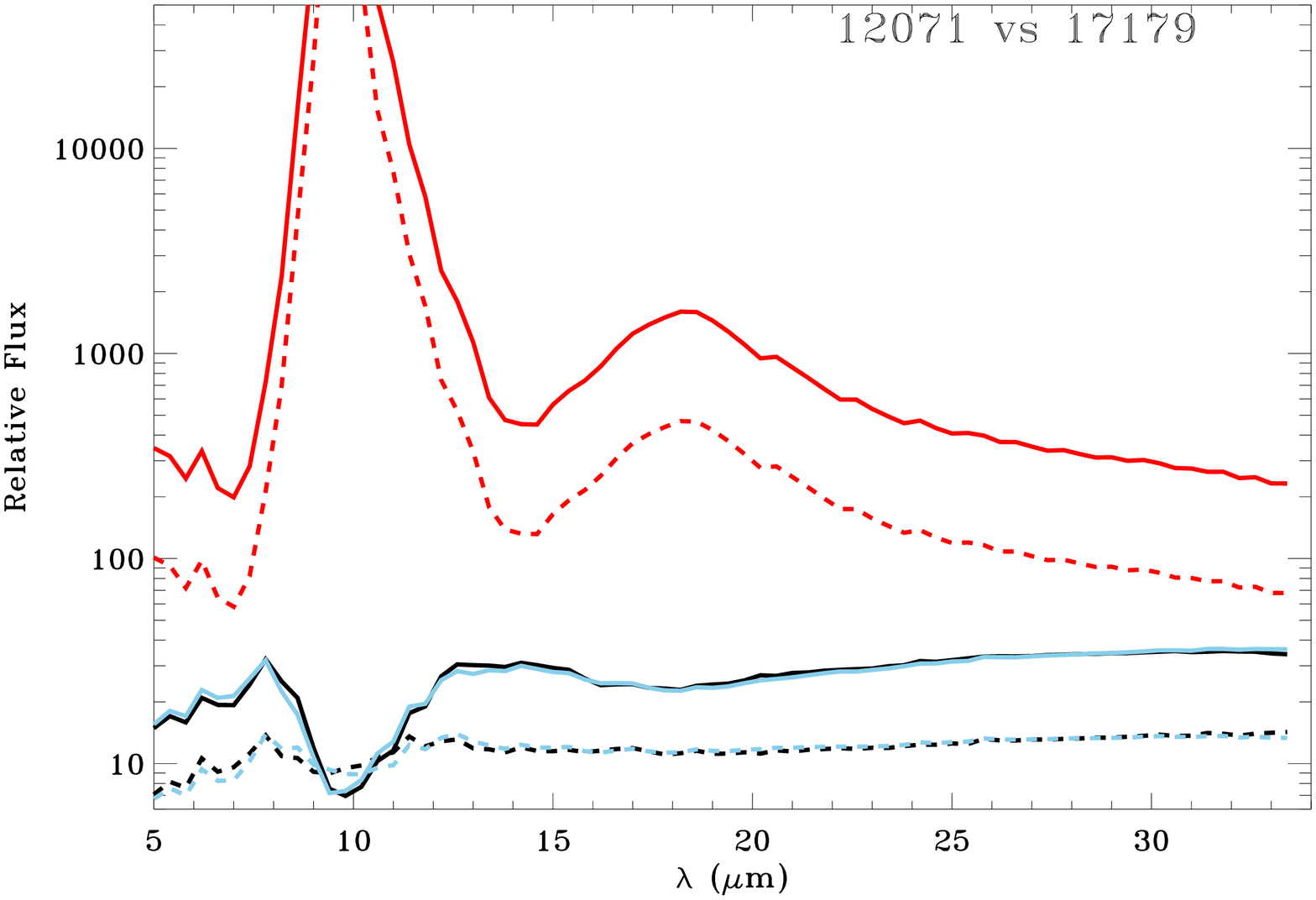}
\includegraphics[width=0.35\textwidth,angle=90.0]{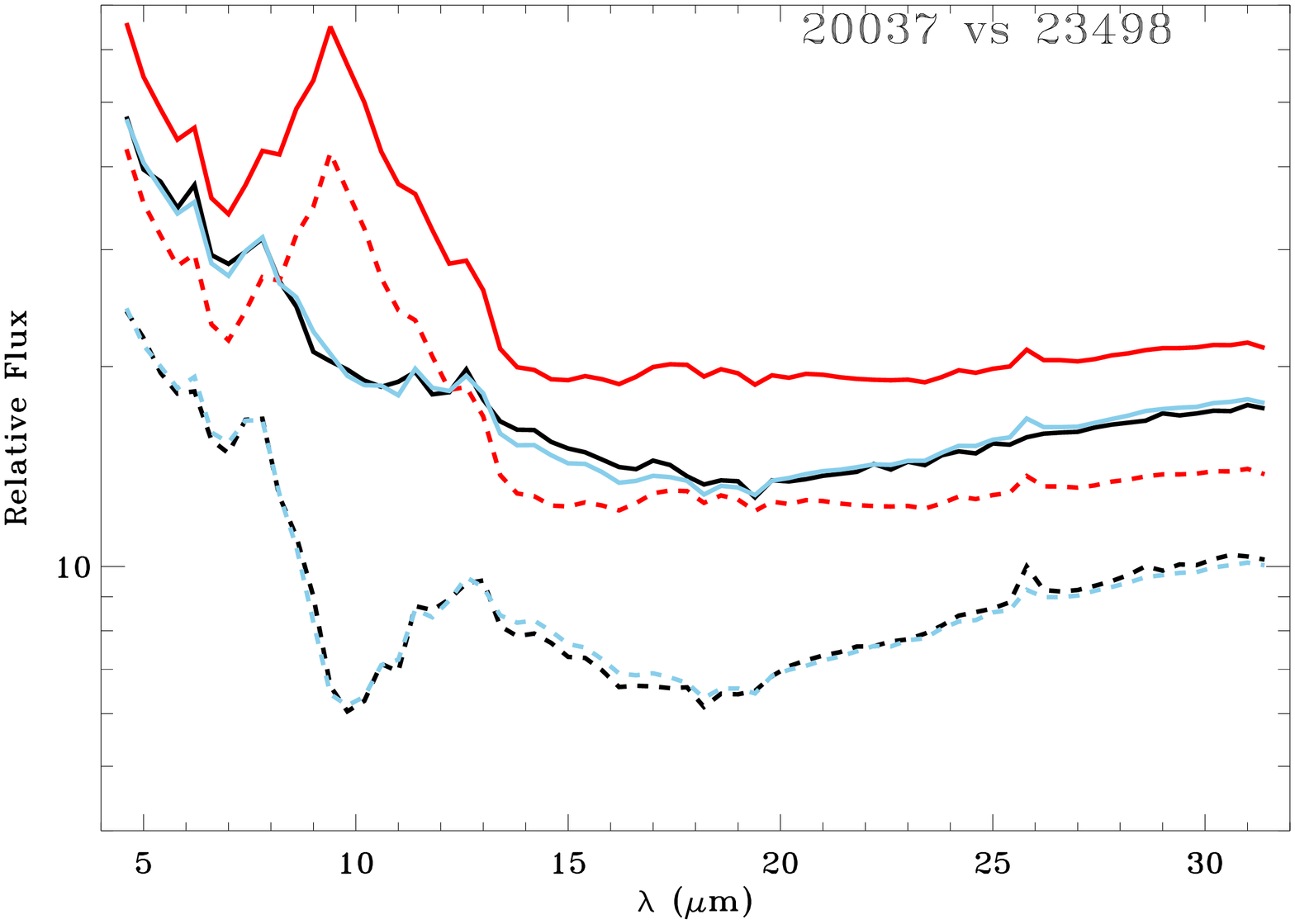}
 \caption{\label{besta}
Six examples of the 'best' fits (for which $-150 < log (\mathcal{R}) < -250$), calculated by minimizing Eqn.~\ref{eqn:loglike}. The solid (dashed) black line is the IRS spectrum for object A (B). The solid (dashed) red line is the predicted intrinsic spectrum (i.e. without foreground extinction) for object A (B). The solid (dashed) blue line is the predicted intrinsic spectrum with cold foreground extinction applied for object A (B). The identical shapes of the solid and dashed red line shows that the intrinsic spectra of objects A and B are the same (they differ by a factor of $fe^{f_eb_e}/(1-f)e^{(1-f_e)b_e}$). The identical shapes of the solid (dashed) black and blue lines shows that the observed and `predicted observed' spectra of object A (B) are the same. We do not constrain the shape of the intrinsic spectrum, and are showing the purely mathematically `best' fits, so some predicted features in the intrinsic spectra, such as the very strong silicate emission feature in the bottom left panel, may not be `real'. 
}
\end{center}
\end{figure*}

\section{Method}\label{methodstuff}

\subsection{The Sample}\label{lesample}
The sample was observed as part of the IRS Guaranteed Time program to obtain mid-infrared spectra of low redshift IR-luminous sources (Spitzer program ID 105), selected from the IRAS 1Jy \citep{kim98} and 2Jy \citep{str90} surveys, and from the FIRST sample \citep{sta00}. The sample is slightly biased towards sources with warm infrared colors, as described in \citet{des07}, but should still be representative of the low-redshift IR-luminous galaxy population. A few of our sample have IR luminosities that fall outside the usual definition of a ULIRG, but for simplicity we refer to all of our sample as ULIRGs for the remainder of this paper. 

Low resolution spectra (5.2$\mu$m - 38.5$\mu$m, R$\sim60-125$) were obtained of 118 objects, and high resolution spectra (9.6$\mu$m - 38.0$\mu$m, R$\sim600$) were obtained of a subset of 53 objects. Data reduction methods and initial results are presented in \citet{arm04,spo04} and \citet{arm06}. Further results are presented in \citet{hig06,spo06,spo07} and \citet{des07}. Atlases of the high and low resolution spectra can be found in \citet{far07b} and Armus et al (2009, in preparation), respectively.

The high resolution spectra contain more elements than the low resolution spectra, but sky background cannot be subtracted from them as we lack dedicated contemporaneous sky observations. Therefore, we use the low resolution spectra. We exclude objects with $z>0.4$, so that we see approximately the same wavelength range for all objects. We also exclude IRAS 11119+3257 and IRAS 23365+3604 as they have poor quality data in the Short-Low modules. This leaves 102 objects, listed in Table \ref{sample}. We smooth the spectra to the instrumental resolution using a 0.4$\mu$m boxcar, and assume a 5\% flux error for each resulting resolution element, consistent with the observed variations between individual nod positions. 

\begin{figure}[tbhp]
\begin{center}
\includegraphics[width=0.35\textwidth,angle=90.0]{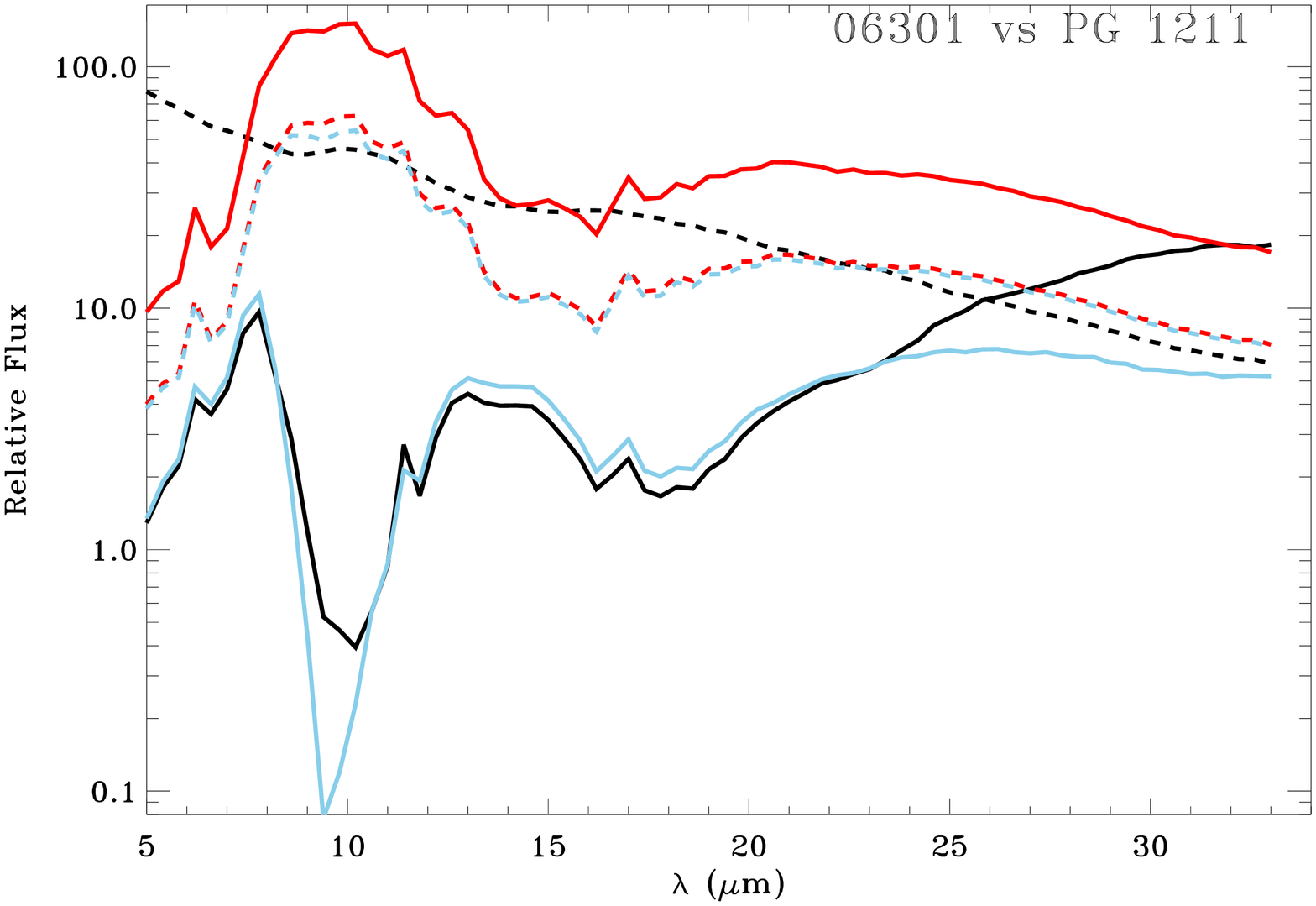}
\includegraphics[width=0.35\textwidth,angle=90.0]{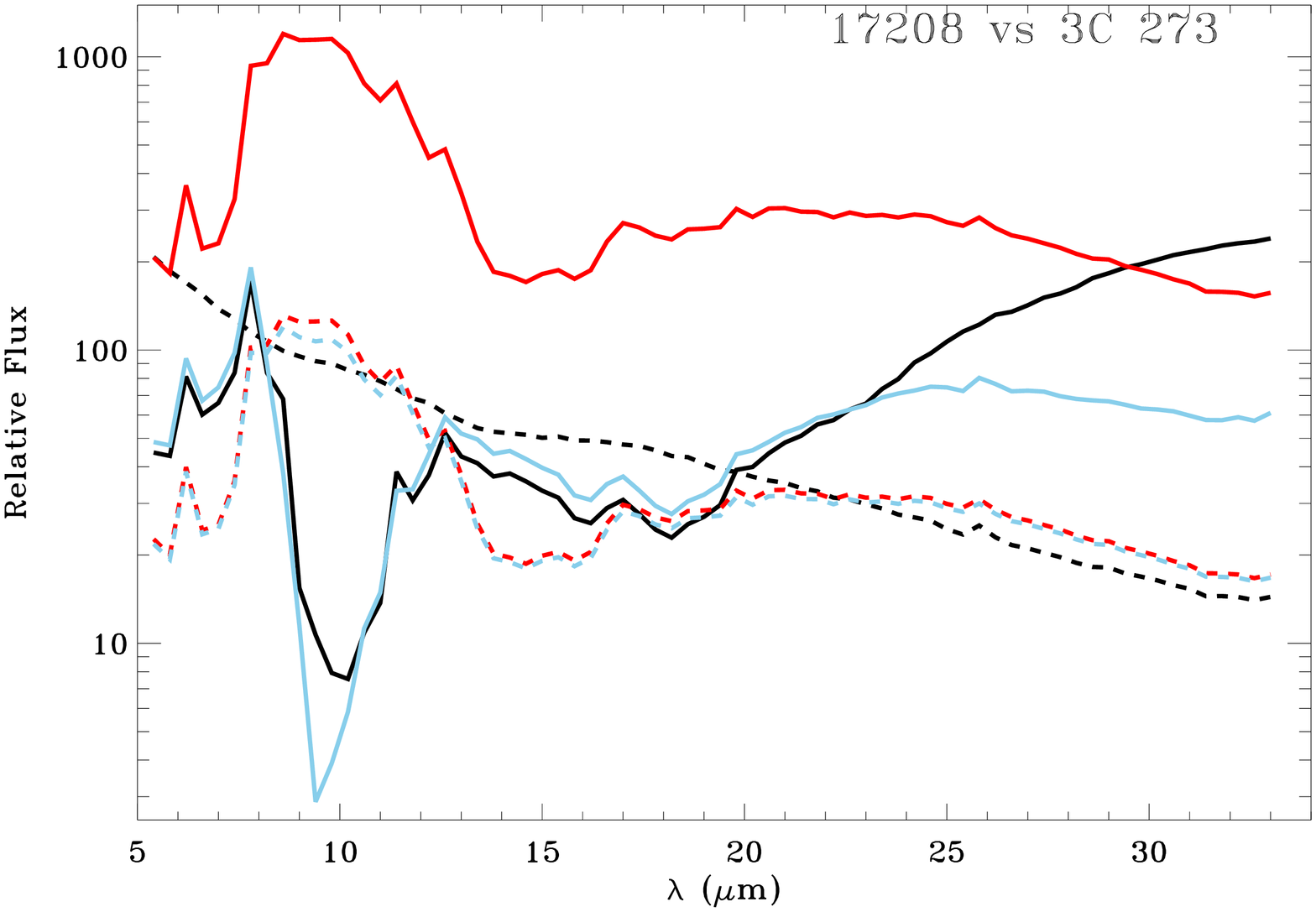}
 \caption{\label{worsta}
For comparison with Figure \ref{besta}; two examples of the `worst' fits calculated by minimizing Eqn.~\ref{eqn:loglike}, where $log (\mathcal{R}) \simeq3000$. Here the the solid and dashed red lines are identical, but the blue and black lines are completely different for both objects, showing that the IRS spectrum and the `predicted' IRS spectrum do not match each other.
}
\end{center}
\end{figure}

\subsection{Analysis}\label{subsanalysis}
\subsubsection{Bayesian Measures of Resemblance}
To compute the level of resemblance between any pair of (rest-frame) spectra we adopt a Bayesian approach. For any two spectra, A and B, we compute:

\begin{eqnarray}
\label{equation:likelihoodratio}
\mathcal{R}=\frac{P(\rm{A,B}|\rm{different})}{P(\rm{A,B}|\rm{same})}
\end{eqnarray}

\noindent where $P(\rm{A,B}|\rm{same})$ is the probability density that the two spectra are identical, and $P(\rm{A,B}|\rm{different})$ is the probability density that the spectra are different. The quantity $\mathcal{R}$ is thus the Bayes factor\footnote{Arguably, the ratio of posteriors - i.e. $P(\text{different}|\text{A,B})/P(\text{same}|\text{A,B})$ is a more intuitive statistic. However, the calculation of the odds requires one to make prior assumptions about $P(\text{same})$ and $P(\text{different})$, which we prefer to avoid.} \citep{jeffreys,connolly} quantifying the belief\footnote{We use the word `belief' in its Bayesian sense, i.e. the odds of a successful trial of the truth of a given proposition, and not in the colloquial sense} that this pair of spectra arise from sources whose physical properties (or at least those that give rise to the mid-IR emission) are the same. In essence, we are performing a pixel-by-pixel comparison between two spectra, where no spectral region is preferentially weighted. While this method could be used to compare data to models, we are here making no model comparisons, instead comparing the spectra to each other.

The simplest use of this method would involve computing $\mathcal{R}$ for every possible pair of `raw' (i.e. reduced, rest-frame but otherwise unaltered) spectra. This, however, is not enough. There exist two variables that will increase the $\mathcal{R}$ values, but which do not necessarily reflect physical differences. The first is instrument noise; differences between spectra that arise due to Gaussian fluctuations in the measurements should not contribute to the $\mathcal{R}$ values. The second is cold foreground extinction; if object A and object B are intrinsically identical, but object A has a thicker screen of cold dust in front of it, then this should not contribute to the $\mathcal{R}$ values either. We make a third, simplifying assumption that differences in mid-IR luminosity (i.e. multiplicative scalings between spectra) do not reflect `real' differences. Therefore, in calculating the $\mathcal{R}$ values, we marginalize with respect to instrument noise, mid-IR luminosity, and cold foreground extinction. The resulting (fully marginalized) probabilities, $P(\rm{A,B}|\rm{same})$ and $P(\rm{A,B}|\rm{different})$, are thus measures of the {\em evidence} for the hypothesis \citep{sivia}. The full methodology is described in the Appendix.

Finally, we adopt a boundary condition for $\log_{10} (\mathcal{R})$; pairs of spectra with $\log_{10} (\mathcal{R})$ below this boundary are treated as similar, and those pairs with $\log_{10} (\mathcal{R})$ above it are treated as different. We set this boundary at $\log_{10} (\mathcal{R})= 0$. In frequentist terms, this boundary is equivalent to demanding $\chi^{2} \lesssim 0.8$ \citep{sel01}. In Bayesian terms, using the scale given in \citet{jeffreys}, this corresponds to `marginal' strength of evidence. We explore the sensitivity of our results to the error in $\mathcal{R}$ and choice of boundary condition in \S\ref{sectrelia}.

\begin{figure*}[tbhp]
\begin{center}
\includegraphics[width=1.0\textwidth]{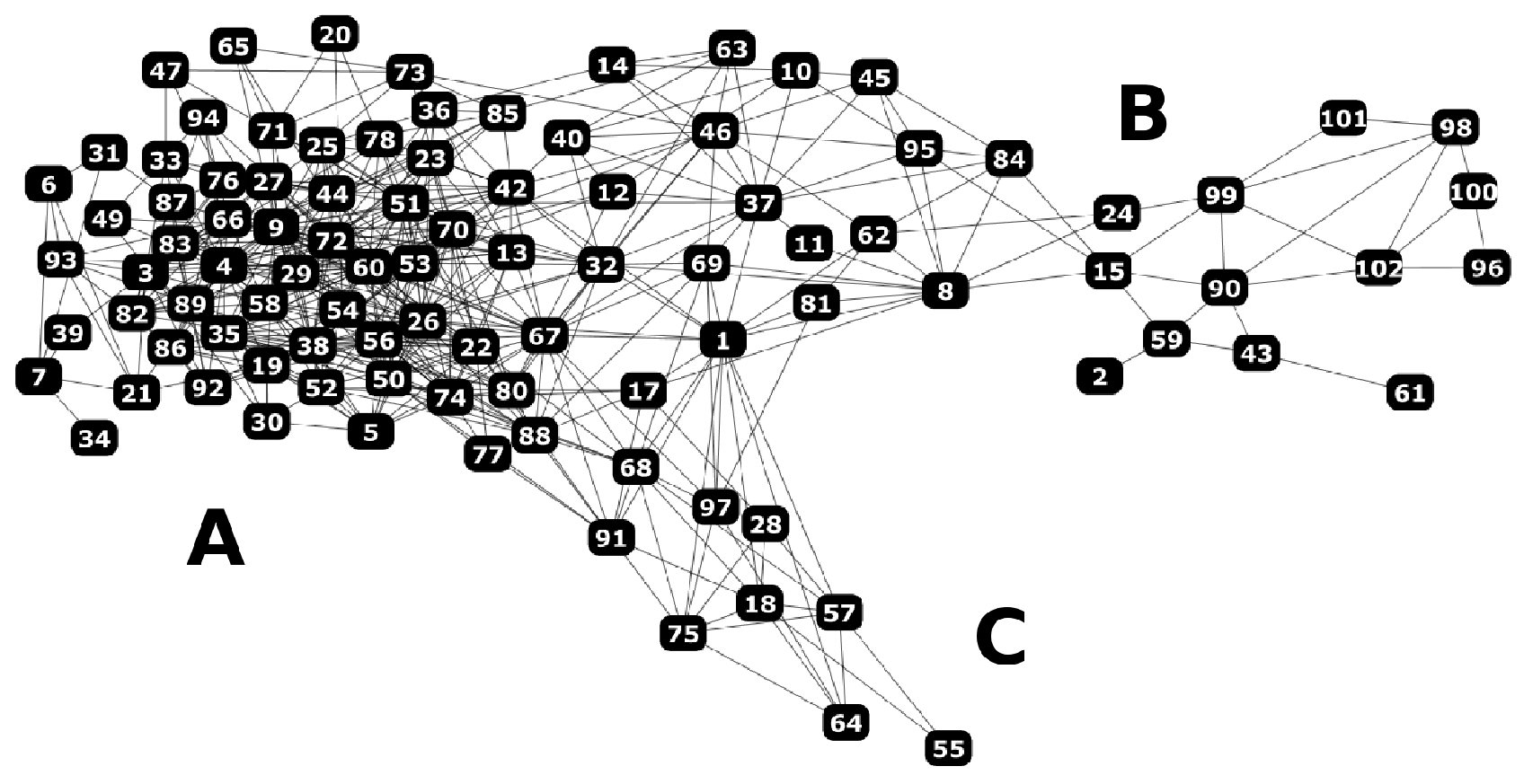}
 \caption{\label{ulirglayout}
The network for our sample (generated using a spring-embedded algorithm within Cytoscape). The numbered points, or `nodes', are the objects in Table \ref{sample}. An edge between two nodes indicates that $log (\mathcal{R}) < 0$ for that pair of objects. 
}
\end{center}
\end{figure*}

\subsubsection{Network Construction}\label{sssectnet}
With a sample of 102 galaxies, we have $^{102}C_{2}=5151$ $\mathcal{R}$ values, one for each possible pair. Our second requirement is a method to study these Bayes factors across the sample.

This is an example of studying a pairwise-connected group of entities. Other examples include a computer network (e.g. \citealt{sig03}), predator-prey relationships among animals, or `social' networks such as friendships between individuals in a group. As such, a common terminology has arisen to describe them. Each entity (e.g. a computer or a person) is a `node'. Connections between nodes are `edges' if they have no direction (for example, if two computers share data) and `arcs' if they have direction (for example, lions eat gazelles but gazelles do not eat lions). The number of edges connecting to a node is the `degree' of that node. The nodes in our network are the ULIRGs. The pairs where $\log_{10} (\mathcal{R}) < 0$ are connected via edges, while those pairs with $\log_{10} (\mathcal{R}) > 0$ are not connected. 

Several methods have been developed to plot nodes and their connections in informative ways. In our case we require a method that (a) places connected nodes close together, and (b) produces as few `crossing' edges as possible. A suitable algorithm for this is a `force-directed' algorithm, in which attractive and repulsive forces govern the arrangement of the nodes and edges \citep{kam89,fru91}. Connected nodes attract each other along edges, while {\it all} nodes repel each other. The attractive force is modeled as if the edges are springs (i.e. a Hookes law type force) while the repulsive force is modeled as if the nodes are electrically charged (i.e. a Coulomb type force). The parameters of the two forces are adjusted, and nodes allowed to move according to the forces acting on them until (a) an equilibrium state is reached in which the positions of the nodes and edges do not change appreciably, and (b) the nodes and edges can be seen simultaneously.

To create the network for our sample we use two software packages; the Network Workbench tool\footnote{This tool is developed jointly by Indiana University and Northwestern University, and is available from http://nwb.slis.indiana.edu}, and {\it Cytoscape}\footnote{Available from http://cytoscape.org/}.

\section{Results}\label{resultsstuff}
\begin{figure}[tbhp]
\begin{center}
\includegraphics[width=0.3\textwidth,angle=90.0]{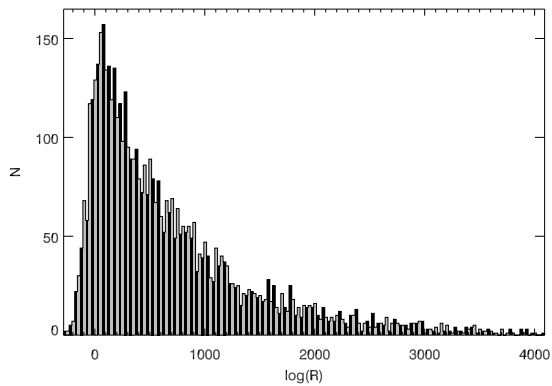}
 \caption{\label{rmiserhisto}
Histogram of the Bayes factors for every possible pair of spectra (5151 in total) in Table \ref{sample}, computed using the $miser$ algorithm.
}
\end{center}
\end{figure}

The pairs of sources for which $\log_{10}(\mathcal{R})<0$ are given in Table \ref{sample}. Examples of pairs where $\log_{10}(\mathcal{R})<0$ are shown in Figure \ref{besta}, and examples of pairs with $\log_{10}(\mathcal{R})>0$ are shown in Figure \ref{worsta}. Presenting all of the $\mathcal{R}$ values would take an unreasonable amount of space, so instead we plot their histogram in Figure \ref{rmiserhisto}. The adjacency matrix, $A$, for our network\footnote{The adjacency matrix, $A$ for an undirected graph with $n$ nodes is defined as the $n\times n$ matrix where $A_{ij}$ is the number of edges from vertex $i$ to vertex $j$, and $A_{ii}$ is the number of loops for vertex $i$} is too large to give in full, but the first few elements of it read:

\begin{equation*}
A = 
\left[
\begin {array}{cccc}
 0  &  0   &  0   &  0   \\
 0  &  0   &  0   &  0   \\
 0  &  0   &  0   &  1   \\
 0  &  0   &  1   & 0
\end {array}
\right]
\end{equation*}

\noindent where the diagonal elements are zero as our network contains no loops. The network for the sample is shown in Figure \ref{ulirglayout}.

\subsection{Structure}\label{discstruct}
Figure \ref{ulirglayout} exhibits a strong degree of connectivity. Only four objects (16,41,48,79, which are not plotted) are not connected to any other. All the other nodes are connected to at least one other node, with most having three or more connections. The average node degree is high, at 10.2, and the average shortest path between any two nodes is short for a 102 node network, at 3.2 edges.

\begin{figure}[tbhp]
\begin{center}
\includegraphics[width=0.50\textwidth,angle=0.0]{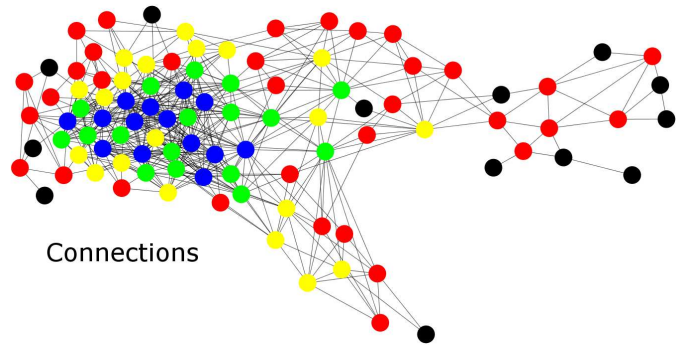}
\caption{\label{ulirglayoutscodea}
Network diagram, with the nodes color-coded by the number of edges connecting them (black=$<$4, red=4-7, yellow=7-14, green=14-20, blue=$>$20).
}
\end{center}
\end{figure}

There is significant variation in node degree across the diagram (Figure \ref{ulirglayoutscodea}). We quantify this by computing the $k$-Nearest-Neighbor distribution \citep{pas} for our network. We find that, as the average degree per node, $k_{mean}$, increases, so does $k_{NN}$; $k_{NN}\simeq0.5$ for $k_{mean}\simeq3$, rising to $k_{NN}\simeq1.2$ for $k_{mean}\simeq15$. So, with the caveat of the small number of nodes, Figure \ref{ulirglayoutscodea} shows a correlation between the degree of a node and that of its neighbors - nodes with a high degree are more likely to be connected to other nodes with a high degree - and is thus an `assortative' network. 

We identify at least two substructures. The first is a strongly interconnected group centered on object 29 (IRAS 03000-2919), accompanied by some outliers on the left hand side, containing $\sim$60\% of the sample. We call this group A. The second is a weakly interconnected group extending in the rightward direction from group A, and containing the remaining 40\% of the sample. It is plausible, given the two `branches' in this second group, that it is composed of two groups; one extending along the top of Figure \ref{ulirglayout} and centered on object 15 (IRAS 00275-2859), and one on the lower side of Figure \ref{ulirglayout}, centered on object 97 (Mrk 1014). We label these two branches groups B and C respectively.  Solely with this diagram to go on however, this subdivision is tentative, as the purported Group C resembles the `outliers' on the left side of group A. We discuss the robustness of this subdivision in \S\ref{evolp}.

\subsection{Node properties}
In studies of networks, insight can be gained by coding the nodes according to some property of the nodes (for example, coding the networks in \citet{lus04} by gender or age revealed clear sub-communities). We adopt this approach in this section. We here present the individual coded diagrams, and interpret them in \S\ref{discussstuff}.

\begin{figure*}[tbhp]
\begin{center}
\includegraphics[width=0.49\textwidth,angle=0.0]{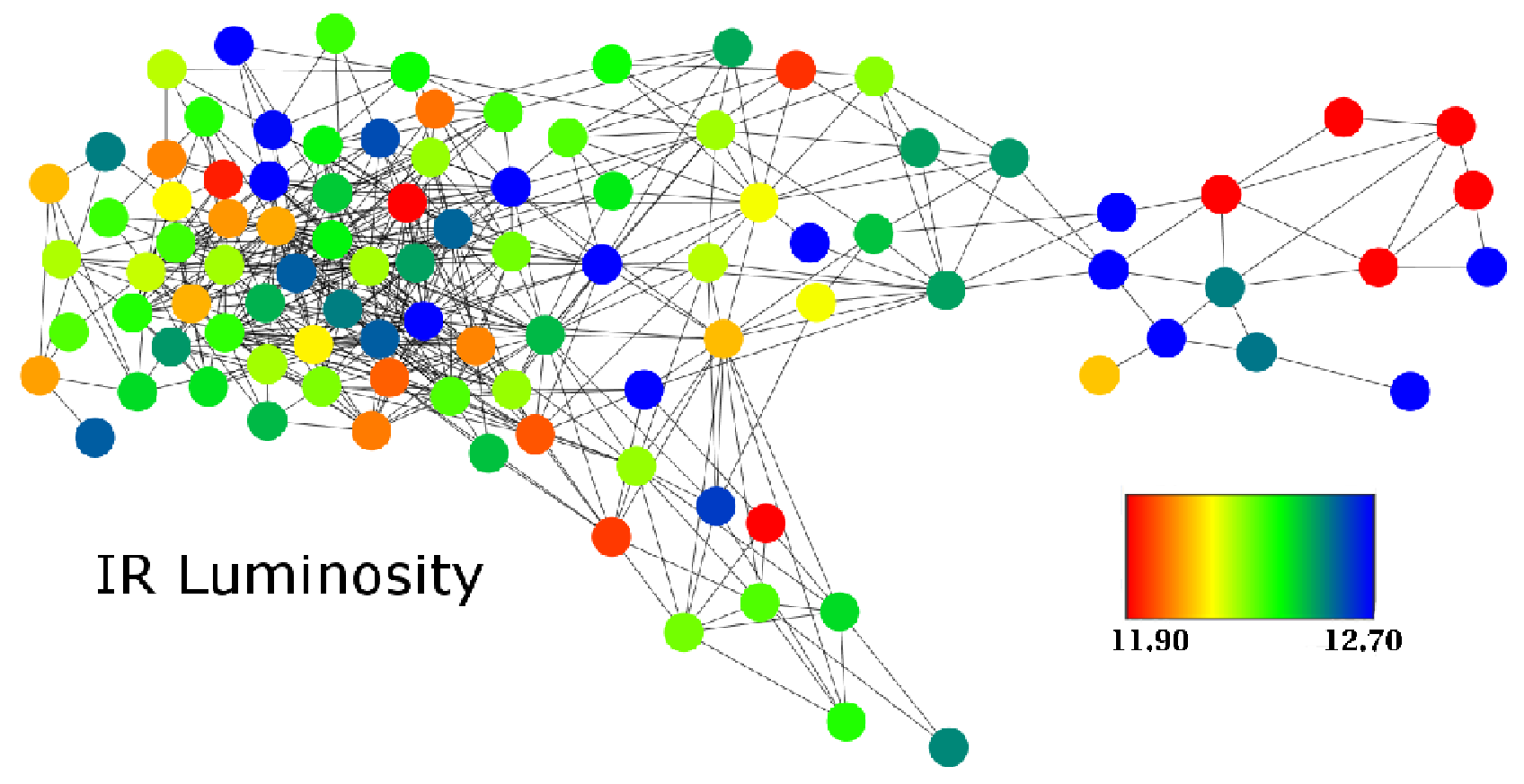}
\includegraphics[width=0.49\textwidth,angle=0.0]{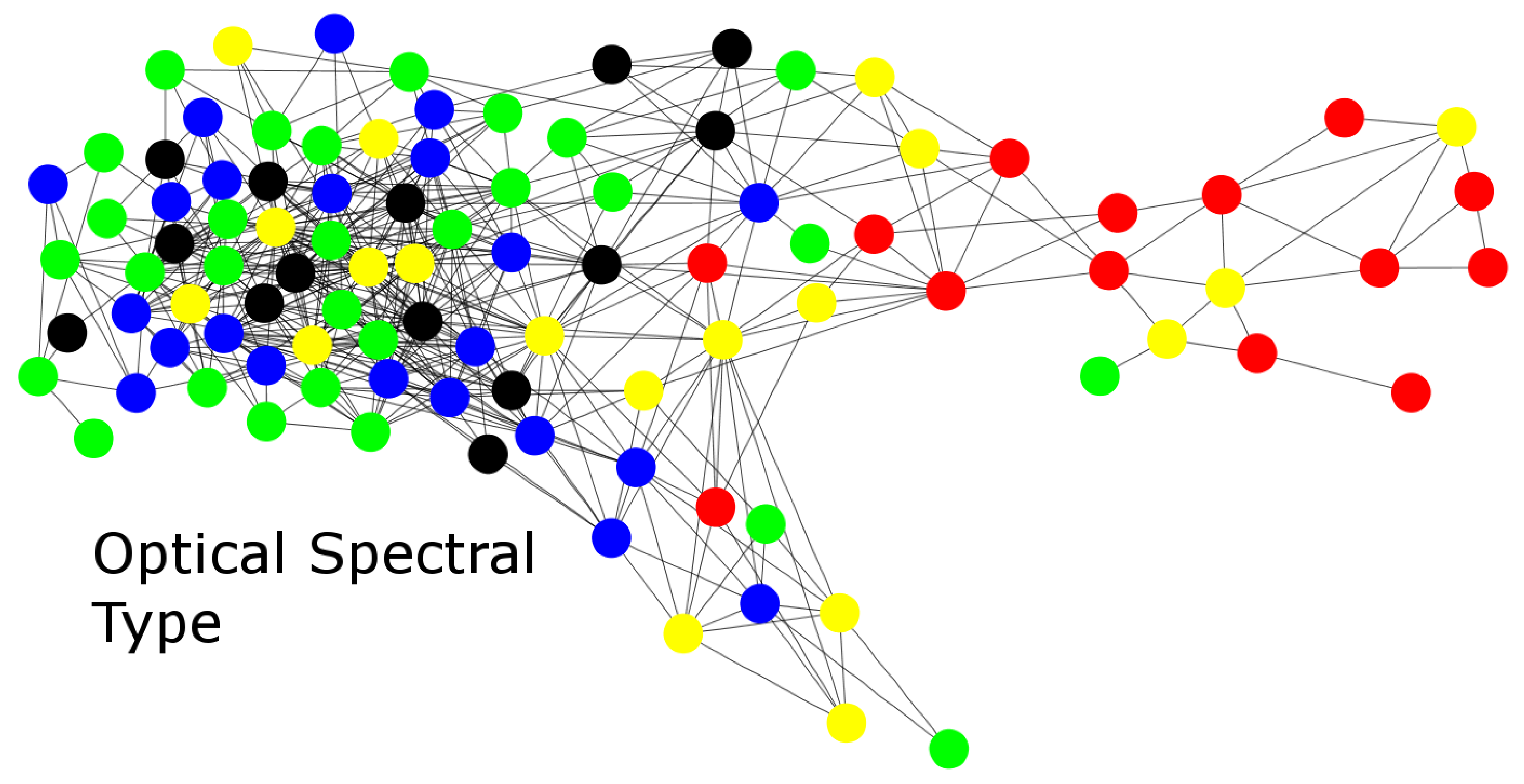}
\caption{\label{ulirglayoutscodec}
Network diagram, with the nodes color-coded by {\it Left Panel:} IR luminosity, and {\it Right Panel:} optical spectral class (black: unknown, blue: HII, green: LINER, yellow: Sy2, red: Sy1)
}
\end{center}
\end{figure*}

\begin{figure*}[tbhp]
\begin{center}
\includegraphics[width=0.49\textwidth,angle=0.0]{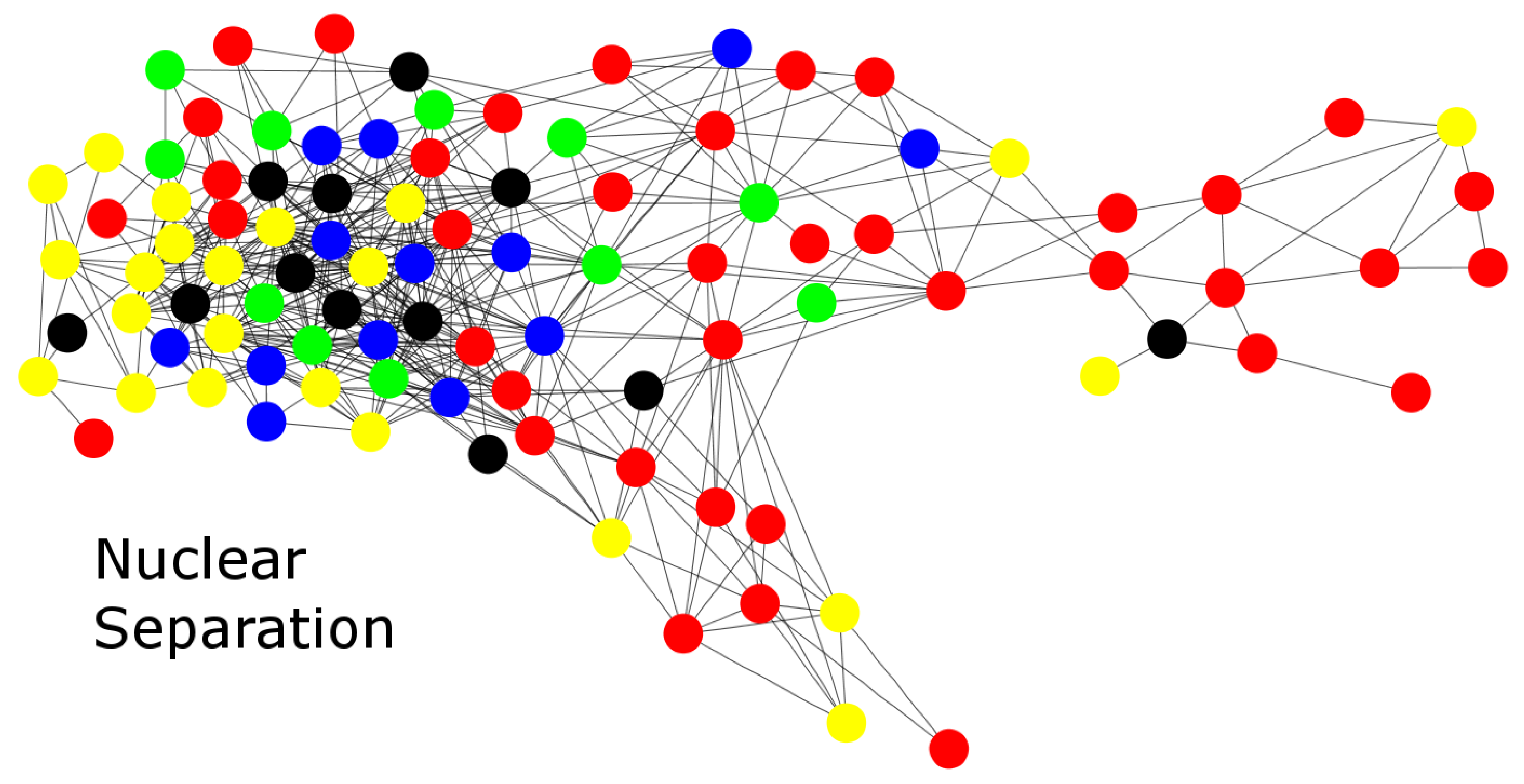}
\includegraphics[width=0.49\textwidth,angle=0.0]{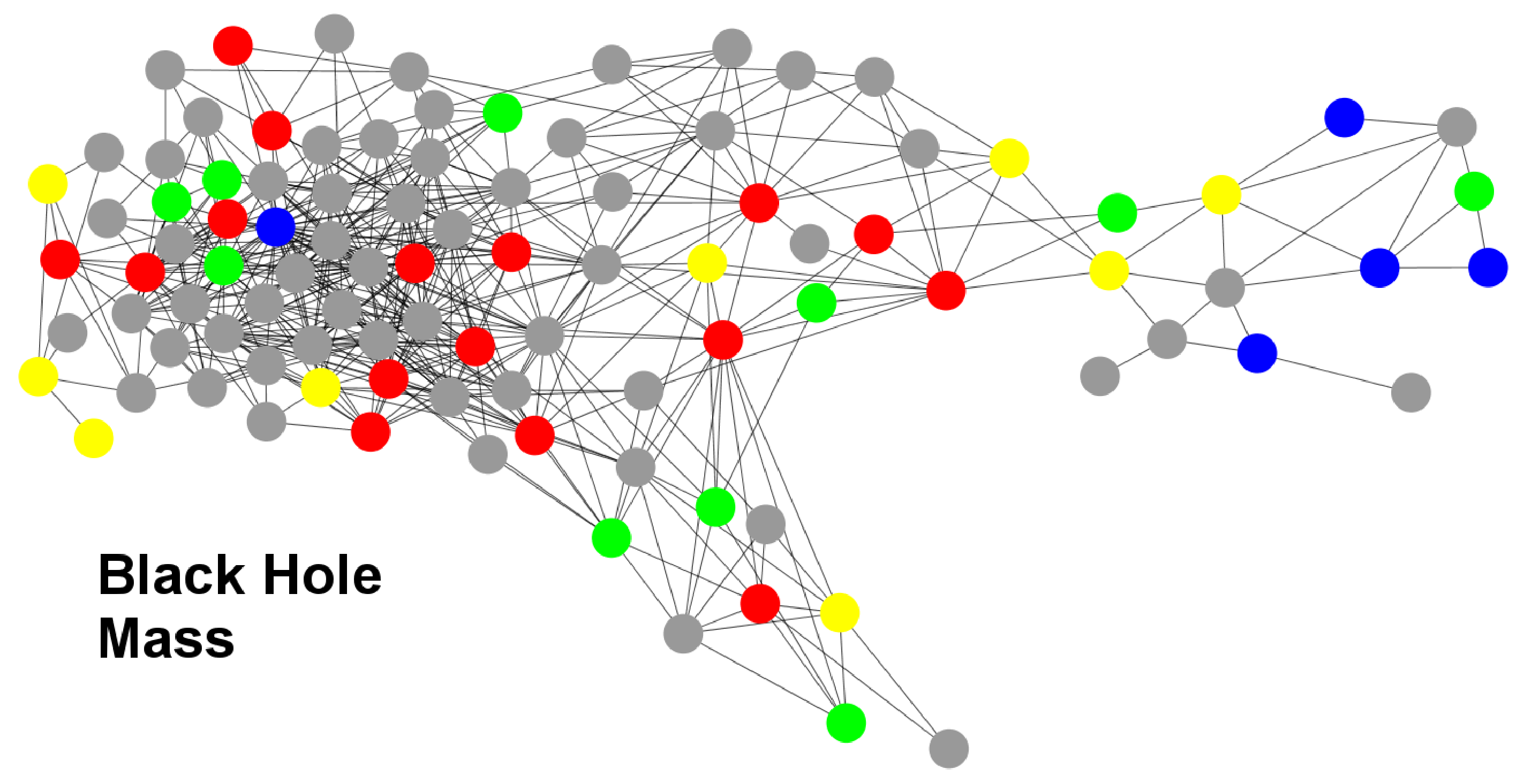}
\caption{\label{ulirglayoutscodeb}
Network diagram, with the nodes color-coded by {\it Left Panel:} projected nuclear separation (black: unknown, blue: $>$12kpc, green: 6-12Kpc, yellow: 0.1-6Kpc, red: single nucleus)  {\it Right Panel:} black hole mass  (black: unknown, blue: $>2.5\times10^{8}$M$_{\odot}$, green: $8.0 < $M$\times10^{7}$M$_{\odot} < 25.0$, yellow: $5.0 < $M$\times10^{7}$M$_{\odot} < 8.0$, red: $<5\times10^{7}$M$_{\odot}$) 
}
\end{center}
\end{figure*}

{\bf Optical Spectral Type}: (Figure \ref{ulirglayoutscodec}) The majority of the objects in group A have HII or LINER optical spectra, with a few Sy2's. This pattern is reversed in group B; most of the objects have Sy2 or Sy1 spectra with a small number of LINERS and HII's, especially towards the right hand side where nearly all the objects are Sy1's. In group C there appear to be approximately equal numbers of all spectral types.

{\bf IR Luminosity}: (Figure \ref{ulirglayoutscodec}) We expect a weakened correlation with 1-1000$\mu$m luminosity, given the large uncertainties caused by the paucity of flux measurements at $>100\mu$m. This expectation appears to be borne out\footnote{A weakening in the correlation should not arise from the marginilization over IR luminosity. Marginilizing over luminosity will remove any dependence on luminosity. However, if luminosity itself depends on (say) spectral shape at long wavelengths then the dependence on luminosity will remain, since we have not margnilized over spectral shape at long wavelengths.}; low and high luminosity systems are found in all three groups in approximately equal numbers, and there are no clear trends. There are no high luminosity systems in group C, though this may be due to the small number of objects in this group.

{\bf Projected Nuclear separation\footnote{Taken from \citealt{rig99,far01,meu01,cui01,bus02,vei02,vei06,bia08} and rescaled to our cosmology where appropriate}}: (Figure \ref{ulirglayoutscodeb}) There are strong caveats in interpreting this network; the imaging is heterogeneous (e.g. ground-based for some, space-based for others), the separations are projected rather than real, premergers that are widely separated can be erroneously identified as single nucleus systems and vice versa, and nuclear separations are degenerate with merger stage \citep{bar92,dub99}. The figure does however show trends. While the single nucleus systems are found in all groups, the majority of them are in group B, including nearly all the objects at the end. Conversely, the widely and moderately separated systems are found almost exclusively in group A, with a few in groups B and C.

{\bf Black Hole Mass\footnote{Taken from \citealt{das06a,das06} and \citealt{kaw07} and rescaled to our cosmology where appropriate}}: (Figure \ref{ulirglayoutscodeb}) Here we use only those BH masses measured via velocity dispersions (see e.g. \citealt{tre02}), but the caveats are even stronger than for the nuclear separations diagram. Only 33 measurements are available, the random uncertainties are large, and the measurements depend critically on the calibration of the $M_{BH} - M_{bulge}$ relation. Most importantly, the quantity that we're really interested in is not $M_{BH}$, but rather the increase in black hole mass during the merger (i.e. $\Delta M_{BH}|_{merger}$). The instantaneous snapshots of black hole mass are of limited use since the black holes in the progenitors can in principle have a wide range of starting values. Therefore, this diagram is of limited interest. The intermediate mass black holes are spread randomly through the diagram, but 4/5 high mass black holes are at the end of group B, and 11/16 low mass black holes are in group A, with the rest lying in the first part of group B, or group C.

\begin{figure*}[tbhp]
\begin{center}
\includegraphics[width=0.49\textwidth,angle=0.0]{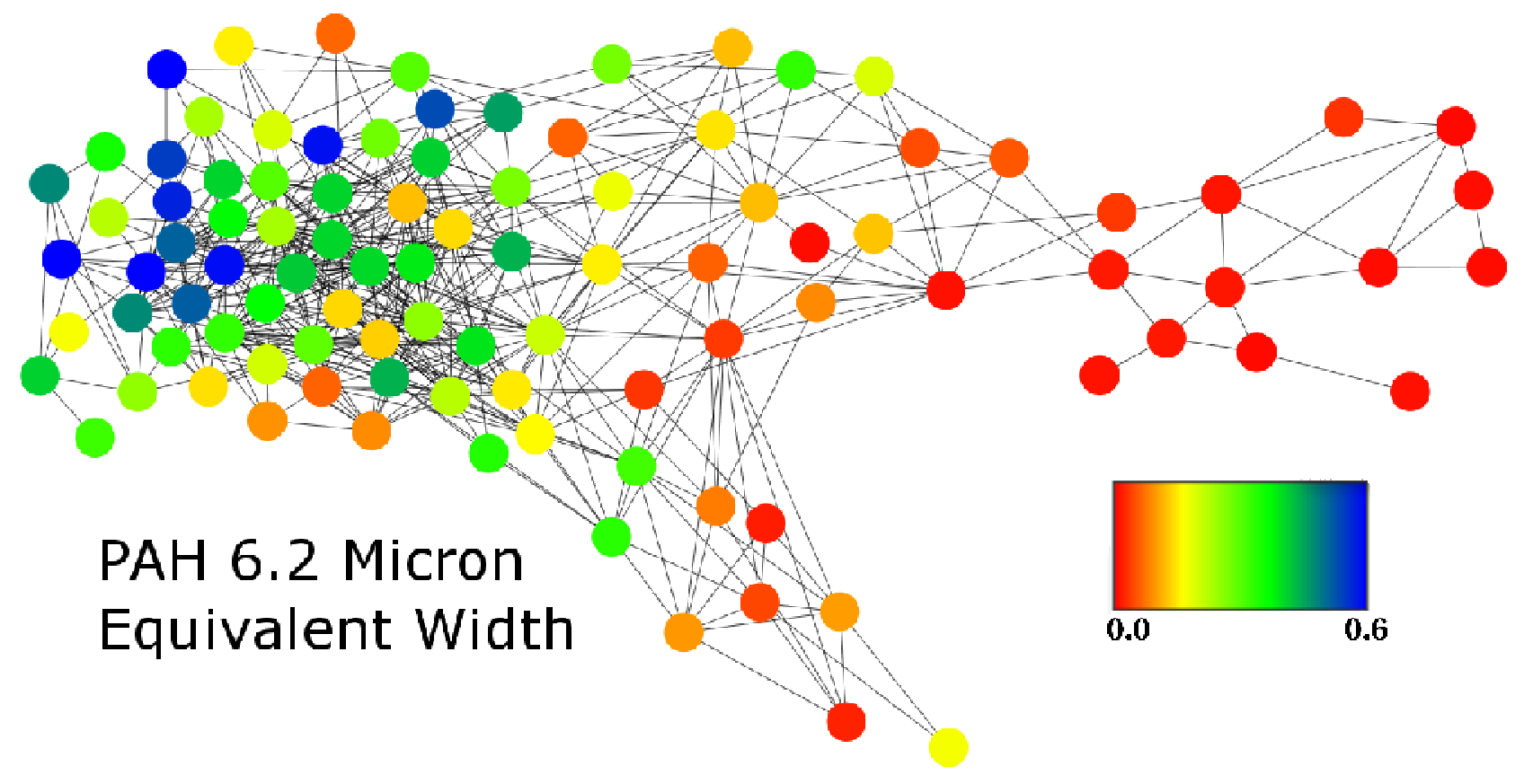}
\includegraphics[width=0.49\textwidth,angle=0.0]{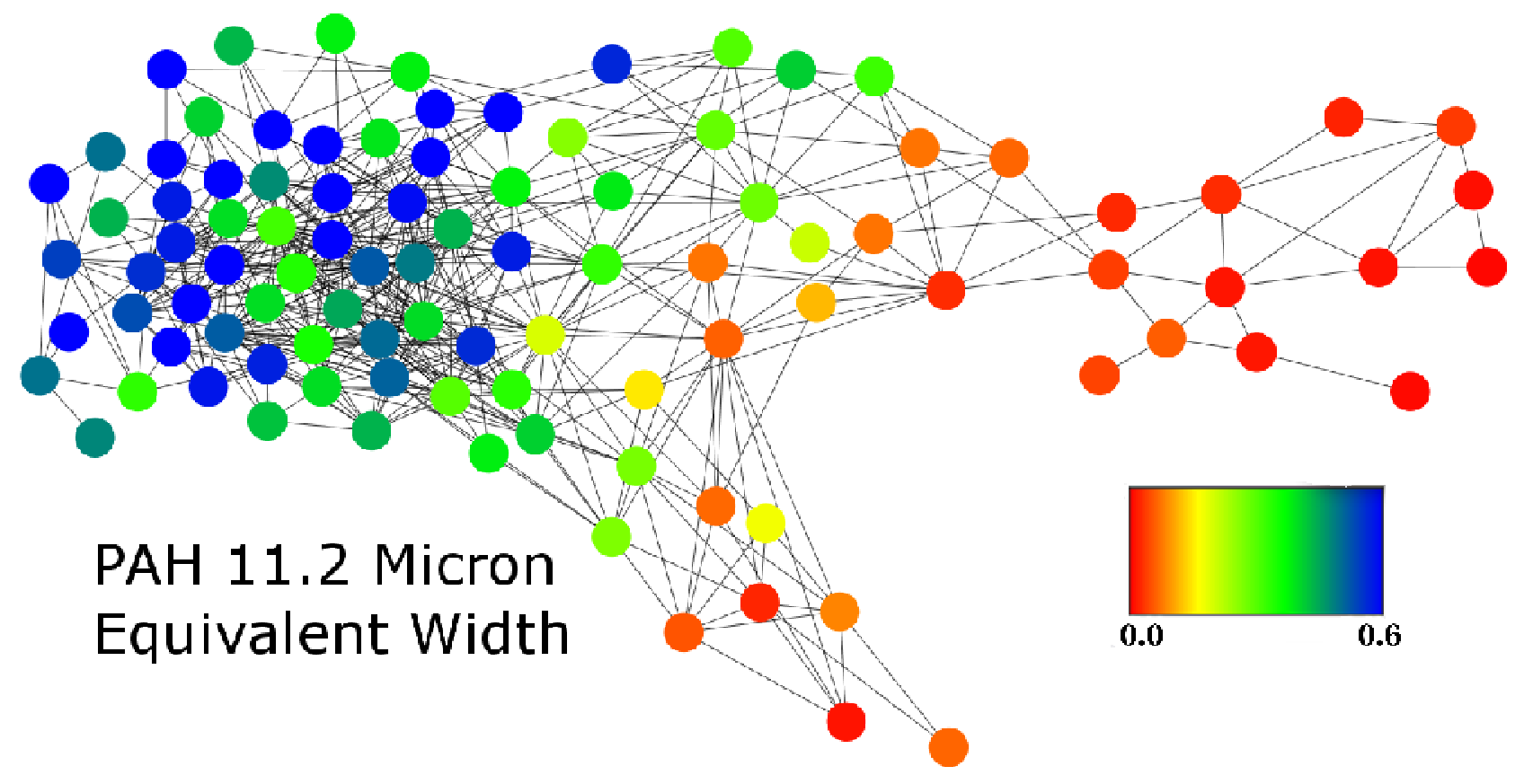}
\caption{\label{ulirglayoutscodee}
Network diagram, with the nodes color-coded by mid-infrared star formation rate indicators. {\it Left Panel:} equivalent width of the  6.2$\mu$m feature. {\it Right Panel:} equivalent width of the 11.2$\mu$m feature. 
}
\end{center}
\end{figure*}

{\bf PAH Equivalent Width}: (Figure \ref{ulirglayoutscodee}) The mid-IR spectra of many ULIRGs show broad emission features at 6.2$\mu$m, 7.7$\mu$m, 8.6$\mu$m, 11.2$\mu$m and 12.7$\mu$m, attributed to bending and stretching modes in neutral and ionized Polycyclic Aromatic Hydrocarbon (PAH) molecules, and it is now accepted that these features indicate ongoing star formation. Therefore, the prominence of PAH features above the continuum, which we quantify via equivalent width, is a crude but reliable measure of the energetic importance of star formation\footnote{As opposed to PAH fluxes, which measure the absolute luminosity of the starburst} (see also \citealt{gen98,rig99,arm06,des07}). As there is still debate over the use of individual PAH features as star formation rate diagnostics, we show networks coded by the 6.2$\mu$m and 11.2$\mu$m PAH features. The 11.2$\mu$m diagram is particularly striking. All the objects in group A have prominent PAHs and there is a high degree of homogeneity in their strengths. The PAHs then decline in prominence as we move left to right through groups B and C, until we reach the right hand side of both groups where the PAHs are negligible. The 6.2$\mu$m diagram shows the same trends, but less obviously; most of the objects in group A have prominent PAHs with some outlying objects showing weakened features, and there is a less pronounced, though still clear decline in PAH strength as we move left to right through groups B and C. 

\begin{figure}[tbhp]
\begin{center}
\includegraphics[width=0.5\textwidth,angle=0.0]{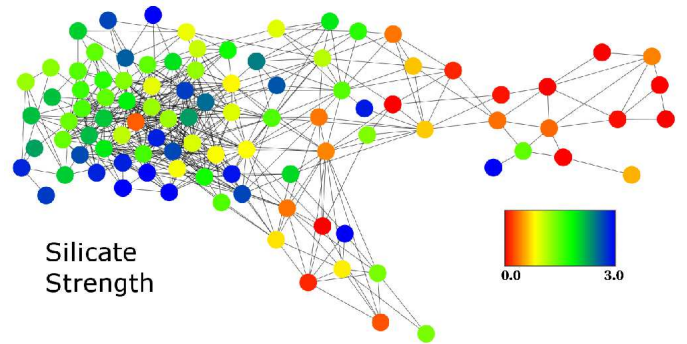}
\caption{\label{ulirglayoutscoded}
Network diagram, with the nodes color-coded by silicate strength.
}
\end{center}
\end{figure}

{\bf Silicate strength}: (Figure \ref{ulirglayoutscoded}, defined in \citealt{spo07} and \citealt{sir08}) The 9.7$\mu$m feature is thought to arise from large silicate dust grains; in absorption when `cold' silicate grains absorb mid-IR continuum emission from a background source, and in emission when the silicate grains are `hot'. It is usually interpreted as a measure of the obscuration towards the central, sub-Kpc nuclear regions. Under this interpretation, a prominent silicate feature is more correlated with AGN activity than star formation - an absorption feature suggests a buried AGN \citep{ima07} and an emission feature suggests an unobscured AGN \citep{hao05}. There is a caveat in interpreting this network though - the silicate feature is broad, substantially more so than a PAH feature, so its contribution to the $\mathcal{R}$ values will be commensurately larger\footnote{As with the IR luminosities, the marginilization over foreground extinction should not affect any correlations with silicate strength}. We cannot reliably gauge the magnitude of this effect, so the conclusions that can be drawn from this diagram are limited. We do however see trends. The silicate strengths of group A are fairly homogeneous; nearly all are moderately to heavily obscured, with (perhaps) slightly higher values for the outliers. Group C and the first part of group B are more varied, with a wide range of silicate strengths. The nodes at the end of group B universally show negligible absorption, or silicates in emission. 

\section{Reliability}\label{sectrelia}
We assess the accuracy and precision, along with possible sources of error, on both the Bayes factors and the network in this section. 

\subsection{The Bayes Factors}\label{subsrelbay}
\begin{figure*}[tbhp]
\begin{center}
\includegraphics[width=0.8\textwidth,angle=90.0]{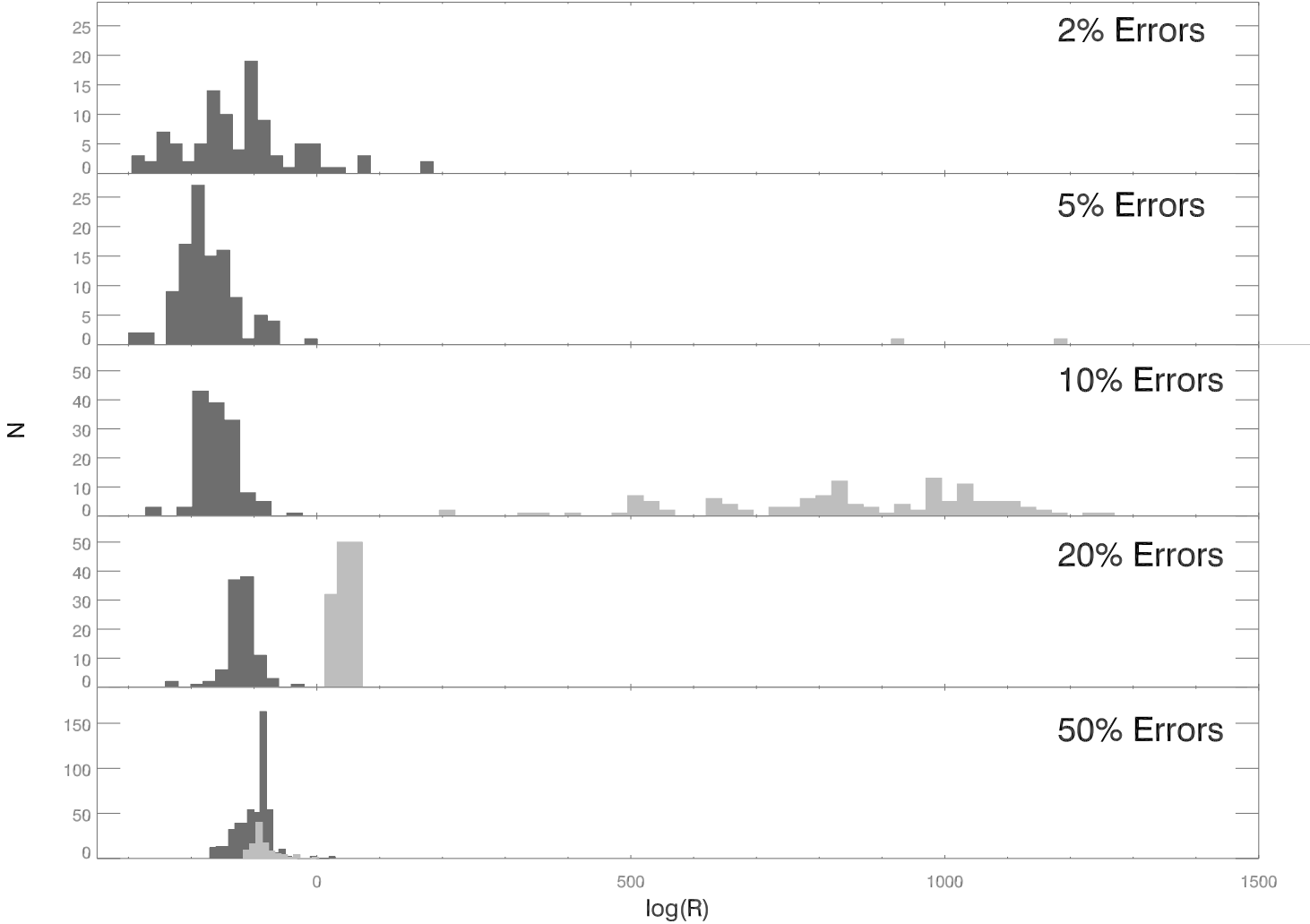}
 \caption{\label{fawkedx}
Histograms of the Bayes factors assuming five different Gaussian bin-to-bin uncertainties. The dark grey histograms are those spectral pairs for which the intrinsic spectra for the two ULIRGs are the same but the dust extinctions are different. The light grey histograms are those ULIRGs for which both the intrinsic spectra and dust extinctions are different. As the uncertainties become smaller, there is a higher probability that $\log_{10}(\mathcal{R})<0$ for those ULIRGs whose intrinsic spectra are the same, and that $\log_{10}(\mathcal{R})>0$ for those ULIRG whose intrinsic spectra are different. Hence, the Bayes factors are behaving as expected. The increased separation between histograms for intrinsically different and intrinsically similar pairs as the errors decrease is somewhat obscured by the limited accuracy/precision of $\mathcal{R}$ from the Monte Carlo integration.
}
\end{center}
\end{figure*}

We use three methods to examine the behavior of the Bayes factors. First, we assess the precision and accuracy of the whole procedure. We compare what happens to the $\log_{10}(\mathcal{R})$ values in two situations; (1) when the input spectra are intrinsically different, and (2) when the input spectra are intrinsically identical but where one has been scaled in luminosity and/or dust extinction. In both situations we vary the bin-to-bin uncertainties from $2\%$ to $50\%$. The results from these simulations are shown in Figure \ref{fawkedx}. This shows that, as the uncertainties become smaller, there is a higher probability that (a) $\log_{10}(\mathcal{R})<0$ for those ULIRGs whose intrinsic spectra are the same, and (b) $\log_{10}(\mathcal{R})>0$ for those ULIRGs whose intrinsic spectra are different. Hence, the Bayes factors are behaving as expected. When the uncertainties are large there is an almost 100\% overlap of those ULIRGs with `same' and `different' intrinsic spectra. However, the distributions are not centered around zero. We have found that when the uncertainties are increased even further ($\sim$10000\%), pairings of same {\it and} different intrinsic spectra have a mean $\log_{10}(\mathcal{R})$ closer to 1, indicating this effect is primarily due to an increasing lack of accuracy in the integration as the uncertainties become smaller and the parameter space becomes more sparse.

\begin{figure}[tbhp]
\begin{center}
\includegraphics[width=0.35\textwidth,angle=90.0]{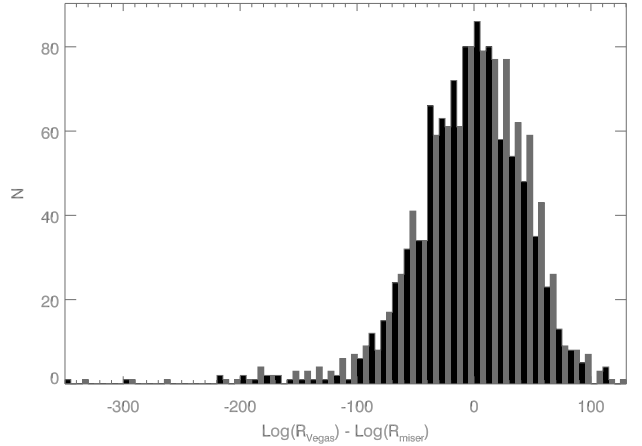}
 \caption{\label{fawked}
A comparison of the log$(\mathcal{R})$ values computed using two Monte-Carlo algorithms; $miser$ (which we use in our analysis), and VEGAS.
}
\end{center}
\end{figure}

Second, we assess the accuracy and precision of the integration algorithm used to compute the Bayes factors. We used the $miser$ algorithm, which is (to our knowledge) the most robust available \citep{numerical}, but as it is a Monte Carlo algorithm, it has uncertainties associated with it which are difficult to compute\footnote{Usually, one can approximate how close one is to the ``true'' integral through the variance of the integrand. However, in our case, the errors are $\sim 100\%$ because the parameter space is both sparse and dynamic. Further, it is known {\it a priori} that the integrand can never be less than 0.} . Therefore, we adopt a conservative approach. We compare in Figure \ref{fawked} the $\log_{10} (\mathcal{R})$ values obtained with the $miser$ algorithm, and an alternative algorithm, VEGAS. The spread is large, with a 1$\sigma$ dispersion on $\log_{10} (\mathcal{R})$ of $\sim50$. The actual uncertainties due to the limited accuracy/precision of the Monte Carlo integration are, however, much lower than this, as the VEGAS algorithm is less robust than $miser$. Furthermore, the effect on Figure \ref{ulirglayout} of uncertainties on individual $\mathcal{R}$ values is likely to be small, as we simply demand that $\log_{10} (\mathcal{R}) < 0$. 

Finally, we examine whether small regions of the spectra dominate the derived $\mathcal{R}$ values. This is important for assessing the reliability of some of the coded variants of Figure \ref{ulirglayout}; the PAH and silicate strength plots repeat (some) information already in the $\mathcal{R}$ values, and so trends may be artifically amplified. We selected 200 pairs at random, and removed from each spectrum a 0.7$\mu$m wide region centered on the 11.2$\mu$m PAH feature. While there was some variation in the derived $\mathcal{R}$ values, we found that, in 97\% of the cases, $\log_{10} (\mathcal{R})$ did not change sign. We repeated this test for 0.7$\mu$m wide regions centered on the 7.7$\mu$m PAH feature and in the continuum at 15$\mu$m, and found similar results. We conclude that, while spectral windows of width $\sim1\mu$m do contribute to the $\mathcal{R}$ values, they do not dominate them. Therefore, we argue that if a spectral feature spans a small wavelength range, and contains information on a specific physical property, then coding Figure \ref{ulirglayout} by that feature tells us how that property varies with network position, while introducing minimal contamination.

\subsection{The Network}\label{networkrobust}
In this section, we assess four possible sources of error in the methods used to create Figure \ref{ulirglayout}. 

\begin{figure*}[tbhp]
\begin{center}
\includegraphics[width=0.45\textwidth,angle=0.0]{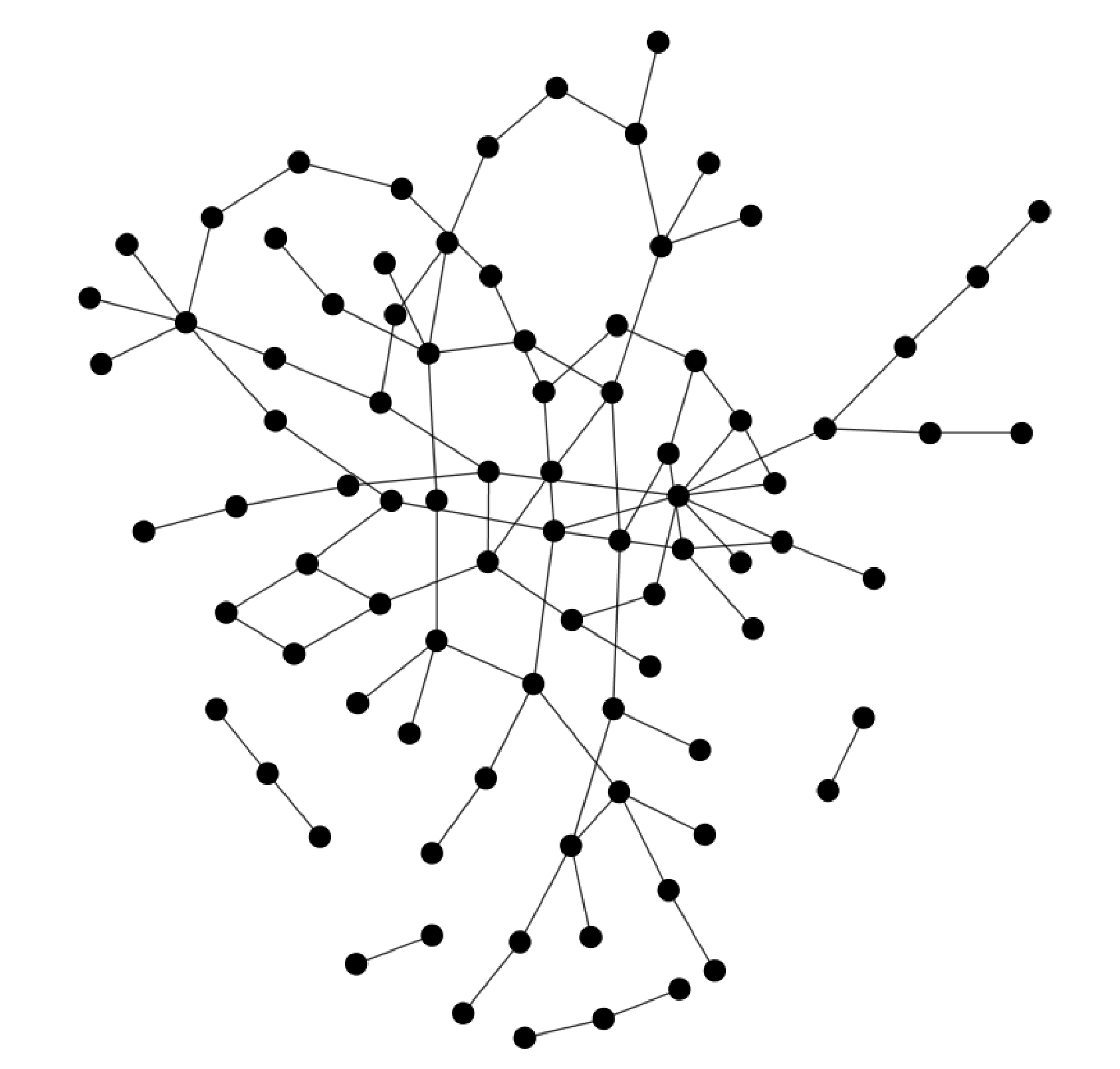}
\includegraphics[width=0.45\textwidth,angle=0.0]{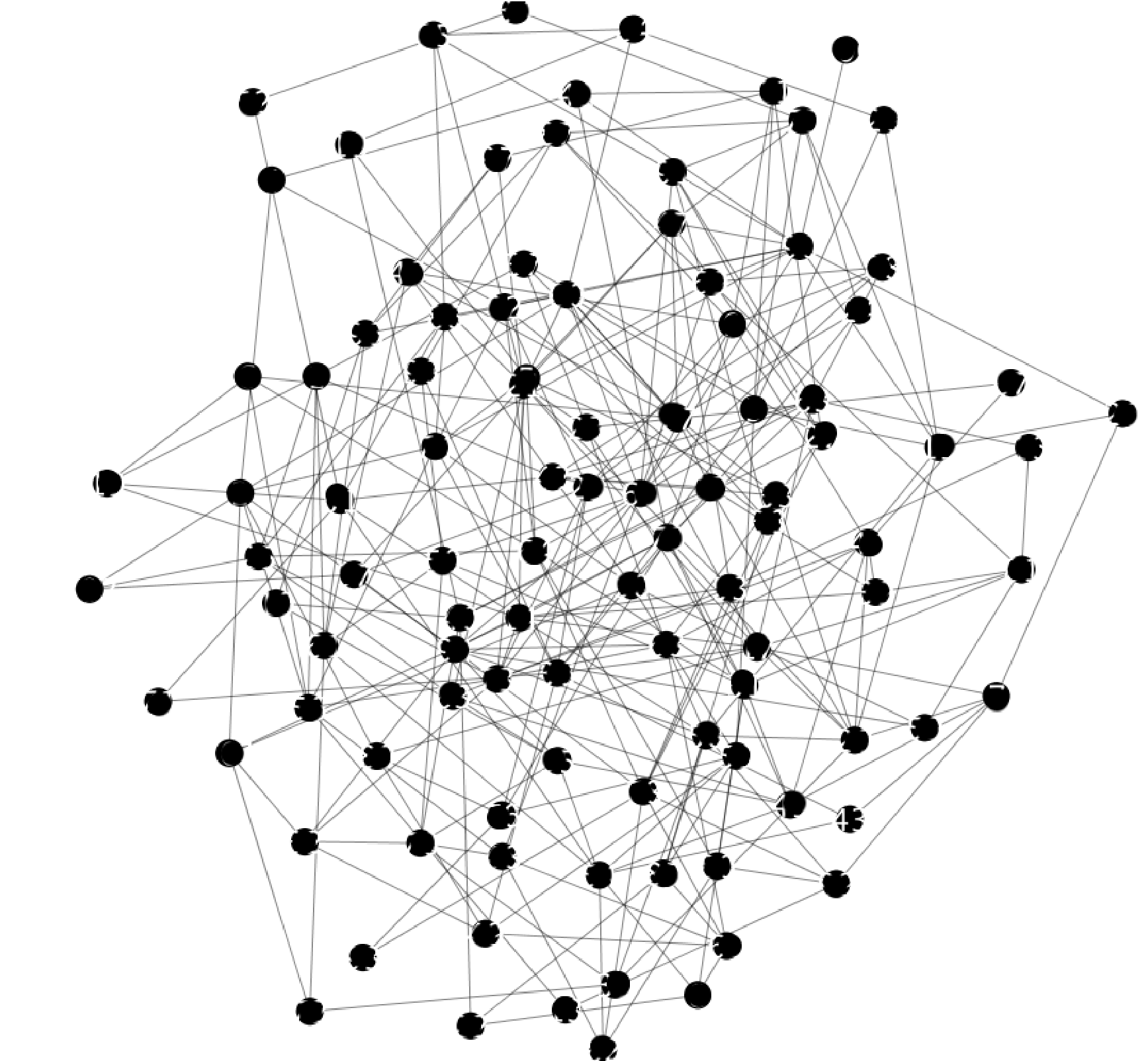}
\includegraphics[width=0.45\textwidth,angle=0.0]{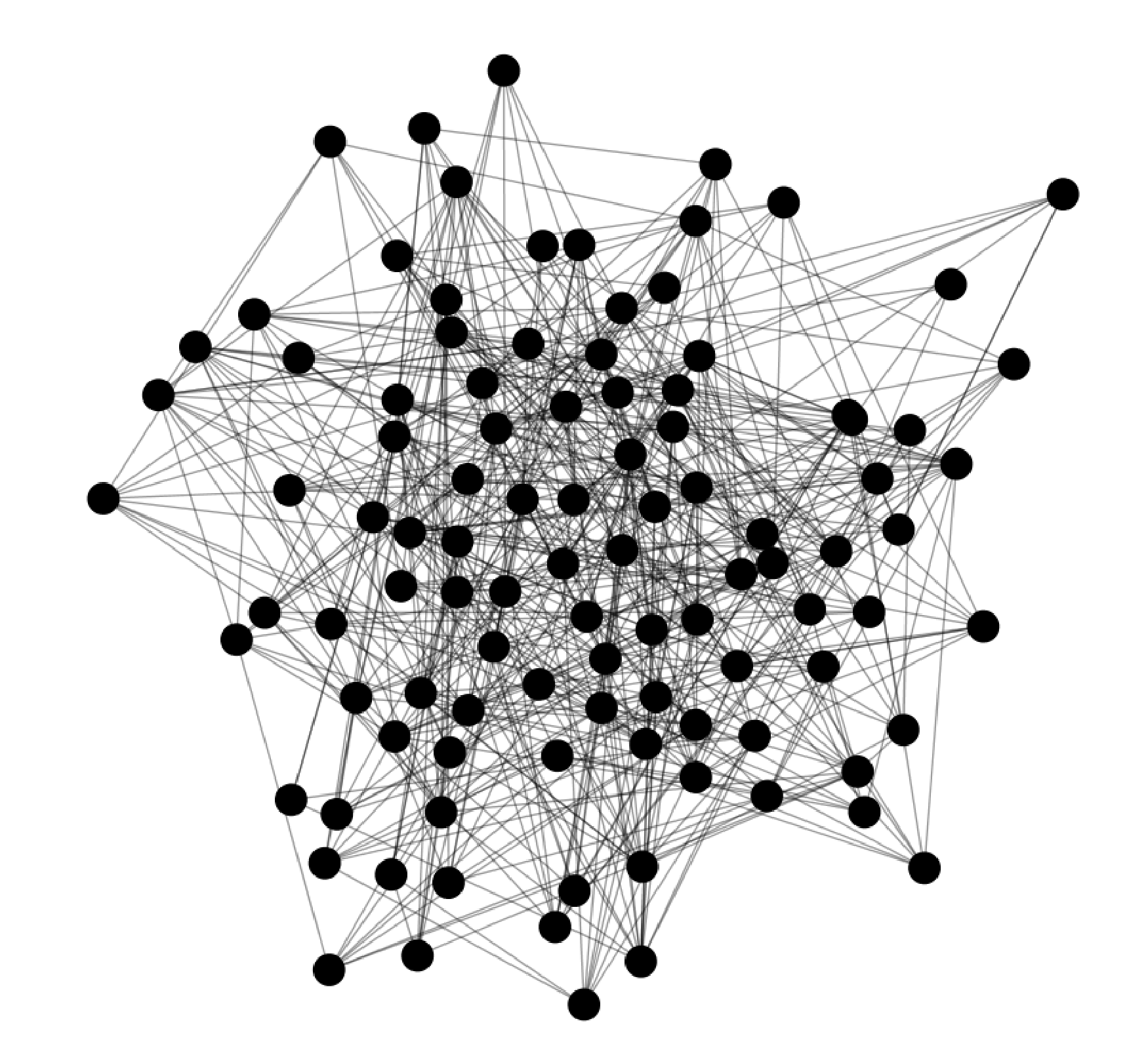}
\includegraphics[width=0.45\textwidth,angle=0.0]{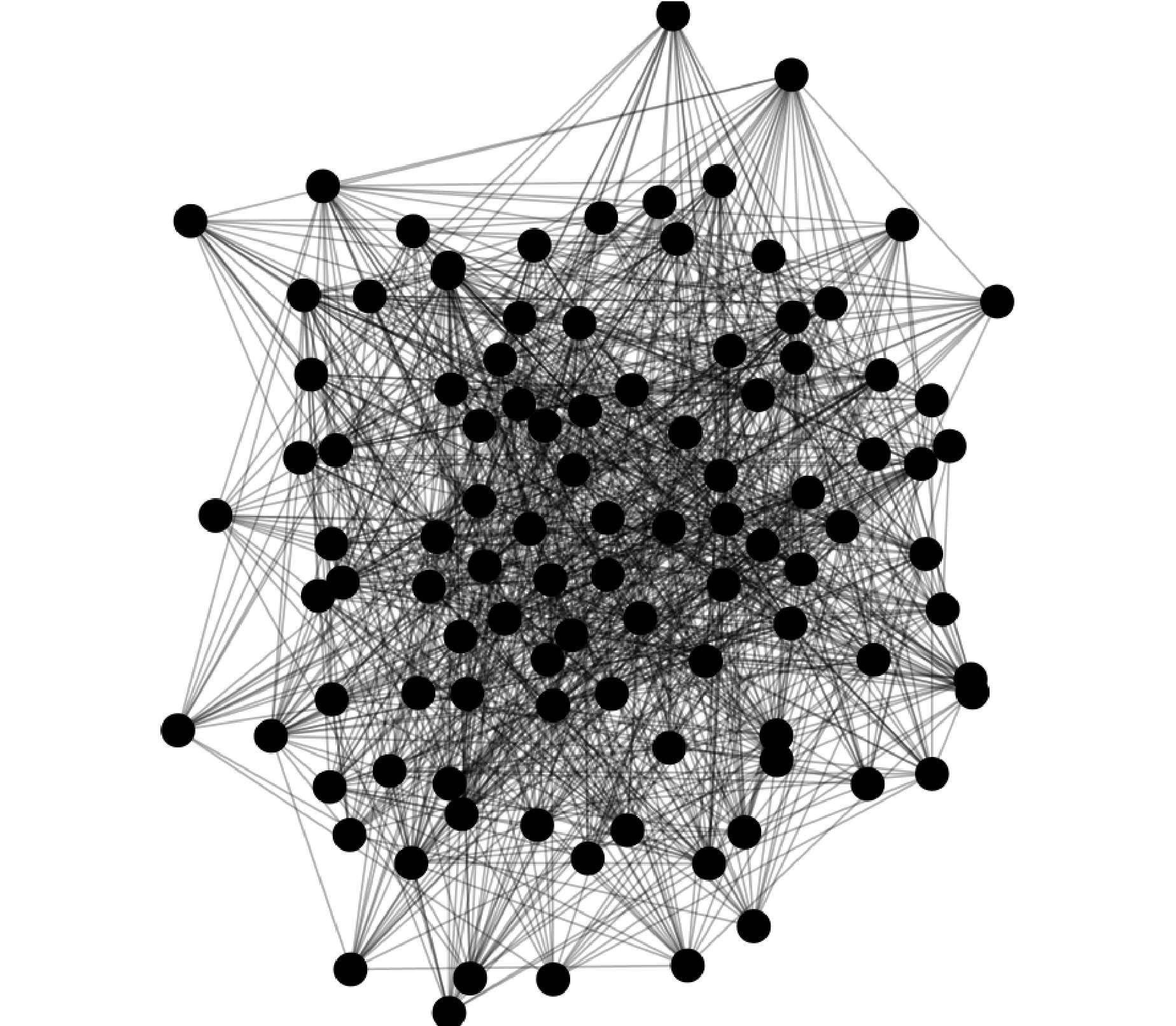}
 \caption{\label{randoms}
102 node random undirected graphs, with probabilities of a connection of 1\% 5\% 10\% and 25\% respectively. Only nodes with at least one edge are plotted. Qualitatively, these networks do not resemble the network in Figure \ref{ulirglayout}. 
}
\end{center}
\end{figure*}

\begin{figure}[tbph]
\begin{center}
\includegraphics[width=0.35\textwidth,angle=90.0]{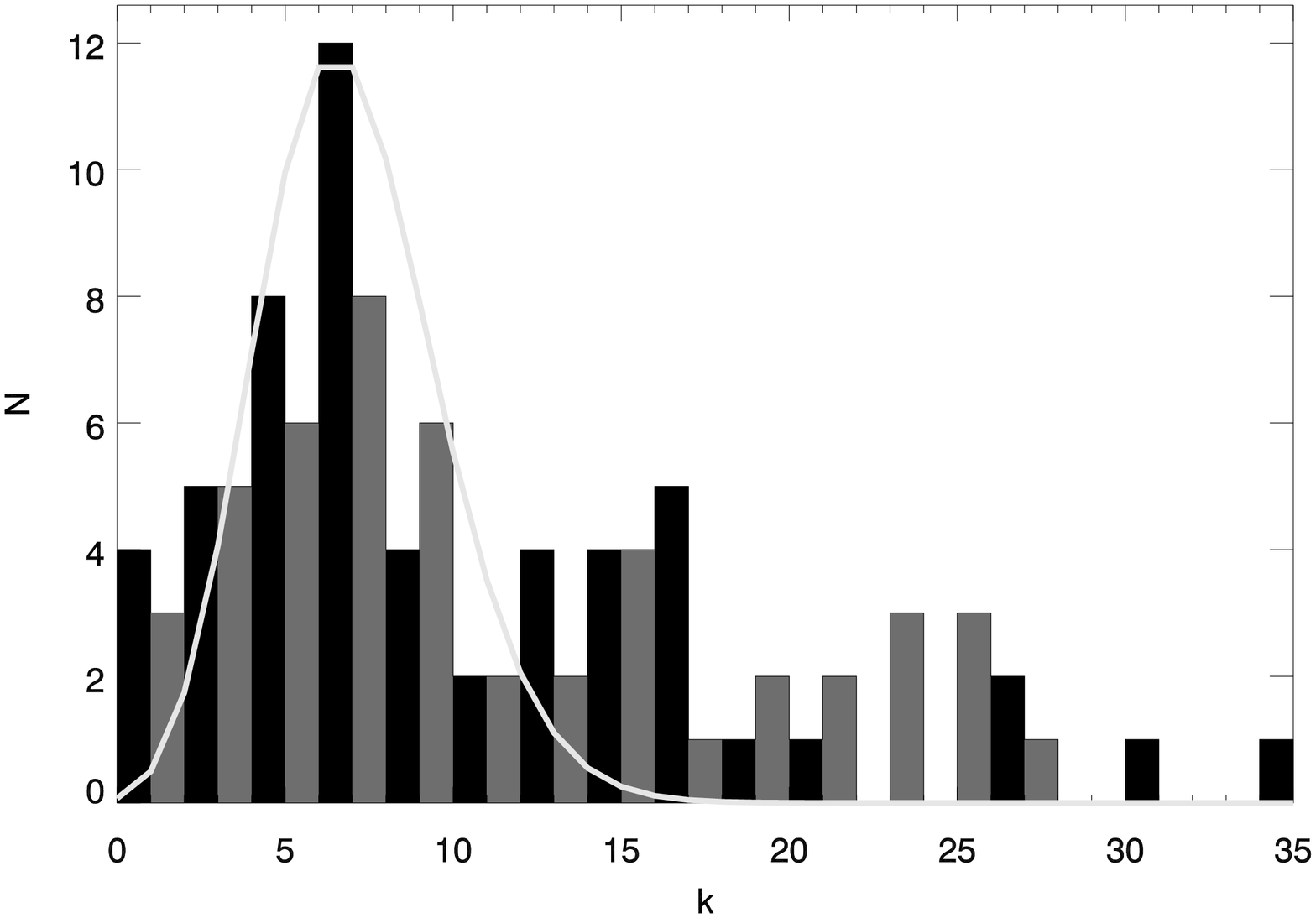}
 \caption{\label{degdisthist} 
The degree distribution of Figure \ref{ulirglayout}, i.e. the histogram of number of nodes with a given number of connections, $k$. The light grey line is a (scaled) Poisson distribution with a mean of 7.  Our degree distribution is clearly not Poissonian, and so does not follow the degree distribution of a random graph. 
}
\end{center}
\end{figure}

First, Figure \ref{ulirglayout} could simply be randomly connected points, and not contain any `real' structure. We assess this in two ways. Qualitatively, Figure \ref{ulirglayout} does not resemble a random network with 102 nodes, examples of which are shown in Figure \ref{randoms}. Quantitatively, it has been shown that randomly connected graphs have a Poissonian degree distribution \citep{erd}. The degree distribution for Figure \ref{ulirglayout} is shown in Figure \ref{degdisthist}. It is not well matched to a Poisson distribution. 

Second is the effect on Figure \ref{ulirglayout} from the limited precision of the Bayes factors. The variation in the $\log_{10}(\mathcal{R})$ values due to the Monte-Carlo integration is difficult to determine, though we showed in \S\ref{subsrelbay} that the upper limit is $\simeq50$. So, to assess this, we show in the top panel of Figure \ref{ulirglayouttests} the layout obtained after randomly changing all the $\log_{10} (\mathcal{R})$ values by $\pm$50. Even when using the upper limit on the errors, we get essentially the same structure. We conclude that Figure \ref{ulirglayout} is robust to within the accuracy and precision of the $\mathcal{R}$ values.

Third, one could argue that the structure of Figure \ref{ulirglayout} is an artifact of the priors, and does not reflect trends in the data. To examine this, we test the sensitivity of Figure \ref{ulirglayout} to the maximum allowed cold foreground dust extinction, which is the only prior we adopt. Figure \ref{ulirglayout} assumes a weak upper limit on the cold foreground extinction of A$_{9.7\mu m} \simeq 50$ (see \S\ref{appextinct}). In the lower panel of Figure \ref{ulirglayouttests} we show the diagram obtained if this limit is relaxed further, to A$_{9.7\mu m} \simeq 80$. It closely resembles Figure \ref{ulirglayout}, though there are differences in the number of edges connecting some nodes. As we are using a Bayesian approach, the relaxation of the constraint on A$_{9.7\mu m}$ can cause the $\mathcal{R}$ values to rise or fall; for example, IRAS 08572+3915 is now not connected to any other node and is not plotted. Overall, we conclude that changing the prior on A$_{9.7\mu m}$ does not substantially affect the structure of Figure \ref{ulirglayout}.

Fourth, it is possible that the algorithm used to generate the network is a source of error. The methods described in \S\ref{sssectnet} start each network from a random `seed' position, and assume parameters for the attractive and repulsive forces. Therefore, the final appearance of a network will differ from case to case, and it may be that the seed position or the force parameters are governing the final appearance, not the input data. To test this, we checked to see if different seed positions or different force parameters gave networks without the structure seen in Figure \ref{ulirglayout} and found they did not. To illustrate, we show in Figure \ref{rawcyto} an example of a `raw' network for our data, made using the default parameters for the spring-embedded algorithm within Cytoscape. The same structures can be seen in Figures \ref{ulirglayout} and \ref{rawcyto}, the only difference being that the nodes in Figure \ref{rawcyto} are too small to identify by number. We conclude that our network is at least reasonably robust to the choice of parameters of the algorithm. This test illustrates a common problem with this type of analysis; the `raw' networks are not always amenable to visual interpretation. This is a particular problem in our case; we want all the nodes to be individually identifiable, but the node labels in the raw networks are invariably too small to see. Therefore, to arrive at Figure \ref{ulirglayout}, we adjusted node/edge properties such that the information in the network was preserved, but in which the individual nodes can be identified. 

It is difficult to be certain of the robustness of Figure \ref{ulirglayout} as this technique has (to our knowledge) not been previously used to examine any astronomical dataset, and so there exists no previous study for comparison. We have however comprehensively tested the robustness of Figure \ref{ulirglayout} and not found any significant problems. We conclude that this possibility is remote, and proceed on the assumption that the diagram reflects real trends in the data.

\section{Discussion}\label{discussstuff}
Figure \ref{ulirglayout} highlights intrinsic similarities in the mid-IR spectra of low redshift ULIRGs. It places little emphasis on individual spectral features (see \S\ref{subsrelbay}), and we have marginalized over IR luminosity, foreground cold dust extinction and detector noise. Furthermore, as our sample selection is almost unbiased (see \S\ref{lesample}), each node is a random snapshot of the ULIRG population. In this section, we explore the conclusions that can be drawn from this diagram and its coded variants. In so doing, we assume that the IRS spectra are a product, on average, of the power sources governing the {\it total} IR emission. This assumption is reasonable for the population as a whole, but may break down for individual objects.

\subsection{The Network}\label{evolp}
The first conclusion we draw is that, while graph-theory based approaches are a powerful tool for visualizing complex and heterogeneous datasets, they require a large number of nodes. Our study has 102 nodes, and yet the conclusions we can draw from Figure \ref{ulirglayout} alone are limited to the subdivision of the sample into two or three groups. Solely from Figure \ref{ulirglayout} we cannot say if groups A, B and C are phases in time, or phases in some other variable. Even if we assume they are temporal phases, then it is impossible to say what the time-ordering of the groups is. If we increased the number of nodes by an order of magnitude, then the resulting increase in resolution of the network may lead to more insight, but we cannot test this hypothesis here. 

Turning to the coded variants of Figure \ref{ulirglayout}, it is clear that we still lack a complete and homogeneous dataset for $z<0.4$ ULIRGs. Several nodes lack morphological and/or optical spectral classifications, and most do not have a dynamical estimate of central black hole mass. This omission is serious, given that local ULIRGs are the most easily accessible templates we have for understanding the high redshift ULIRG population.

We now examine the implications from Figure \ref{ulirglayout} and its coded variants for the ULIRG population. First, we examine the drivers behind the division of the network into groups A, B and C. Three important lines of evidence are; (a) the presence of nearly all the widely and moderately separated systems in group A, (b) the fraction of single nucleus systems in groups B/C is much higher than in group A, and (c) the generally lower black hole masses in group A compared to group B, though the number of black hole mass measurements is too small for this to have much weight. We therefore propose that groups A and B/C represent temporal phases and are not significantly determined by other factors. 

Next, we examine the likelihood of group C being a separate entity to groups A and B. Group C is unlikely to be outliers to group A as its nuclear separations and PAH strengths are different from the outliers on the left hand side of group A. Instead, it is similar to the first `half' of group B. Based on the lack of connections between groups B and C, we {\it tentatively} propose that group C is a distinct stage from groups A and B, but stress that this is not robust. For example, it is conceivable that a source could move from group A to group C and then `jump' to group B, although the lack of a bridge node connecting C to B means such a jump phase is likely short; as we have $\sim 100$ nodes, the lack of a bridge node suggests a jump phase length of order 2\% or less of the total ULIRG lifetime\footnote{If we assume a total ULIRG lifetime of $10^8$ years then the jump phase would be $\lesssim 2\times10^{6}$ years long. This is short but feasible; for example, some Wolf-Rayet stars are expected to live approximately this long.}. 

Overall, we propose that groups A, B, and C are distinct but overlapping evolutionary phases, with A occurring first, followed by B and/or C. If a merger remains a ULIRG for most of the duration of the merger, then we can also estimate timescales based on the number of objects in each group; phase A lasts just over half the lifetime of a ULIRG, and phases B and C last approximately half and one third the duration of group A, respectively. We do {\it not} claim that a ULIRG starts at the left end of A and goes gradually to the right. Instead, we propose that ULIRGs start in group A, with position and intragroup movement determined by unknown factors, and then proceed to B or C. We also note that the tests of reliability in \S\ref{networkrobust} involving randomization of the $\mathcal{R}$ values and varying the extinction still gave a clear subdivision of the network into groups A and B/C, but groups B and C were somewhat `blended' and less distinct. We conclude that the division of the network into `A vs. B/C' is more robust than the division `A vs. B vs. C'. 

If the network structure is governed by temporal evolution, we can use the purely network based metric of Betweenness Centrality to make testable predictions. The Betweenness Centrality (which we term $\mathcal{B}$) of a node is the number of shortest paths between other pairs of nodes that pass through that node \citep{fre79,bra01}. A low $\mathcal{B}$ means the node is inconsequential, while a high $\mathcal{B}$ means the node is an important junction. A suitable analogy would be airports; a regional airport would have a low $\mathcal{B}$, while an international hub would have a high $\mathcal{B}$. It is therefore plausible that a node in the network with a high $\mathcal{B}$ is an archetype of a `transitional' phase that many ULIRGs pass through. Most of the nodes in Figure \ref{ulirglayout} have $\mathcal{B}$ values in the range $100< \mathcal{B}< 400$. Fifteen nodes have $\mathcal{B}$ values in the range $400< \mathcal{B}< 800$, while the four unconnected nodes have $\mathcal{B}=0$. Six nodes however have what appear to be unusually high $\mathcal{B}$ values; Mrk 231 ($\mathcal{B}=1850$), IRAS 00275-2859 (1620), IRAS 03538-6432 (1420), IRAS 05189-2524 (1260), and IRAS 14348-1447 and Mrk 273 (both $\sim$1000). We propose that these six objects are examples of key evolutionary phases. Based on their positions in Figure \ref{ulirglayout}, we speculate that Mrk 273 and IRAS 14348-1447 are templates of group A objects, IRAS 03538-6432 is a template of an object transitioning from phase A to phase B, IRAS 05189-2524 is a classic example of an object on the boundary between groups B and C, and Mrk 231 and IRAS 00275-2859 are prime early B to late B type objects.

\begin{figure*}[tbhp]
\begin{center}
\includegraphics[width=0.75\textwidth,angle=0.0]{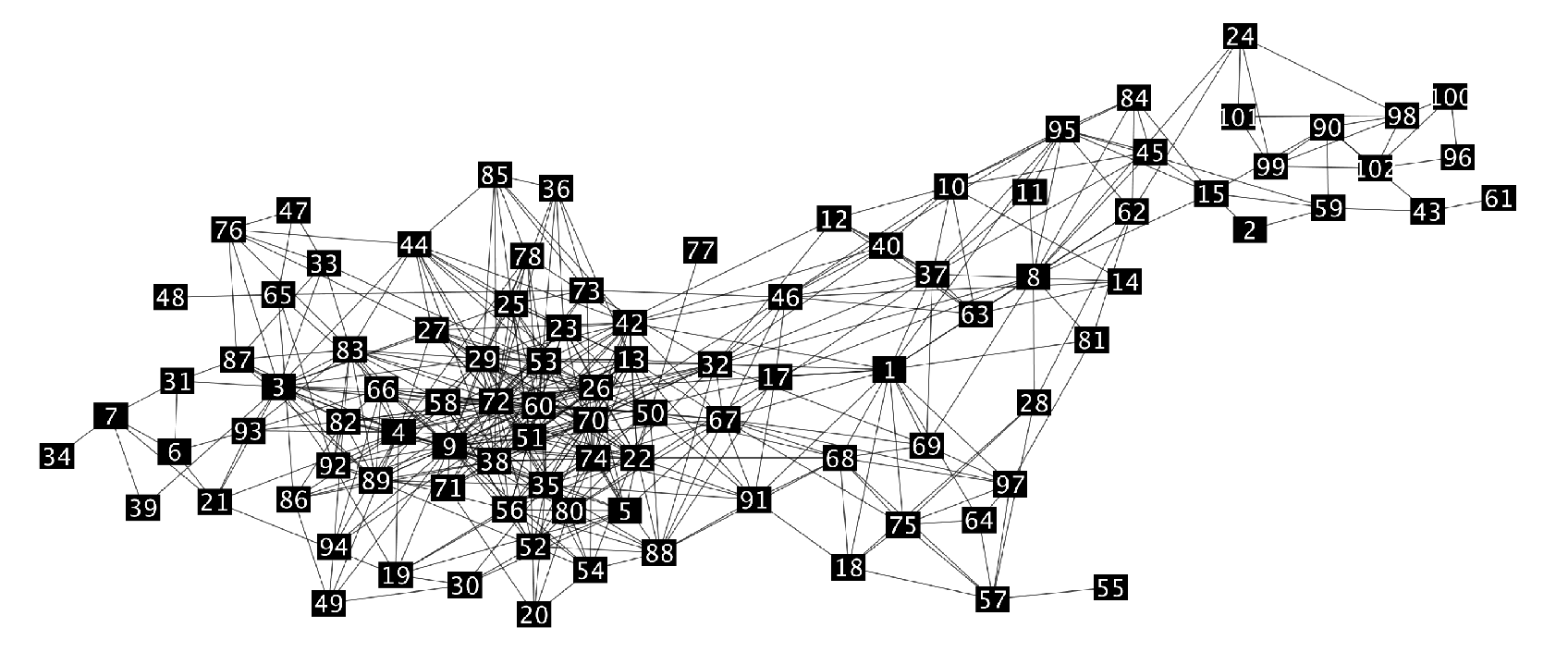}
\includegraphics[width=0.75\textwidth,angle=0.0]{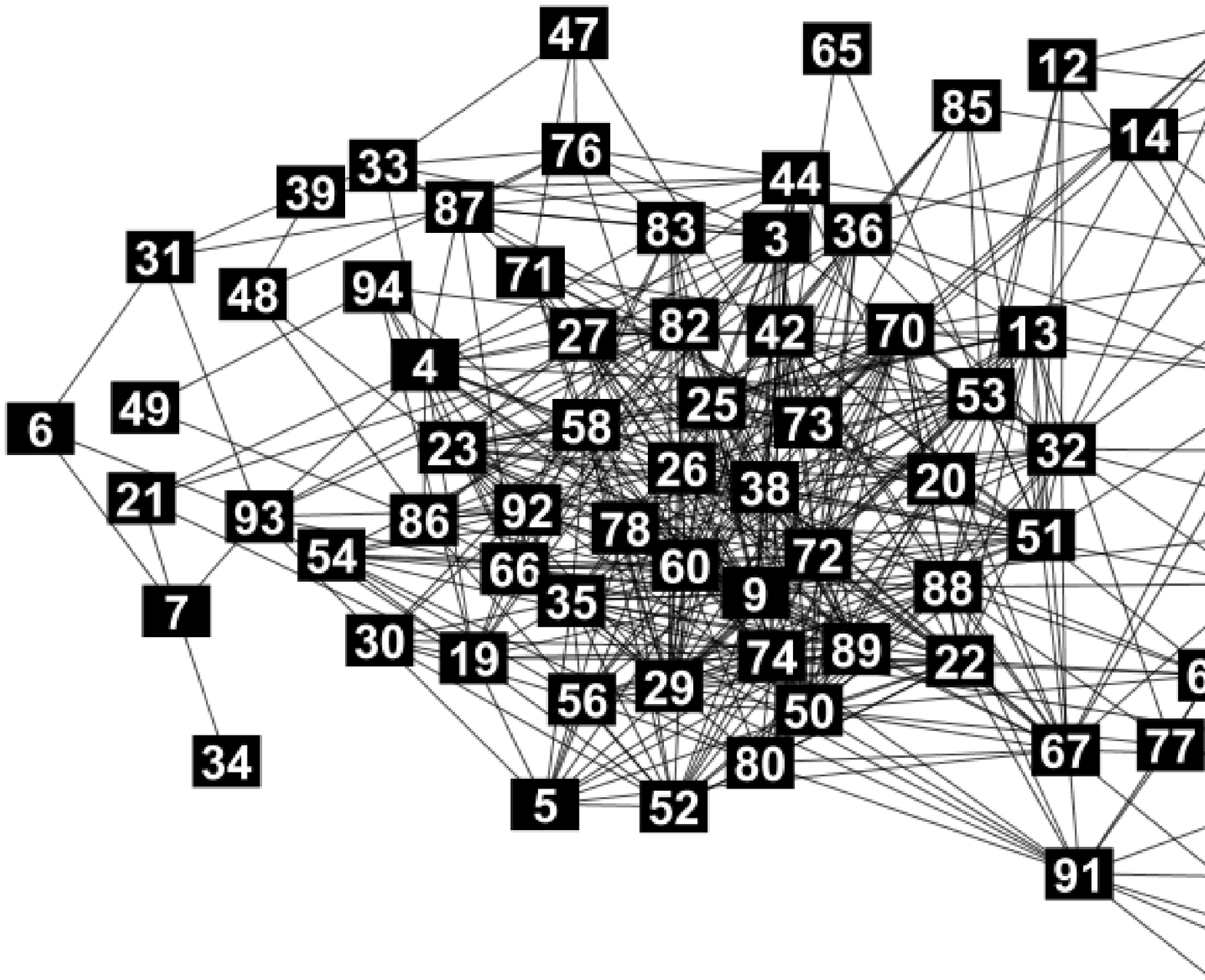}
 \caption{\label{ulirglayouttests}
Tests of robustness. The top panel shows the network diagram obtained if all 5151 $\mathcal{R}$ values are randomly changed by $\pm50$. The bottom panel shows the network diagram obtained if the $\mathcal{R}$ values are computed with a relaxed limit on the maximum foreground extinction of A$_{9.7\mu m} \simeq 80$, instead of A$_{9.7\mu m} \simeq 50$.
}
\end{center}
\end{figure*}

\begin{figure}[tbhp]
\begin{center}
\includegraphics[width=0.45\textwidth,angle=0.0]{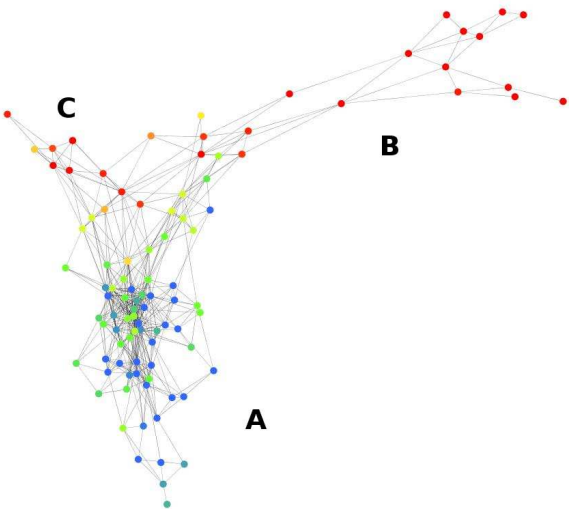}
 \caption{\label{rawcyto} 
 An example of a `raw' Network diagram. This figure was made using the same procedures as for Figure \ref{ulirglayout}, but the positions of the nodes have not been subsequently adjusted. The nodes are coded by  11.2$\mu$m Equivalent Width, as in Figure \ref{ulirglayoutscodee}
}
\end{center}
\end{figure}

\begin{figure*}[tbhp]
\begin{center}
\includegraphics[width=0.8\textwidth,angle=0.0]{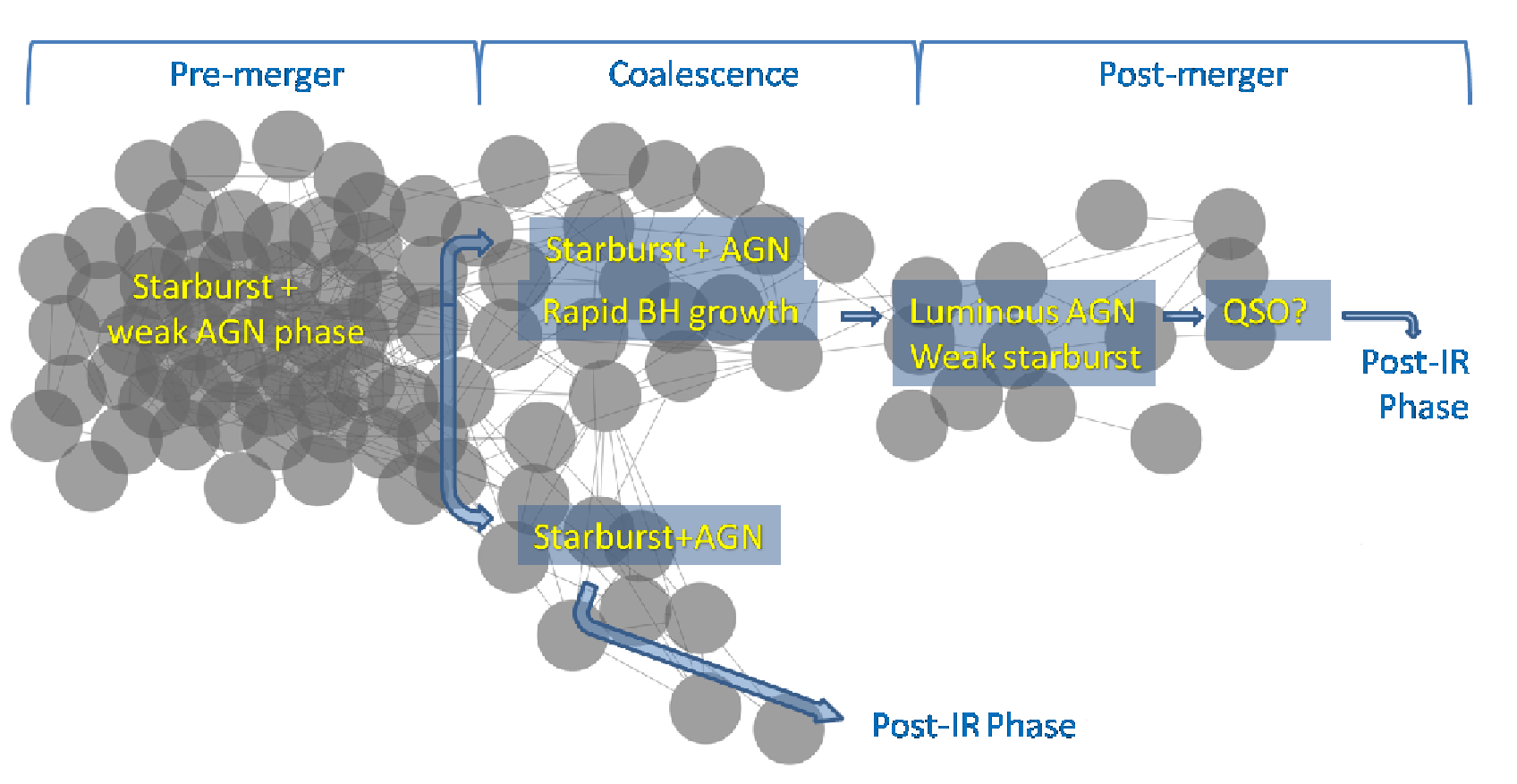}
 \caption{\label{ulirglayoutcart}
Schematic of the evolutionary scheme described in \S\ref{evolp}.
}
\end{center}
\end{figure*}

\subsection{The Groups}\label{evolpg}
We now turn to the properties of groups A, B and C. The duration of group A is hard to quantify, but a reasonable estimate would be from the start of the merger to around the time the progenitor galaxies physically coalesce. This is based on the broad range of projected nuclear separations in this group, from widely separated to single nucleus. The almost universally prominent PAH features suggest the IR emission is powered mainly by star formation, though this does not preclude the presence of a luminous AGN. 

Group A is also highly interconnected, and, as described in \S\ref{discstruct}, is assortative. In other fields where assortative networks are seen, neighboring nodes tend to identify as common members of a group, and/or have similar properties (see \citet{lus04} for an interesting example). We therefore propose that starbursts in group A are similar, at least to the extent to not give rise to significant differences in the IRS spectra. We propose that outliers to this group are instead caused by heavy intrinsic obscuration (see for example \S\ref{notables}), and speculate that there are no large variations in stellar IMF or metallicity from ULIRG to ULIRG. 

Phase B follows and overlaps with phase A. Based on the fact that nearly all the objects on the right hand side of group B have single nuclei, group B likely ends some time after the nuclei of the progenitors have coalesced. We see three interesting trends as we move from left to right in this group. First, the PAHs decline in prominence, becoming negligible (with respect to the continuum) by the right hand side. Second, silicate absorption varies from strong to weak on the left side, but is universally weak or in emission on the right side. Third, the optical spectral types are varied in the first half but almost universally Seyferts in the second half. We therefore propose that the relative contribution from star formation to the mid-IR emission declines as we move from left to right, while the contribution from an AGN increases\footnote{In contrast to phase A, where we make no claims relating a ULIRGs intragroup position to its evolutionary stage}, until some objects at the end of this phase briefly become optical QSOs. This is consistent with previous studies which show that the AGN fraction increases with increasing merger age (e.g. \citealt{vei02,vei06}, though see also \citealt{rig99}), and suggests either that the accretion rate has increased upon moving into group B, and/or that the central black hole has reached a `threshold' mass for luminous AGN activity. We further propose that the heterogeneity of phase B arises from two factors. First is varying amounts of gas/dust driven into the nuclear regions. Second is AGN feedback; a luminous AGN can generate nuclear or galactic-scale winds, and the effects of these winds will vary substantially from case to case. 

Two examples lend support to our proposal that AGN become more luminous and less obscured as we move through phase B. First, IRAS 19254-7245 (object 81, also known as the Superantena) is located where we expect the AGN to be intrinsically luminous but still deeply buried, an expectation that appears to be borne out by recent {\it Suzaku} observations \citep{brai09}. Second, Mrk 231 (object 8) is located where an AGN-driven wind may be expected, and indeed this object is thought to contain a starburst, an energetic AGN, and a nuclear outflow \citep{lip05}.

Assuming that phase C is a separate evolutionary stage, then it is difficult to interpret as it contains a small number of objects. It seems to have similar properties to the first half of phase B, except perhaps for the nuclear separations, which {\it may} be smaller, on average, in phase C. Therefore, we propose that this phase also follows phase A, and that it is characterized mainly by waning star formation. We do not see evidence for a substantial increase in AGN activity in this group, and so propose that this phase is shorter in time than group B, and that its members are unlikely to become optical QSOs. 

We show a cartoon type diagram of this scheme in Figure \ref{ulirglayoutcart}.

If all ULIRGs start off in group A, then the obvious question is what determines if they go to phase B or phase C?\footnote{Assuming they cannot do both, but see \S\ref{evolp}} Broadly, there are two possible drivers; the dynamics/morphology of gas and dust (e.g. how much is available, and how efficiently it is channeled into the nuclear regions), or seed black hole mass (models suggest the minimum seed black hole mass for AGN activity in ULIRGs is $\sim10^{7}$M$_{\odot}$ \citep{tan99}). If the latter is the driver, and the end-product of a ULIRG is an elliptical galaxiy (e.g. \citealt{gen01,das06}) then the antecedents of phase C should have smaller mass bulges than phase B. As there is no plausible evidence for a bimodality in BH/bulge masses in ellipticals, we think it more likely that the end products of B and C are similar, except that some aspect of the merger dynamics of objects in B allows some of them to go through a brief optical QSO phase. 

This evolutionary picture fits well with recent studies of IR-luminous galaxies. The interconnected nature of Figure \ref{ulirglayout} implies that the starburst and AGN activity arises from a common physical mechanism, which tallies with imaging studies of ULIRGs, which show them all to be mergers \citep{sur00,cui01,far01,bus02,vei02,vei06}. The location of all the QSOs in the sample at the end of group B, and their low $\mathcal{B}$ values, suggests that few ULIRGs pass through a phase where they are simultaneously ULIRGs and Quasars, and/or that the ULIRG-QSO phase is brief, in agreement with recent work \citep{far01,tac02,kaw06,kaw07}. Our picture takes the idea that IR-luminous starbursts are present in most ULIRGs, while IR-luminous AGN are present in just under half \citep{gen98,rig99,tra01,kla01,far03,fra03,veg08} and extends it by providing (1) a single diagrammatic representation of the ULIRG evolutionary plane, (2) groupings into evolutionary phases, and (3) descriptions of the properties of these phases, including homogeneity, timescales, and power source.

Finally, we note a peculiar aspect of the diagrams in Figure \ref{ulirglayoutscodee}; the remarkable homogeneity of the 11.2$\mu$m PAH strengths in group A, and the smoother gradient of 11.2$\mu$m PAH strengths through groups B and C, in comparison to the 6.2$\mu$m PAH strengths. We do not have a plausible explanation for this. It could for example be highlighting an important part of the way in which PAHs diagnose star formation rates, or a subtle systematic error in our calculations. We do not consider this point further here, but highlight it as an interesting avenue to pursue in future work.

\begin{figure}[tbhp]
\begin{center}
\includegraphics[width=0.45\textwidth,angle=0.0]{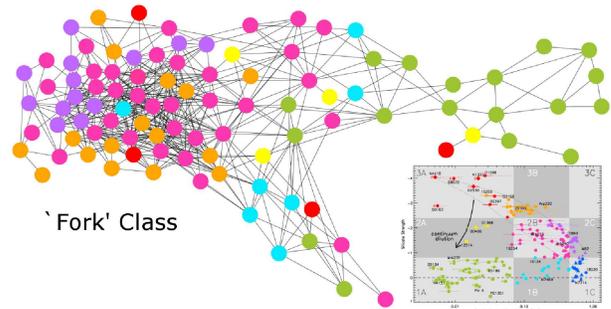}
\caption{\label{ulirglayoutscodedxyz}
Network diagram, with the nodes color-coded by their `Fork' classification \citep{spo07}. A copy of the Fork diagram has been embedded for reference.
}
\end{center}
\end{figure}

\begin{figure*}[tbhp]
\begin{center}
\includegraphics[width=0.7\textwidth]{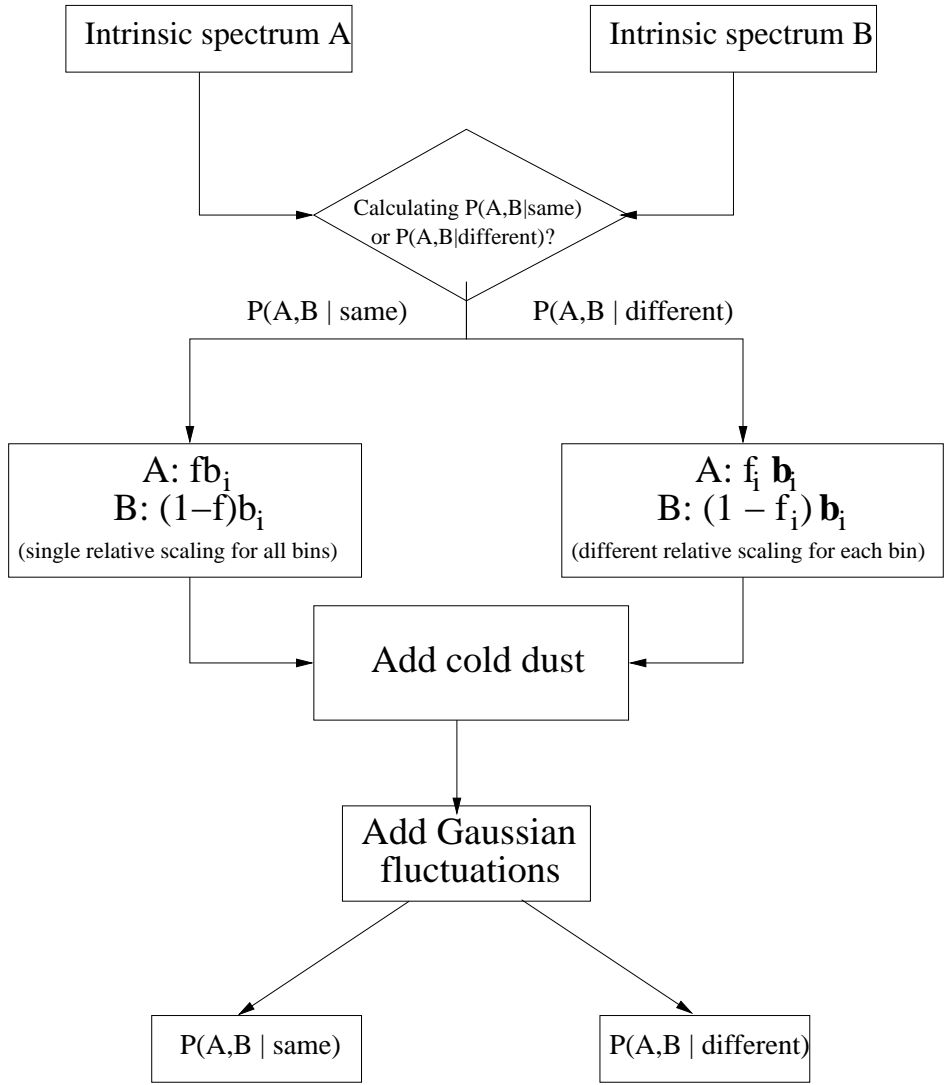}
 \caption{\label{fig:flowchart}
 Flowchart of the method used to calculate $\mathcal{R}$. For each pair of spectra, the flowchart is followed twice; once to calculate P$(\rm{A,B}|\rm{different})$, and once to calculate P$(\rm{A,B}|\rm{same})$.
}
\end{center}
\end{figure*}

\subsection{Comparison to other Mid-IR classification schemes}
As our evolutionary framework is based mostly on mid-IR spectroscopy, it is interesting to compare it to previous work in this field. A recent example is that of \citet{spo07} who published a mid-infrared based classification scheme (the `Fork' diagram, their figure 1) for IR-luminous galaxies. The `Fork' diagram and our network diagram overlap significantly in information use, as the Fork diagram uses both the silicate feature and the 6.2$\mu$m PAH feature, but our diagram includes every other part of the spectrum, and is constructed using a fundamentally different methodology. Our intention though is to compare the predictions from the two schemes, not to perform an independent check. 

In Figure \ref{ulirglayoutscodedxyz} we code each node by its classification in the Fork diagram. We see a clear delineation. With a single exception, all the objects in group A reside in the upper branch of the Fork diagram (classes {\it 2A/2B/2C}, hereafter, Fork classes are given in italics). Group B on the other hand contains all the class {\it 2A} objects, nearly all the class {\it 1A} objects, and a few {\it 1B/2B/3A} objects. Finally, group C contains the remaining class {\it 2A}'s, some {\it 1B}'s, and a few {\it 2B}'s and {\it 3A}'s.

This indicates that the two schemes are crudely in agreement. \citet{spo07} suggest an evolutionary picture in which sources move up the diagonal branch in their figure 1 ({\it 2C $->$ 2B $->$ 3B $->$ 3A}), before either dropping vertically downwards ({\it 3A $->$ 1A}) or via the slanted branch back to {\it 1C} and on from there to {\it 1A/1B}. In Figure \ref{ulirglayoutscoded} we see a trend where the {\it 2C/2B/3B} sources lie on the left hand side, with the {\it 1A} sources lying on the right hand side. Figure \ref{ulirglayoutscodedxyz} however provides a more nuanced diagnostic. From it, we can discern two distinct evolutionary paths, and that the {\it 1B} and {\it 2A} classes are likely starburst/AGN transition classes, rather than just the {\it 2A} class.

\begin{figure}[thpb]
\begin{center}
\includegraphics[width=0.35\textwidth,angle=90.0]{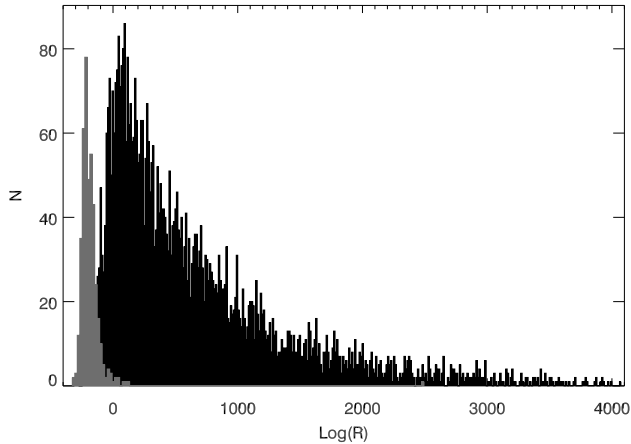}
 \caption{\label{fig:errchange}
Distributions of $\mathcal{R}$ calculated on spectral pairs with the same intrinsic spectrum but different dust extinctions.  The black histogram is for those randomly generated spectral pairs with 5\% Gaussian uncertainties while the grey histogram is for those pairs with 2\% uncertainties. As expected, those pairs with ``true'' uncertainties less than those used in the calculation of $\mathcal{R}$ are systematically shifted to smaller values.}
\end{center}
\end{figure}

\begin{figure}[hpbt]
\begin{center}
\includegraphics[width=0.37\textwidth,angle=90.0]{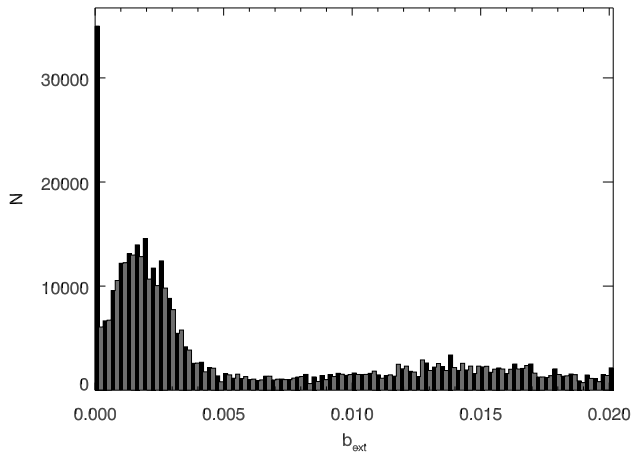}
 \caption{\label{fig:b_ext}
The distribution of values of $b_{e}$ resulting from maximizing Eqn.~\ref{eqn:likelihood} for all possible pairs of spectra. $b_{e}$ is defined as the best estimate of the sum of the column densities of the two ULIRGs that obtains the best fitting intrinsic spectra.}
\end{center}
\end{figure}

\begin{figure}[hbpt]
\begin{center}
\includegraphics[width=0.35\textwidth,angle=90.0]{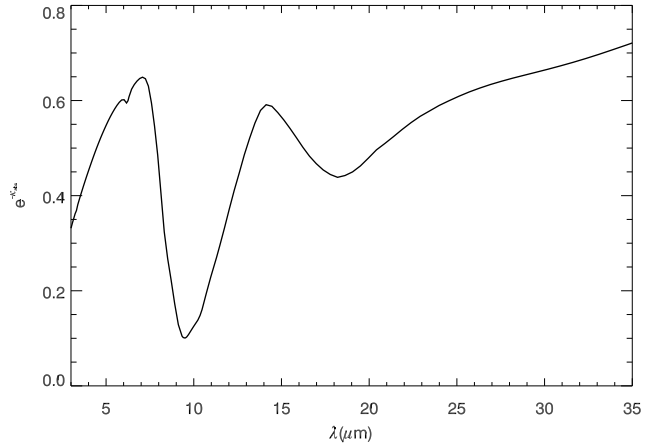}
 \caption{\label{fig:dust}
 The dust attentuation factor, $e^{-k_{abs}(\lambda)x}$, for an arbitrarily chosen value of $x$. }
\end{center}
\end{figure}

\subsection{Notable Objects}\label{notables}
The positions of some famous objects are interesting. We describe some of them in this section.

Arp 220 (object 7) is frequently used as a template for high redshift objects. Its presence in group A marks it as young. It is connected by only 4 edges, making it an atypical example of a local ULIRG. This does not preclude its use as a template, but suggests it is not suitable to use if one is interested in determining if high redshift ULIRGs resemble local ones. From \S\ref{evolp}, its outlier status arises due to heavy intrinsic obscuration, which agrees with the properties of its mid-IR spectra; it shows strong PAHs, but also strong silicate absorption, and a steeply rising continuum at $>10\mu$m.

Next, we consider IRAS F00183-7111 (object 11). This object is atypical (only two edges), and harbors an obscured AGN that is close to burning through its dust cocoon. An independent study comes to similar conclusions; \citet{spo08}, using the high-resolution IRS spectra to look for outflow signatures in fine structure lines (i.e. information that is not contained in the low-resolution spectra) show that this object contains an obscured nuclear outflow driven by an AGN.

Finally, the four objects in Table \ref{sample} that are not plotted in Figure \ref{ulirglayout} (16,41,48,79, one could also include object 2 here as its single connection depends on the assumptions made when computing $\mathcal{R}$) are interesting because their lack of connection is unusual. We defer a complete discussion on these sources to a future paper, and here only briefly describe some possibilities. Part of the reason for this may be luminosity dependent; object 16 (IRAS 00397-1312) is the brightest in the sample, albeit by a small margin, and object 41 (IRAS 07145-2914) is the faintest by over an order of magnitude. It may also be due to a combination of unusual spectral properties; object 16 (IRAS 00397-1312) has the deepest CO absorption feature of any object in the sample, and object 79 (IRAS 18030+0705) has extraordinarily strong PAH features. It is possible therefore that these objects represent either a very brief and/or very rare evolutionary stage, or that the merger dynamics are highly atypical in some way.

\section{Conclusions}\label{concstuff}
We have taken a large mid-infrared spectroscopic database of low redshift ULIRGs, and applied to it two novel analysis methods; (1) a Bayesian based estimator of similarities between pairs of spectra that takes into account every spectral resolution element, marginalizing over luminosity, foreground obscuration and instrument noise, and (2) a visualization algorithm based on force-directed networks that efficiently presents these similarities across the sample simultaneously. We combine these results with archival data to propose, with some reserve, the following evolutionary description for ULIRGs in the low-redshift Universe: 

1 - The IR emission in ULIRGs is consistent with being driven by a single underlying physical process. We see no evidence for multiple, separate evolutionary tracks. There is however evidence for at least two and possibly three evolutionary sub-phases.

2 - The first phase (phase A), through which all ULIRGs go, lasts from the initial encounter until approximately coalescence. The IR emission arises mainly from star formation, with a contribution from an AGN in some cases. The highly interconnected, assortative nature of phase A suggests that there is little variation in starburst parameters from object to object, with observed variations in the IRS spectra instead caused largely by differing foreground obscuration. 

3 - At around the time the progenitors start to coalesce, a ULIRG can branch off into one of two phases. We suggest that the track a ULIRG takes depends primarily on the initial impact parameters and dynamics of the merger and the availability of gas/dust, and to a lesser extent on the masses of the central black holes in the merger progenitors. 

4 - The first of these two phases (phase B) lasts approximately half the length of phase A. The relative contribution from star formation to the mid-IR emission declines as we move from left to right, while the contribution from an AGN increases, until some objects near the end of this phase briefly become optical QSOs. Phase B is less interconnected and more heterogeneous than phase A, implying that more than just foreground obscuration is driving the shape of the IRS spectra. We propose that this increased heterogeneity arises from two factors; (1) varying amounts of gas/dust driven into the nuclear regions by differing merger dynamics, and (2) feedback effects from AGN-driven winds.

5 - The second phase (C) lasts about one third the length of phase A. It is similar to phase B in that the mid-IR spectra are heterogeneous, but the decline in luminosity of the starburst relative to the AGN is less pronounced. Few or no systems on this track pass through a QSO phase.

6 - We use the graph-theory based metric of node betweenness centrality to identify six ULIRGs that may be archetypes of key points in this evolutionary cycle. We propose that Mrk 273 and IRAS 14348-1447 are examples of phase A objects, IRAS 03538-6432 is a prime example of an object transitioning from phase A to phase B, IRAS 05189-2524 is an example of an object on the boundary between phase B and phase C, and Mrk 231 and IRAS 00275-2859 are prime early B to late B type objects. 

7 - The 11.2$\mu$m PAH feature appears to be a remarkably good diagnostic of the evolutionary phase of a ULIRG, more so than the 6.2$\mu$m PAH feature or the 9.7$\mu$m silicate absorption feature. Even though we are using the entire spectral range to construct the network, all the objects in group A have prominent, homogeneous 11.2$\mu$m Equivalent Widths, which then smoothly decline as we move left to right through groups B and C, until we reach the right hand side of both groups where the PAHs are negligible.

\acknowledgements
We thank the referees for very helpful reports that greatly improved the clarity of the paper, and Steve Young at Hamilton College for computer technical support. This work is based on observations made with the Spitzer Space Telescope, and has made extensive use of the NASA/IPAC Extragalactic Database, both of which are operated by the Jet Propulsion Laboratory, California Institute of Technology under contracts with NASA. This project was supported in part by NSF grant CHE-0116435 as part of the MERCURY supercomputer consortium\footnote{http://mercury.chem.hamilton.edu}. Support for the IRS GTO team at Cornell University was provided by NASA through Contract Number 1257184 issued by JPL/Caltech. DF and SO thank the Science and Technologies Facilities Council for support. Funding for Cytoscape is provided by a federal grant from the U.S. National Institute of General Medical Sciences (NIGMS) of the National Institutes of Health (NIH) under award number GM070743-01 and the U.S. National Science Foundation (NSF). Corporate funding is provided through a contract from Unilever PLC. Network Workbench was supported in part by the NSF IIS-0513650 award.

\appendix

\section{Mathematical Details}

\subsection{Introduction}
The formalism for determining the degree of similarity between two data vectors using a Bayes factor was first developed in detail in \citet{jeffreys}, who quantifies the belief that two flux measurements using different detectors with different acceptances are measuring the same flux. This, in essence, reduces to whether or not the differences between the two measured fluxes are what is expected if the assumptions about the acceptances of the detectors are correct.

We here extend this formalism in two ways; (1) we extend the calculation to involve not just one measurement per experiment, but many measurements from many wavelength bins (\emph{i.e.} a spectrum), (2) we incorporate the ability to marginalize over external variables (e.g. luminosity, dust extinction) so that contributions from those variables to the $\mathcal{R}$ values can be accounted for. The procedure is shown as a flow chart in Figure \ref{fig:flowchart}.

\subsection{Method}
We start by defining the flux in the $i^{\rm th}$ wavelength bin of spectrum A as $f_{i}^\prime b_i$, where $b_i$ ranges from 0 to $\infty$, and $f_{i}^\prime$, from 0 to 1. If the fluxes are given in photon counts, then $b_i$ is the mean number of photons in the $i^{\rm th}$ bin of both spectra, $f_{i}^{\prime}$ is the probability that the photons are emitted from source A, and $1 - f_{i}^{\prime}$ is the probability that the photons are emitted from source B. The number of photons in the $i^{\rm th}$ bin of spectrum B is therefore $(1-f_{i}^\prime )b_i$. An analogy would be collecting balls into two receptor bins with different probabilities of accepting an individual ball; in this case, $b_i$ is the mean number of balls that will enter both bins, $f^{\prime}_{i}$ is the relative acceptance of one bin, and $(1-f^{\prime}_{i})$ is the relative acceptance of the other.

If the two spectra are identical, then we will see equal contributions from the two spectra to each bin:

\begin{eqnarray}
f_{ i}^\prime b_i = (1-f_{i}^\prime)b_i,
\end{eqnarray}

\noindent and therefore it follows that:

\begin{eqnarray}
f_{i}^\prime = f=\frac{1}{2},
\end{eqnarray}

\noindent where $f$ and ($1-f$) are the `true' fluxes emitted by sources A and B, respectively. In other words, to ask whether or not the sources have the same intrinsic spectrum is equivalent to asking whether or not $f_{i}^\prime = f$.

Now, if the two spectra differ, but only by a multiplicative scaling factor, and if $N_A$ and $N_B$ are the normalizations for spectra A and B, respectively, then:

\begin{eqnarray}
f_{ i}^\prime = f = \frac{N_A}{N_A+N_B}.
\end{eqnarray}

\noindent Note that $f$ would be the same for all the bins if the two sources were the same (that is, if they differed only by a normalization factor). We will not assume a specific value of $f$ {\it a priori}; hence it must be marginalized (\emph{i.e.} integrated out, thereby accounting for all possibilities for $f$). It is important to note that if the sources are intrinsically different, then $f_{ i}^{\prime}$ is not constrained by $f$.

It is now straightforward to expand Eqn.\,\ref{equation:likelihoodratio} in terms of $f$, $f_{i}^\prime$ and $b_i$:

\begin{eqnarray}
\label{equation:likelihoodratio_expanded}
\mathcal{R}&=&\frac{P(\rm{A,B}|\rm{different})}{P(\rm{A,B}|\rm{same})}\notag\\
&=&\frac{\int_0^1 d\vec{f}^\prime \int_0^\infty d\vec{b} P(\text{A,B}|\vec{f}^\prime, \, \vec{b})P(\vec{f}^\prime,\vec{b}|\rm{different})}
{\int_0^1df \int_0^\infty d\vec{b} P(\text{A,B}|f, \vec{b})P(f, \vec{b}|\rm{same})},
\end{eqnarray}

\noindent where $P(f, \vec{b}|\rm{same})$ and $P(\vec{f}^\prime,\vec{b}|\rm{different})$ are the prior densities, which encode any information that might be known about $\vec{b}$, $\vec{f}^{\prime}$, and $f$ before the data are taken. For a Bayes factor to be well-defined the priors that enter the calculation must be proper probability densities; that is, they must integrate to one. However, as noted below, for this problem we are able to finesse this restriction.

Let us consider each of the terms in Eqn.~\ref{equation:likelihoodratio_expanded} in turn. If the counts in the $i^{\rm th}$ bins of the two spectra follow Poisson statistics, then:

\begin{eqnarray}
\label{eq:Poisson_num}
P(\text{A,B}|\vec{f}^\prime,\vec{b})=\prod_{i=1}^D\frac{(f_{i}^\prime b_i)^{N_i}e^{-f_{i}^\prime b_i}}{N_i!}\frac{((1-f_{i}^\prime)b_i)^{M_i}e^{-(1-f_{i}^\prime)b_i}}{M_i!},
\end{eqnarray}

\noindent where $D$ is the number of wavelength bins in both spectra and $f_{i}^{\prime} b_i$ and $(1-f_{i}^\prime) b_i$ are interpreted as the mean photon count in the $i^{\rm th}$ bin of spectrum A and B, respectively; and $N_i$ and $M_i$ are the numbers of photons that are present in this bin of spectra A and B, respectively. Note that $f^\prime _i$ is allowed to vary independently in each bin. If the spectra are different, as is assumed to be the case when we are calculating $P(\text{A,B}|\vec{f}^\prime,\vec{b})$, then there is no constraint on the relative normalizations of the two spectra from bin to bin.

When calculating the probability of obtaining the two spectra given that they are the same, we must consider:

\begin{eqnarray}\label{eq:Poisson_denom}
P(\text{A,B}|f,\vec{b})=\prod_{i=1}^D\frac{(f b_i)^{N_i}e^{-f b_i}}{N_i!}\frac{((1-f)b_i)^{M_i}e^{-(1-f)b_i}}{M_i!}.
\end{eqnarray}

\noindent In contrast to Eqn.\,\ref{eq:Poisson_num}, the spectra here are only expected to have different overall normalizations and fluctuate according to Poisson statistics.

We now consider the priors in Eqn.~\ref{equation:likelihoodratio_expanded} where we shall assume flat priors for $f_i^\prime$, $f$ and $\vec{b}$, defined initially on compact sets. We also assume that $f_i^\prime$ and $\vec{b}$ are independent \emph{a priori} as are $f$ and $\vec{b}$. That is:

\begin{eqnarray}
P(f,\vec{b}|\text{same})&=&P(f|\vec{b},\text{same})P(\vec{b}|\text{same})\\
&=&P(f|\text{same})P(\vec{b}|\text{same}).
\end{eqnarray}

\noindent We take the prior for $f$ to be:

\begin{eqnarray}
P(f|\text{same})=\frac{1}{f_{max}-f_{min}} ,
\end{eqnarray}

\noindent where $f_{min}\rightarrow 0$ and $f_{max}\rightarrow 1$\footnote{We take limits of the relevant variables following the advice in \citet{jaynes}. One reason for doing so is to handle variables which in the limit are defined on $[0,\infty)$. A flat prior is improper in this limit, that is, it does not integrate to one.}. Similarly:

\begin{eqnarray}
P(\vec{b}|\text{same})=\frac{1}{(b_{max}-b_{min})^D}
\end{eqnarray}

\noindent where $b_{min}\rightarrow 0$ and $b_{max}\rightarrow \infty$. And so:

\begin{eqnarray}
\label{eqn:same_prior}
P(f,\vec{b}|\text{same})=\frac{1}{(f_{max}-f_{min}) (b_{max}-b_{min})^D}.
\end{eqnarray}

Similarly:

\begin{eqnarray}
P(\vec{f}^\prime,\vec{b}|\text{different})&=&P(\vec{f}^\prime|\vec{b},\text{different})P(\vec{b}|\text{same})\\
&=&P(\vec{f}^\prime|\text{different})P(\vec{b}|\text{same}),
\end{eqnarray}

\noindent and likewise for $f_i^\prime$:

\begin{eqnarray}
P(\vec{f}^\prime|\text{different})=\frac{1}{(f^\prime _{max}-f^\prime _{min})^D}
\end{eqnarray}

\noindent where $f_{min}^\prime \rightarrow 0$ and $f_{max}^\prime \rightarrow 1$. Finally, we assume:

\begin{eqnarray}
P(\vec{b}|\text{different})=\frac{1}{(b _{max}-b _{min})^D}.
\end{eqnarray}

Therefore:

\begin{eqnarray}
\label{eqn:different_prior}
P(\vec{f}^\prime,\vec{b}|\text{different})=\frac{1}{(f^\prime _{max}-f^\prime _{min})^D (b _{max}-b _{min})^D}
\end{eqnarray}

As noted above, in principle the priors should be proper. However, for this problem the normalization factors in the densities $P(f,\vec{b}|\text{same})$ and $P(\vec{f}^\prime,\vec{b}|\text{different})$ cancel in Eqn.\,\ref{equation:likelihoodratio_expanded} and we can finesse the issue of improper priors. But this is only because $f_{max}=f^\prime _{max}\rightarrow 1$ and $f_{min}=f^\prime_{min}\rightarrow 0$.

In general, one wants to be sure that in higher dimensional problems such as the one considered here, $\mathcal{R}$ behaves properly. For instance, suppose that the upper limit on $f$ was not 1 but $\rho$ and, similarly, the upper limit on $f_i^\prime$ was $\xi$. The priors on $\vec{b}$ would cancel, but we would be left with an overall constant $\rho/\xi^D$ that depends on the number of dimensions, $D$. There are ways to ensure that $\mathcal{R}$ behaves as would be expected. For instance, one can use a method proposed by \citet{ber01}, where the priors are such that:

\begin{eqnarray}
\label{eqn:integrateToOne}
\int \int df  d\vec{b} P(f,\vec{b}|\text{same})
=\int \int d\vec{f}^\prime d\vec{b} P(\vec{f}^\prime,\vec{b}|\text{different}
)=1
\end{eqnarray}

\noindent and therefore $\mathcal{R}$ is guaranteed to make sense. Another is simply to check that $\mathcal{R}$ exhibits the behavior that would be expected from a Bayes factor. For instance, one would expect that as the uncertainties decrease ({\it i.e.} $N_i\rightarrow \infty$ and $M_i\rightarrow \infty$), a larger fraction of pairs with different intrinsic spectra would have $\mathcal{R}<1$ and a larger fraction of pairs with similar intrinsic spectra would have $\mathcal{R}>1$. We will come back to this point later when we show that the $\mathcal{R}$ used to calculate the similarity of spectra indeed behaves properly as the flux uncertainties go to 0\footnote{If $\mathcal{R}$ were {\it not} a Bayes factor, this would not affect the results presented in this work as $\mathcal{R}$ is only used as a (very good) discriminant between spectral pairs whose intrinsic spectra are different and those whose intrinsic spectra are the same (see Fig.\,\ref{fawkedx}).}.

Collecting the terms, the full Bayes factor for the Poisson case becomes:

\begin{small}
\begin{eqnarray}
\label{eq:like_full}
\mathcal{R}&=&\frac{P(\rm{A,B}|\rm{different})}{P(\rm{A,B}|\rm{same})} \\ \nonumber
&=&\frac{   \prod_{i=1}^D \int_0^1  df_{i}^{\prime} \int_0^\infty db_i  \,\, \frac{(f_{i}^\prime b_i)^{N_i}e^{-f_{i}^\prime b_i}}{N_i!}\frac{((1-f_{i}^\prime)b_i)^{M_i}e^{-(1-f_{i}^\prime)b_i}}{M_i!}} {\int_0^1 df \prod_{i=1}^D   \int_0^\infty db_i \,\, \frac{(f b_i)^{N_i}e^{-f b_i}}{N_i!}\frac{((1-f)b_i)^{M_i}e^{-(1-f)b_i}}{M_i!}} .
\end{eqnarray}
\end{small}

\noindent which can be evaluated to be:

\begin{eqnarray}
\mathcal{R} & = &\frac{\prod_{i=1}^D \frac{1}{N_i+M_i+1}}
	{\int_0^1df \prod_{i=1}^D \binom{N_i+M_i}{M_i} f^{N_i}(1-f)^{M_i}},
\nonumber \\
	& = & \frac{\prod_{i=1}^D B(N_i+1, M_i+1)}{B(N+1, M+1)},
\label{eq:like_Poisson}
\end{eqnarray}

\noindent where $B(n,m)$ is the beta function $B(n,m)  \equiv \Gamma(n) \Gamma(m) / \Gamma(n+m)$,
$N \equiv \sum_{i=1}^D N_i$ and $M \equiv \sum_{i=1}^D M_i$.

\subsection{Gaussian Errors}
In many cases (including ours) the data are in units of flux, not photon counts, and the flux distribution for the $i^{\rm th}$ bin will follow a Gaussian distribution. In this case, the likelihood in Eqn.\,\ref{eq:Poisson_num} is re-expressed as:

\begin{eqnarray}
P(\text{A,B}|\vec{f}^\prime,\vec{b})=\prod_{i=1}^D\frac{1}{\sqrt{2\pi} \sigma_{A\,i}} e^{-\frac{[f_{i}^\prime b_i-F_{A\,i}]^2}{2\sigma_{A\,i}^2}} \frac{1}{\sqrt{2\pi} \sigma_{B\,i}} e^{-\frac{[(1-f_{i}^\prime )b_i-F_{B\,i}]^2}{2\sigma_{B\,i}^2}}
\label{eq:gaussian_num}
\end{eqnarray}

\noindent and the likelihood in Eqn.\,\ref{eq:Poisson_denom} as

\begin{eqnarray}
P(\text{A,B}|f,\vec{b})=\prod_{i=1}^D\frac{1}{\sqrt{2\pi} \sigma_{A\,i}} e^{-\frac{[f b_i-F_{A\,i}]^2}{2\sigma_{A\,i}^2}} \frac{1}{\sqrt{2\pi} \sigma_{B\,i}} e^{-\frac{[(1-f )b_i-F_{B\,i}]^2}{2\sigma_{B\,i}^2}}
\label{eq:gaussian_denom}
\end{eqnarray}

\noindent where, for the $i^{th}$ wavelength bin, $fb_i$ and $f_{i}^{\prime} b_i$ are the mean fluxes for source A; $(1-f) b_i$ and $(1-f_{i}^{\prime}) b_i$ are the mean fluxes expected from sources B; $F_{A\,i}$ and $F_{B\,i}$ are the measured fluxes; and $\sigma_{A\,i}$ and $\sigma_{B\,i}$ are the errors on the measured fluxes, all in the $i^{\rm th}$ bin. Plugging Eqn.\,\ref{eq:gaussian_num} and \ref{eq:gaussian_denom} into Eqn.~\ref{eq:like_full}:

\begin{eqnarray}
\mathcal{R}&=&\frac { \prod_{i=1}^D \int_0^1  df_{i}^{\prime} \int_0^\infty db_i  \,\, \frac{1}{2\pi \sigma_{A\,i}\, \sigma_{B\,i}} e^{-\frac{[f_{i}^{\prime}b_i-F_{A\,i}]^2}{2\sigma_{A\,i}^2}}  e^{-\frac{[(1-f_{i}^{\prime})b_i-F_{B\,i}]^2}{2\sigma_{B\,i}^2}}}{ \int_0^1 df \prod_{i=1}^D   \int_0^\infty db_i \,\,
\frac{1}{2\pi \sigma_{A\,i}\, \sigma_{B\,i}} e^{-\frac{[fb_i-F_{A\,i}]^2}{2\sigma_{A\,i}^2}}  e^{-\frac{[(1-f)b_i-F_{B\,i}]^2}{2\sigma_{B\,i}^2}}},
\label{eq:like_gaus}
\end{eqnarray}

\noindent where now $b_i$ can be integrated semi-analytically (see Section\,\ref{section:integrate_b}).

\subsection{Including Extinction}\label{appextinct}
We also wish to account for the possible presence of a screen of cold dust in front of our sources. To do this, we use the carbonaceous-silicate grain model of~\citet{wein}, their model `A', with $R_{V}=5.5$ and the grain abundances per H increased by a factor of 1.42, although any dust extinction law can in principle be used. The effect of dust extinction is accounted for by a simple exponential factor, $e^{-k(\lambda)x}$, where $k(\lambda)$ is the absorption cross-section per mass of dust ($cm^2/g$) as a function of wavelength, and $x$ is the column density ($g/cm^2$).

Figure\,\ref{fig:dust} shows the distribution of $e^{-k(\lambda)x}$ for an arbitrarily chosen value for $x$.  If $x_A$ and $x_B$ are the column densities for sources A and B, respectively, they are parametrized by $x_A = f_{e}b_{e}$ and $x_B = (1-f_{e})b_{e}$, where the parameters $f_{e}$ and $b_{e}$ are analogous to the parameters $f_{i}^{\prime}$ and $b_i$ for the flux distribution considered above. That is, $b_{e}$ is the sum of the dust columns in front of sources A and B, and $f_{e}$ is the fraction of the dust column in front of source A. The Bayes factor, Eqn.~\ref{equation:likelihoodratio_expanded}, is then modified to
include these new parameters:

\begin{small}
\begin{eqnarray}
\mathcal{R}=\frac{\int_0^1df_{e}\int_0^\infty db_{e}\int_0^1 d\vec{f}^\prime \int_0^\infty d\vec{b} \, P(\text{A,B}|f_{e},b_{e},\vec{f}^\prime, \vec{b})P(f_{e},b_{e},\vec{f}^\prime,\vec{b}|\rm{different})}
{\int_0^1df_{e}\int_0^\infty db_{e}\int_0^1df \int_0^1 \int_0^\infty d\vec{b} P(\text{A,B}|f_{e},b_{e},f, \vec{b}) \, P(f_{e},b_{e},f, \vec{b}|\rm{same})},
\label{eq:like_gaus_dust1}
\end{eqnarray}
\end{small}

\noindent and the likelihood of parameters $f_e$, $b_e$, $\vec{f}^\prime$ and $\vec{b}$ is now:

\begin{eqnarray}
P(\text{A,B}|f_e,b_e,\vec{f}^\prime ,\vec{b})=\prod_{i=1}^D
\frac{1}{\sqrt{2\pi} \sigma_{A\,i}} e^{-\frac { [e^{-k_if_{e}b_{e}} \,  f_{
i}^{\prime} b_i-F_{A\,i} ]^2} {2\sigma_{A\,i}^2}}  \frac{1}{\sqrt{2\pi}
\sigma_{B\,i}} e^{-\frac{[-e^{-k_i(1-f_{e})b_{e}}\, (1-f_{i}^{\prime})
b_i-F_{B\,i} ]^2  } {2\sigma_{B\,i}^2}}\label{eq:gaussian_num_ext}
\end{eqnarray}

\noindent while the likelihood of parameters $f_e$, $b_e$, $f$ and $\vec{b}$ is:

\begin{eqnarray}
P(\text{A,B}|f_e,b_e,f,\vec{b})=\prod_{i=1}^D
\frac{1}{\sqrt{2\pi} \sigma_{A\,i}} e^{-\frac { [e^{-k_if_{e}b_{e}} \,  f b_i-F_{A\,i} ]^2} {2\sigma_{A\,i}^2}}  \frac{1}{\sqrt{2\pi}
\sigma_{B\,i}} e^{-\frac{[-e^{-k_i(1-f_{e})b_{e}}\, (1-f)
b_i-F_{B\,i} ]^2  } {2\sigma_{B\,i}^2}}\label{eq:gaussian_denom_ext}
\end{eqnarray}

\noindent where $k_i$ is the absorption coefficient in the $i^{th}$ wavelength bin. In other words, the mean flux obtained from source A with no dust is $f_{i}^{\prime}b_i$ and $fb_i$ for the `different' and `same' hypotheses, respectively.  If dust is included\footnote{We have parametrized the dust in such a way that one could in principle ask the question whether or not two sources have the same column density.  Under the `same' hypothesis, $f_{e}=1/2$, while under the `different' hypothesis $f_{e}$ would be marginalized. However, we do not consider this question here.}, this flux would be attenuated by an additional factor, $e^{-k_if_{e}b_{e}}$.

In practice, to calculate $\mathcal{R}$, Eqn.~\ref{eq:like_gaus_dust1} needs to be constrained as it has too many degrees of freedom. Let us assume that some maximum, $b_{e\,\rm{max}}$, has been chosen so that $b_{e}<b_{e\,\rm{max}}$. The priors are then modified:

\begin{eqnarray}
P(f_{e},b_{e},f,\vec{b}|\text{same}) \rightarrow \frac{\Theta(b_{e\,\rm{max}}-b_{e})}{b_{e\,\rm{max}}(f_{max}-f_{min}) (b_{max}-b_{min})^D}
\end{eqnarray}

\noindent and:

\begin{eqnarray}
P(f_{e},b_{e},\vec{f}^{\prime},\vec{b}|\text{different}) \rightarrow \frac{\Theta(b_{e\,\rm{max}}-b_{e})}{b_{e\,\rm{max}}(f^\prime _{max}-f^\prime _{min})^D (b _{max}-b _{min})^D},
\end{eqnarray}

\noindent where $\Theta$ is defined as:

\begin{equation}
 \Theta(b_{e\,\rm{max}}-b_{e}) =
 \begin{cases}
  1, & b_{e\,\rm{max}}\ge b_{e} \\
  0, & b_{e\,\rm{max}}< b_{e}\\
 \end{cases}
\end{equation}

Putting everything together, the full likelihood then becomes:

\begin{equation}  
\mathcal{R} = \frac{\int_0^1df_{e}\int_0^\infty db_{e} \int_0^1 d\vec{f}^\prime \int_0^\infty d\vec{b} \, P(\text{A,B}|f_{e},b_{e},\vec{f}^\prime, \vec{b})P(f_{e},b_{e}, \vec{f}^\prime,\vec{b}|\rm{different})} {\int_0^1df_{e}\int_0^\infty db_{e}\int_0^1df \int_0^1 \int_0^\infty d\vec{b}\, P(\text{A,B}|f_{e},b_{e},f, \vec{b})P(f_{e},b_{e},f, \vec{b}|\rm{same})}  \label{eqn:full_like} 
\end{equation}

\noindent or:

\begin{equation} 
\mathcal{R}  = \frac { \int_{0}^1 df_{e} \int_0^{b_{e\,\rm{max}}} db_{e} \prod_{i=1}^D \int_0^1  df_{i}^{\prime} \int_0^\infty db_i  \,\, \rm{exp}\left( -\frac{[e^{-k_i f_{e} b_{e}} \, \, f_{i}^{\prime}b_i-F_{A\,i}]^2}{2\sigma_{A\,i}^2}\right)  }
{  \int_{0}^1 df_{e}   \int_0^{b_{e\,\rm{max}}} db_{e} \int_0^1 df \prod_{i=1}^D   \int_0^\infty db_i \,\,  \rm{exp}\left( -\frac{[e^{-k_if_{e}b_{e}} \, \, fb_i-F_{A\,i}]^2}{2\sigma_{A\,i}^2}\right)  } \notag \times\frac
 { \rm{exp}\left( -\frac{[e^{-k_i (1-f_{e}) b_{e}} \,\,   (1-f_{i}^{\prime})b_i-F_{B\,i}]^2}{2\sigma_{B\,i}^2}\right)}
{ \rm{exp}\left(-\frac{[ e^{-k_i (1-f_{e}) b_{e}} \, \, (1-f)b_i-F_{B\,i}]^2}{2\sigma_{B\,i}^2}\right)}  .
\end{equation}

Parameter $b_i$ can be integrated semi-analytically as shown in Section~\ref{section:integrate_b}. Both the numerator and the denominator of Eqn.~\ref{eqn:full_like} are calculated using adaptive Monte Carlo integration methods, VEGAS and \emph{miser}~\citep{numerical}.

There is a degeneracy in Eqn.\,\ref{eqn:full_like}. Suppose we are comparing two spectra which both have the shape of the extinction curve shown in Fig.\,\ref{fig:dust}. In this case, the best estimate of the parameters can come from an `intrinsic' spectrum ($\sim\vec{b}$) that looks like the extinction curve and is not propagated through any dust, or that is flat and is propagated through dust (assuming that dust follows the extinction curve in Fig.\,\ref{fig:dust}). In practice, this situation is exceedingly unlikely, as the intrinsic spectra (which enter through $\vec{b}$), the relative normalizations (that enter through $f$) and the extinction correction all do the best they can to make a good fit. There are however examples where effects of the source distribution taking on characteristics of the dust to make the best fit can be seen; see {\it e.g.} the comparison between IRAS 12071-0444 and IRAS 17179+5444, where the extinction is close to zero, but the spectra follow the features of the extinction curve in Fig.\,\ref{fig:dust}.

Ideally we would like a model for the intrinsic dust distribution in each ULIRG, such that {\it all} the dust, hot and cold, can be placed where it should be (in the extinction parameters $f_{e}$ and $b_{e}$) and the source where it should be (in $f$ and $\vec{b}$). Such a model does not however exist, and so the best we can do is attempt to account for differences solely due to cold foreground dust. While this is undoubtedly an improvement on not attempting to account for dust at all, it does mean that (without constraints on the shape of the source distribution) the contributions of dust and the source may not be proportioned correctly.

\subsection{Behavior of $\mathcal{R}$}
As discussed previously, $\mathcal{R}$ must behave as a Bayes factor should - {\it i.e.} as $\sigma_{A\,i}\rightarrow 0$ and $\sigma_{B\,i}\rightarrow 0$, pairs with like intrinsic spectra should all have $\mathcal{R}$'s less than 1 and pairs with unlike intrinsic spectra should have $\mathcal{R}$'s larger than 1. To this end, Fig.\,\ref{fawkedx} shows that the spectra with lower errors systematically have a larger separation in $\mathcal{R}$ than those with larger errors, indicating the likelihood is behaving as expected, although this separation is somewhat obscured by the decreasing accuracy/precision of the Monte-Carlo integration as the uncertainties decrease.

To gather further intuition about the behavior of $\mathcal{R}$, we study the case where the Gaussian (statistical) fluctuations in the data are smaller than what is assumed in the calculation. We generate pairs of spectra with the same intrinsic spectra but different dust extinctions. The two spectra are then fluctuated independently assuming $2\%$ Gaussian uncertainties; $\mathcal{R}$ is then calculated for each pair assuming 5\% uncertainties. As there would be less variation in the data than expected by the calculation, one would expect the spectra to look more similar than if the fluctuations in the data were, say, $\sim 5\%$.  Therefore, one would expect $\mathcal{R}$ to be less than if the fluctuations in the data were indeed $\sim 5\%$. Fig.\ref{fig:errchange} shows that $\mathcal{R}$ behaves as we expect where those data with $2\%$ uncertainties are systematically shifted to lower values than those with $5\%$ uncertainties.

\subsection{Maximum A Posteriori (MAP) Probability} \label{section:maximum_likelihood}
We would like to maximize the {\it a posteriori} probability defined as the unmarginalized denominator in Eqn.~\ref{eqn:full_like} (which accounts for the probability that the two spectra come from the sources with the same intrinsic spectrum).  This is done for several reasons. First, maximizing this likelihood is one way of finding out how similar the two spectra are. Second, by maximizing {\it a posteriori} probability with respect to $f_{e}$, $b_{e}$, $f$ and $\vec{b}$, one can obtain a best estimate of the mean distributions of these parameters for spectra $A$ and $B$. Third, by plotting the best estimate of $b_{e}$ for all pairing of the spectra, we get an estimate for the upper limit for $b_{e}$.

The third point is worth describing in more detail. The rationale for calculating the upper limit on $b_{e}$ in this way is as follows.  We start by (falsely) supposing that the intrinsic sources are the same, and that the only thing that differentiates them is an overall flux normalization and the amount of dust in front of them. We then calculate the best estimate for $b_{e}$ by finding the MAP defined later in the text (Eqn.~\ref{eqn:likelihood}), thereby creating a distribution of the best estimates for $b_{e}$ for our data sample. The maximum $b_{e}$ found from this distribution is an estimate of the maximum $b_{e}$ that might be found for any of the pairs of objects.  If our assumption that the sources were the same was incorrect, the differences in the sources would randomize the $b_{e}$ distribution to some degree, thus effectively only pushing the upper limit on $b_{e}$ higher. That is, the ``true'' value of $b_{e}$ is likely to be covered in the integration of $b_e$.

One could pose two objections to this method; (1) for a given pair of objects, we are using it in the distribution of the best estimates for $b_{e}$ to find the upper limit of $b_{e}$ and then subsequently using that upper limit to calculate $\mathcal{R}$ with the very same pair; and (2) we are not using the shape of the distribution of $b_{e}$ as a prior when calculating $\mathcal{R}$. The first objection can be addressed by realizing that in our sample there are $5151$ pairings, and the effect of using our knowledge of $b_{e}$ from a given pair twice (once to develop the distribution of the best estimates for $b_{e}$, and once for calculating the result using our knowledge on the upper limit of $b_{e}$) is small.  As for the latter objection, one can argue that the distribution is skewed by the fact that the intrinsic source spectra for the pairs are different and therefore a flat prior on $b_{e}$ is at least as good an estimate of our {\it a priori} knowledge of $b_{e}$ as a prior that takes into account the shape of the $b_{e}$ distribution.

Turning back to the {\it a posteriori} probability defined as:

\begin{eqnarray}
\label{eqn:likelihood}
L(\text{A,B}|f_{e},b_{e},f,\vec{b})
\equiv P(\text{A,B}|f_{e},b_{e},f,\vec{b})
P(f_{e},b_{e},f,\vec{b}|\text{same}).
\end{eqnarray}

Maximizing Eqn.\,\ref{eqn:likelihood} is equivalent to minimizing:

\begin{eqnarray}
\label{eqn:loglike}
-{\text log}\left( L({\text A,B} | f_{e},b_{e},f,\vec{b})\right)& \equiv &
-{\text log}\left[ P(\text{A,B}|f_{e},b_{e},f,  \vec{b}) P(f_{e},b_{e},f,\vec{b}|\text{same})\right] \\ \nonumber
&=&\sum_{i=1}^D  \, \frac{[e^{-k_if_{e}b_{e}} \, \, fb_i-F_{A\,i}]^2}{2\sigma_{A\,i}^2}  +\frac{[ e^{-k_i (1-f_{e}) b_{e}} \, \, (1-f)b_i-F_{B\,i}]^2}{2\sigma_{B\,i}^2} \notag\\ \nonumber
&&+ {\text{log}}(2\pi \sigma_{A\,i} \sigma_{B\,i}).
\end{eqnarray}

Maximizing Eqn.\,\ref{eqn:likelihood} (or minimizing Eqn.~\ref{eqn:loglike}) yields a distribution of $b_{e}$ from all the pairs of objects considered in this work, from where one can empirically determine the limit on $b_{e}$, $b_{e\,\rm{max}}$. The distribution is shown in Fig.\,\ref{fig:b_ext}.  We chose $b_{e\,\rm{max}}=0.02$ as the nominal value as $b_e<b_{e\,max}$ contains the bulk of spectral pairs, and $b_{e\,\rm{max}}=0.03$ to test the dependence of the network diagrams on dust extinction (see Fig.\,\ref{ulirglayouttests}).

The column densities found by minimizing Eqn.\,\ref{eqn:loglike} should not be regarded as good estimates of the `true' column densities because there is degeneracy in the intrinsic shapes of the spectra and the effects of extinction. For instance, two spectra that are intrinsically the same will yield the same measured spectra (within statistical uncertainties) provided there is no dust extinction.  However, two ULIRGs may have different intrinsic spectra and suffer different amounts of dust extinctions, but the combination might conspire to make the measured spectra look similar.  Also, as discussed above, ULIRGs with similar column densities might, in fact, render a low value for $b_e$ while the intrinsic spectra take on characteristics of the dust as there are no constraints imposed on them.

\subsection{Integrations With Respect to $\vec{b}$}\label{section:integrate_b}
Integrating the Gaussians in Eqn.~\ref{eqn:full_like} with respect to $b_i$~\citep{Gradshteyn:2000}, we obtain:

\begin{tiny}
\begin{eqnarray}
\mathcal{R}&=&
 \frac{ \int_{0}^1 df_{e} \int_0^{b_{e\,\rm{max}}} db_{e} \prod_{i=1}^D \int_0^1  df_i^{\prime}
 \sqrt{\frac{\pi}{2\left[ \frac{f_i^{\prime\,2}e^{-2k_i f_{e} b_{e}}}{\sigma_{A\,i}^2} + \frac{(1-f_i^{\prime})^2e^{-2k_i (1-f_{e}) b_{e}}}{\sigma_{B\,i}^2}\right]}}
 }
{  \int_{0}^1 df_{e}   \int_0^{b_{e\,\rm{max}}} db_{e} \int_0^1 df \prod_{i=1}^D
 \sqrt{\frac{\pi}{2\left[ \frac{f^2e^{-2k_i f_{e} b_{e}}}{\sigma_{A\,i}^2} + \frac{(1-f)^2e^{-2k_i (1-f_{e}) b_{e}}}{\sigma_{B\,i}^2}\right]}}
}\notag \\
&&
\frac{ \rm{exp}\left[\frac{ \left( \frac{f^\prime _ie^{-k_i f_{e} b_{e}}F_{A\,i}}{\sigma_{A\,i}^2} +
 \frac{(1-f^\prime _i)e^{-k_i (1-f_{e}) b_{e}}F_{B\,i}}{\sigma_{B\,i}^2} \right)^2 }{2\frac{f_i^{\prime\,2}e^{-2k_i f_{e}b_{e}}}{\sigma_{A\,i}^2} + \frac{(1-f_i^{\prime})^2e^{-2k_i (1-f_{e}) b_{e}}}{\sigma_{B\,i}^2}}
 - \frac{F_{A\,i}^2}{2\sigma_{A\,i}^2} - \frac{F_{B\,i}^2}{2\sigma_{B\,i}^2} \right]
}
{
 \rm{exp}\left[\frac{ \left( \frac{fe^{-k_i f_{e} b_{e}}F_{A\,i}}{\sigma_{A\,i}^2} +
 \frac{(1-f)e^{-k_i (1-f_{e}) b_{e}}F_{B\,i}}{\sigma_{B\,i}^2} \right)^2 }{2\frac{f^2e^{-2k_i f_{e} b_{e}}}{\sigma_{A\,i}^2} + \frac{(1-f)^2e^{-2k_i (1-f_{e}) b_{e}}}{\sigma_{B\,i}^2}}
  - \frac{F_{A\,i}^2}{2\sigma_{A\,i}^2} - \frac{F_{B\,i}^2}{2\sigma_{B\,i}^2}\right]
} \notag\\
&&
\frac
{
 \left( 1-\rm{erf}\left(-\frac{\frac{f^\prime _ie^{-k_i f_{e} b_{e}}F_{A\,i}}{\sigma_{A\,i}^2} + \frac{(1-f^\prime _i)e^{-k_i (1-f_{e}) b_{e}}F_{B\,i}}
 {\sigma_{B\,i}^2} }
 {\sqrt{2\left[ \frac{f_i^{\prime\,2}e^{-k_i f_{e} b_{e}}}{\sigma_{A\,i}^2} + \frac{(1-f_i^{\prime})^2}{\sigma_{B\,i}^2}\right]}}\right) \right)
}
{
 \left( 1-\rm{erf}\left(-\frac{\frac{fe^{-k_i f_{e} b_{e}} F_{A\,i}}{\sigma_{A\,i}^2} + \frac{(1-f)e^{-k_i (1-f_{e}) b_{e}}F_{B\,i}}
 {\sigma_{B\,i}^2} }
 {\sqrt{2\left[ \frac{f^2e^{-2k_i f_{e} b_{e}}}{\sigma_{A\,i}^2} + \frac{(1-f)^2}{\sigma_{B\,i}^2}\right]}}\right) \right)
}
\label{eq:like_gaus_dust}
\end{eqnarray}
\end{tiny}

\noindent where $\rm{erf}(u)$ is defined as:

\begin{eqnarray}
\rm{erf}(u)\equiv \frac{2}{\sqrt{\pi}}\int_0^ue^{-x^2}dx.
\end{eqnarray}

\clearpage
 \LongTables
\begin{landscape}
\begin{deluxetable*}{llcccccl}
\tablecolumns{8}
\tablecaption{The Sample \label{sample}}
\tablehead{
\colhead{ID}&\colhead{Galaxy}&\colhead{RA(J2000)}&\colhead{Dec}&\colhead{$z$}&\colhead{Class\tablenotemark{a}}&\colhead{$L_{ir}$\tablenotemark{b}}&\colhead{Connections\tablenotemark{c}}
}
\startdata
1 & IRAS 05189-2524  & 05 21 01.5 & -25 21 45.4 & 0.043 & S2  & 12.11 & 12,17,29,32,35,38,51,52,54,56,67,70,74,88 \\
2 & IRAS 08572+3915  & 09 00 25.4 & +39 03 54.4 & 0.058 & L  & 12.12 & 59 \\
3 & IRAS 12112+0305  & 12 13 46.0 & +02 48 38.0 & 0.073 & L  & 12.23 & 4,21,25,27,29,33,35,39,44,58,66,72,76,78,82,83,87,89,92,93 \\
4 & IRAS 14348-1447 & 14 37 38.4 & -15 00 22.8 & 0.083 & L  & 12.26 & 3,9,19,21,26,27,29,33,35,38,49,51,53,58,60,66,72,78,82,83,86,87,89,92,93,94 \\ 
5 & IRAS 15250+3609  & 15 26 59.4 & +35 58 37.5 & 0.055 & L                  & 12.04 & 19,22,26,30,35,38,52,53,56,58,67,70,74 \\
6 & IRAS 22491-1808  & 22 51 49.3 & -17 52 23.5 & 0.078 & H                  & 12.11 & 7,21,31,93 \\
7 & Arp 220          & 15 34 57.1 & +23 30 11.5 & 0.018 & L                  & 12.08 & 6,21,31,34,39 \\
8 & Mrk 231          & 12 56 14.2 & +56 52 25.2 & 0.042 & S1                 & 12.51 & 1,11,15,17,24,32,45,62,69,81,84,95  \\
9 & Mrk 273    & 13 44 42.1 & +55 53 12.7 & 0.038 & S2                 & 12.09 & 4,13,19,22,23,25-27,29,32,35,38,42,49,51,52-54,56,58,60,65-67,70-72,74,82,83,85,86,89,94  \\
10   & UGC 5101               & 09 35 51.7 & +61 21 11.3 & 0.039 & L & 11.96 & 12,14,37,40,45,95 \\
11   & IRAS F00183-7111  & 00 20 34.7 & -70 55 26.7 & 0.327 & L   & 12.91 & 8,37 \\
12   & IRAS F00188-0856  & 00 21 26.5 & -08 39 26.3 & 0.128 & L                                 & 12.42 & 10,37,40,53,70,80 \\
13   & IRAS 00199-7426   & 00 22 07.0 & -74 09 41.7 & 0.096 & H? & 12.30 & 9,22,23,26,29,32,36,42,50,51,56,60,67,70,72,74 \\
14   & IRAS 00275-0044   & 00 30 09.1 & -00 27 44.2 & 0.242 & ?                                  & 12.39 & 10,36,37,46,63 \\
15   & IRAS 00275-2859   & 00 30 04.2 & -28 42 25.0 & 0.278 & S1 & 12.72 & 8,59,84,90,95,99 \\
16   & IRAS 00397-1312   & 00 42 15.5 & -12 56 02.8 & 0.262 & H                  & 13.02 & -- \\
17   & IRAS 00406-3127   & 00 43 03.2 & -31 10 49.5 & 0.342 & S2& 12.78 & 1,8,28,74,80,88,91 \\
18   & IRAS 01003-2238   & 01 02 50.0 & -22 21 57.5 & 0.118 & H                  & 12.33 & 1,28,55,57,64,68,75,91 \\
19   & IRAS 01199-2307   & 01 22 20.9 & -22 52 06.7 & 0.156 & H                  & 12.26 & 4,5,9,21,30,35,52,56,86,89,92 \\
20   & IRAS 01298-0744   & 01 32 21.4 & -07 29 08.1 & 0.136 & H                  & 12.35 & 51,54,71 \\
21   & IRAS 01355-1814   & 01 37 57.4 & -17 59 20.6 & 0.192 & H                  & 12.44 & 3,4,6,7,19,93 \\
22   & IRAS 01388-4618   & 01 40 55.9 & -46 02 53.6 & 0.090 & H & 12.05 & 5,9,13,23,26,29,35,38,50,51,52,56,58,60,68,70,72,74,77,88,91 \\
23   & IRAS 01494-1845   & 01 51 51.4 & -18 30 46.4 & 0.152 & H & 12.27 & 9,13,22,25,26,29,36,42,44,50,51,60,70,72  \\
24   & IRAS 02054+0835   & 02 08 06.8 & +08 50 02.0 & 0.345 & S1 & 13.08 & 8,62,99 \\
25   & IRAS 02113-2937   & 02 13 33.0 & -29 23 39.5 & 0.192 & L & 12.41 & 3,9,23,29,36,44,51,60,70,72,73,85 \\
26   & IRAS F02115+0226  & 02 14 10.3 & +02 39 59.7 & 0.399 & ?                  & 12.82 & 1,4,5,9,13,22,23,27,29,32,35,38,42,44,50,51,53,54,56,58,60,67,70,72,88 \\
27   & IRAS F02455-2220  & 02 47 51.3 & -22 07 37.8 & 0.296 & ?                  & 12.74 & 3,4,9,26,29,42,44,51,60,72 \\
28   & IRAS 02530+0211   & 02 55 34.4 & +02 23 41.4 & 0.028 & L & 11.10 & 17,18,57,75 \\
29   & IRAS 03000-2719   & 03 02 11.4 & -27 07 26.3 & 0.221 & ?                  & 12.59 & 3,4,9,13,22,23,25,26,27,35,38,42,44,50,51,53,60,67,70,72,74,80,83 \\
30   & IRAS 03158+4227   & 03 19 12.4 & +42 38 28.0 & 0.134 & L?& 12.48 & 5,19,35,49,52 \\
31   & IRAS 03521+0028   & 03 54 42.1 & +00 37 03.4 & 0.152 & L & 12.55 & 6,7,87 \\
32   & IRAS 03538-6432   & 03 54 25.2 & -64 23 44.7 & 0.301 & ?                  & 12.73 & 1,8,9,13,26,37,38,40,42,46,51,53,63,67,69,70,80,88 \\
33   & IRAS 04114-5117   & 04 12 44.2 & -51 09 40.8 & 0.125 & ?                  & 12.05 & 3,4,44,47,76,87 \\
34   & IRAS 04313-1649   & 04 33 37.1 & -16 43 31.5 & 0.268 & L & 12.59 & 7 \\
35   & IRAS 04384-4848   & 04 39 50.8 & -48 43 17.4 & 0.203 & H & 12.36 & 3,4,5,9,19,22,26,29,30,38,44,50,52,53,54,56,58,60,70,72,74,80,86,88,89 \\
36   & IRAS 06009-7716   & 05 58 37.1 & -77 16 39.0 & 0.117 & H?                 & 12.03 & 13,14,23,25,42,60,72,73,85 \\
37   & IRAS 06035-7102   & 06 02 54.0 & -71 03 10.2 & 0.079 & H & 12.19 & 1,10,11,12,14,32,40,45,46,51 \\
38   & IRAS 06206-6315   & 06 21 01.2 & -63 17 23.5 & 0.092 & S2 & 12.17 & 4,5,9,22,26,29,32,35,42,44,51,52,53,54,56,58,60,67,70,71,72,74,80,82,86,89 \\
39   & IRAS 06301-7934   & 06 26 42.5 & -79 36 31.0 & 0.156 & ?                  & 12.33 & 3,7 \\
40   & IRAS 06361-6217   & 06 36 35.8 & -62 20 33.1 & 0.160 & L & 12.33 & 10,12,32,37,42,46,63 \\
41   & IRAS 07145-2914   & 07 16 31.2 & -29 19 28.8 & 0.006 & S2& 10.08 & -- \\
42   & IRAS 07449+3350   & 07 48 10.6 & +33 43 27.1 & 0.355 & L & 12.84 & 1,9,13,23,26,27,29,32,36,38,40,44,46,51,67,70,72,78,85 \\
43   & IRAS 07598+6508   & 08 04 33.1 & +64 59 48.6 & 0.148 & S1                 & 12.56 & 59,61,90 \\
44   & IRAS F08208+3211  & 08 23 54.6 & +32 02 12.0 & 0.396 & H & 12.46 & 3,23,25,26,27,29,33,35,38,42,53,72,76,83,85 \\
45   & IRAS 08559+1053   & 08 58 41.8 & +10 41 21.9 & 0.148 & S2                 & 12.28 & 8,10,37,46,84,95 \\
46   & IRAS 09022-3615   & 09 04 12.7 & -36 27 01.1 & 0.060 & ?                  & 12.26 & 14,32,37,40,42,45,62,63,67,69,70,73,95 \\
47   & IRAS 09463+8141   & 09 53 00.5 & +81 27 28.4 & 0.155 & L                  & 12.24 & 33,71,73,76 \\
48   & IRAS 10091+4704   & 10 12 16.7 & +46 49 43.5 & 0.246 & L                  & 12.61 & -- \\
49   & IRAS 10378+1109   & 10 40 29.2 & +10 53 18.3 & 0.136 & L                  & 12.35 & 4,9,30,94 \\
50   & IRAS 10565+2448   & 10 59 18.1 & +24 32 34.3 & 0.043 & H & 12.01 & 13,22,23,26,29,35,51,56,58,60,67,68,72,74,77,91 \\
51   & IRAS F11038+3217  & 11 06 35.7 & +32 01 46.4 & 0.130\tablenotemark{d}& ?  & 11.45 & 4,9,13,20,22,23,25,26,27,29,32,37,38,42,50,53,56,70,71,72,74,80,85 \\
52   & IRAS 11095-0238   & 11 12 03.4 & +02 04 22.4 & 0.107 & L                  & 12.29 & 5,9,19,22,30,35,38,54,56,58,70,72,74,80,86,88 \\
53   & IRAS 11223-1244   & 11 24 50.1 & -13 01 13.5 & 0.199 & S2                 & 12.51 & 4,5,9,12,26,29,32,35,38,44,51,60,67,72,78 \\
54   & IRAS 11582+3020   & 12 00 46.8 & +30 04 14.8 & 0.223 & L                  & 12.55 & 9,20,26,35,38,52,56,74,80,88,92 \\
55   & IRAS 12018+1941   & 12 04 24.5 & +19 25 10.3 & 0.169 & L                  & 12.54 & 18,57 \\
56   & IRAS 12032+1707   & 12 05 47.7 & +16 51 08.0 & 0.217 & L & 12.59 & 5,9,13,19,22,26,35,38,50,51,52,54,58,60,70,74,80,88,92 \\
57   & IRAS 12071-0444   & 12 09 45.1 & -05 01 13.9 & 0.128 & S2                 & 12.44 & 1,18,28,55,64,68,75 \\
58   & IRAS 12205+3356   & 12 23 00.3 & +33 39 28.9 & 0.263 & ?                  & 12.49 & 3,4,5,9,22,26,35,38,50,52,56,60,66,72,82,83 \\
59   & IRAS 12514+1027   & 12 54 00.8 & +10 11 12.4 & 0.319 & S2 & 12.72 & 2,15,43,90 \\
60   & IRAS 13120-5453   & 13 15 06.4 & -55 09 22.7 & 0.031 & S2?                & 12.26 & 4,9,13,22,23,25,26,27,29,35,36,38,50,53,56,58,65,66,67,70,72,74,78,82,83,89,94 \\
61   & IRAS 13218+0552   & 13 24 19.9 & +05 37 04.7 & 0.205 & S1                 & 12.73 & 43 \\
62   & IRAS 13342+3932   & 13 36 24.1 & +39 17 31.1 & 0.179 & S1                 & 12.47 & 1,8,24,46,81,84 \\
63   & IRAS 13352+6402   & 13 36 50.7 & +63 47 03.0 & 0.237 & ?                  & 12.50 & 14,32,37,40,46,85 \\
64   & IRAS 13451+1232   & 13 47 33.3 & +12 17 24.2 & 0.121 & S2                 & 12.37 & 1,18,57,75,97 \\
65   & IRAS 14070+0525   & 14 09 31.3 & +05 11 31.8 & 0.264 & S2                 & 12.88 & 9,60,71,73  \\
66   & IRAS 14378-3651   & 14 40 59.0 & -37 04 32.0 & 0.068 & L & 12.07 & 3,4,9,58,60,72,82,83,87,89,93,94 \\
67   & IRAS 15001+1433   & 15 02 31.9 & +14 21 35.1 & 0.163 & S2     & 12.48 & 1,5,9,13,26,29,32,37,38,42,46,50,53,60,68,69,70,72,74,80,88,91,97 \\
68   & IRAS 15206+3342   & 15 22 38.0 & +33 31 35.9 & 0.124 & H                  & 12.27 & 1,18,22,50,57,67,69,74,75,88,91,97 \\
69   & IRAS 15462-0450   & 15 48 56.8 & -04 59 33.6 & 0.100 & S1                 & 12.24 & 1,8,32,37,46,67,68,88,97 \\
70   & IRAS 16090-0139   & 16 11 40.5 & -01 47 05.6 & 0.134 & L                  & 12.58 & 5,9,12,13,22,23,25,26,29,32,35,38,42,46,51,52,56,60,67,71,72,73,74,78,80 \\
71   & IRAS 16300+1558   & 16 32 21.4 & +15 51 45.2 & 0.242 & L                  & 12.69 & 9,20,38,47,51,65,70,73 \\
72   & IRAS 16334+4630   & 16 34 52.6 & +46 24 52.8 & 0.191 & L                  & 12.41 & 3,4,9,13,22,23,25-27,29,35,36,38,42,44,50-53,58,60,66,67,70,74,78,82,83,85,89 \\
73   & IRAS F16576+3553  & 16 59 24.7 & +35 49 01.7 & 0.371 & L & 12.39 & 25,36,46,47,65,70,71,85 \\
74   & IRAS 17068+4027   & 17 08 32.1 & +40 23 28.2 & 0.179 & H                  & 12.33 & 5,9,13,17,22,29,35,38,50,51,52,54,56,60,67,68,70,72,77,80,88 \\
75   & IRAS 17179+5444   & 17 18 54.2 & +54 41 47.3 & 0.147 & S2                 & 12.30 & 1,18,28,57,64,68,91,97 \\
76   & IRAS 17208-0014   & 17 23 22.0 & -00 17 00.9 & 0.043 & H & 11.94 & 3,33,44,47,83,87,89 \\
77   & IRAS F17252+3659  & 17 26 57.8 & +36 56 39.5 & 0.365 & ?                  & 12.47 & 22,50,74,91 \\
78   & IRAS 17463+5806   & 17 47 05.6 & +58 05 18.0 & 0.309 & S2 & 12.61 & 3,4,42,53,60,70,72 \\
79   & IRAS 18030+0705   & 18 05 27.1 & +07 05 57.5 & 0.146 & ?                  & 12.16 & -- \\
80   & IRAS 18443+7433   & 18 42 54.8 & +74 36 21.0 & 0.135 & ?                  & 12.27 & 12,17,29,32,35,38,51,52,54,56,67,70,74,88 \\
81   & IRAS 19254-7245S  & 19 31 21.6 & -72 39 22.0 & 0.063 & S2 & 12.19 & 1,8,62,97 \\
82   & IRAS 19297-0406   & 19 32 21.3 & -03 59 56.3 & 0.086 & H & 12.37 & 3,4,9,38,58,60,66,72,83,86,87,89,92,93,94 \\
83   & IRAS 19458+0944   & 19 48 15.7 & +09 52 05.0 & 0.100 & ?                  & 12.34 & 3,4,9,29,44,58,60,66,72,76,82,87,89,92 \\
84   & IRAS 20037-1547   & 20 06 31.7 & -15 39 08.0 & 0.192 & S1 & 12.52 & 8,15,37,45,62,95 \\
85   & IRAS 20087-0308   & 20 11 23.9 & -02 59 50.7 & 0.106 & L & 12.34 & 9,25,36,42,44,51,63,72,73 \\
86   & IRAS 20100-4156   & 20 13 29.5 & -41 47 34.9 & 0.130 & H & 12.52 & 4,9,19,35,38,52,82,89,92   \\
87   & IRAS 20414-1651   & 20 44 18.2 & -16 40 16.2 & 0.087 & H                  & 12.18 & 3,4,31,33,66,76,82,83,89  \\
88   & IRAS 20551-4250   & 20 58 26.8 & -42 39 00.3 & 0.043 & H & 12.00 & 17,22,26,32,35,52,54,56,67,68,69,74,80,91 \\
89   & IRAS 21272+2514   & 21 29 29.4 & +25 27 50.0 & 0.151 & S2 & 12.10 & 3,4,9,19,35,38,60,66,72,76,82,83,86,87,92,93 \\
90   & IRAS 23060+0505   & 23 08 33.9 & +05 21 29.9 & 0.173 & S2                 & 12.55 & 15,43,59,98,99,102  \\
91   & IRAS 23128-5919   & 23 15 46.8 & -59 03 15.6 & 0.045 & H & 11.97 & 1,17,18,22,50,67,68,75,77,88  \\
92   & IRAS 23129+2548   & 23 15 21.4 & +26 04 32.2 & 0.179 & L & 12.43 & 3,4,19,54,56,82,83,86,89 \\
93   & IRAS 23230-6926   & 23 26 03.6 & -69 10 18.8 & 0.106 & L & 12.25 & 3,4,6,21,66,82,89 \\
94   & IRAS 23253-5415   & 23 28 06.1 & -53 58 31.0 & 0.130 & H & 12.37 & 4,9,49,60,66,82  \\
95   & IRAS 23498+2423   & 23 52 26.0 & +24 40 16.7 & 0.212 & S2                 & 12.51 & 8,10,15,37,45,46,84  \\
96   & 3C 273            & 12 29 06.7 & +02 03 08.6 & 0.158 & S1 & 12.83 & 100,102  \\
97   & Mrk 1014          & 01 59 50.2 & +00 23 40.6 & 0.163 & S1 & 12.63 & 1,64,67,68,69,75,81  \\
98   & Mrk 463           & 13 56 02.9 & +18 22 19.1 & 0.050 & S2 & 11.80 & 90,99,100,101,102  \\
99   & PG 1119+120       & 11 21 47.1 & +11 44 18.3 & 0.050 & S1 & 11.29 & 15,24,90,98,101,102   \\
100 & PG 1211+143        & 12 14 17.7 & +14 03 12.6 & 0.081 & S1 & 11.76 & 96,98,102   \\
101 & PG 1351+640        & 13 53 15.8 & +63 45 45.4 & 0.088 & S1 & 11.88 & 98,99   \\
102 & PG 2130+099        & 21 32 27.8 & +10 08 19.5 & 0.063 & S1 & 11.60 & 90,96,98,99,100   
\enddata
\tablenotetext{a}{Optical classification, taken mainly from \citet{vei99}, and also from \citealt{ber86,arm89,all91,mir91,sek93,lee94,vei95,duc97,law99,sta00,ver01,kew01,meu01,rup05,dar06} }
\tablenotetext{b}{IR luminosities are either taken from \citet{far03}, or calculated from the IRAS fluxes using the same methods.
Units are the logarithm of the rest-frame 1-1000$\mu$m luminosity, in Solar luminosities.} 
\tablenotetext{c}{ID numbers of the ULIRGs that `pair' with this object, i.e. that have log$(\mathcal{R})<0$}
\tablenotetext{d}{Measured from the IRS spectrum. The redshift given by \citet{sta00} does not agree with the IRS data.}
\end{deluxetable*}
\clearpage
\end{landscape}


\begin{thebibliography}{}

\bibitem[Alexander et al.(2005)]{ale05}
Alexander, D.~M., Smail, I., Bauer, F.~E., Chapman, S.~C., Blain, A.~W., Brandt, W.~N., \&
Ivison, R.~J.\ 2005, \nat, 434, 738

\bibitem[Allen et al.(1991)]{all91} 
Allen, D.~A., Norris, R.~P., Meadows, V.~S., \& Roche, P.~F.\ 1991, \mnras, 248, 528 

\bibitem[Armus et al.(1989)]{arm89} 
Armus, L., Heckman, T.~M., \& Miley, G.~K.\ 1989, \apj, 347, 727 

\bibitem[Armus et al.(2004)]{arm04}
Armus, L., et al.\ 2004, \apjs, 154, 178

\bibitem[Armus et al.(2006)]{arm06}
Armus, L., et al.\ 2006, \apj, 640, 204

\bibitem[Armus et al.(2007)]{arm07}
Armus, L., et al.\ 2007, \apj, 656, 148

\bibitem[Barger et al.(1998)]{bar98}
Barger, A.~J., Cowie, L.~L., Sanders, D.~B., Fulton, E., Taniguchi, Y., Sato, Y., Kawara, K.,
\& Okuda, H.\ 1998, \nat, 394, 248

\bibitem[Berger \& Pericchi(2001)]{ber01}
Berger, J., Pericchi, L., 2001, `Model Selection' (P. Lahiri, Ed), Institute of Mathematical Statistics Lecture Notes, Monograph Series Vol. 38

\bibitem[Barnes \& Hernquist(1992)]{bar92}
Barnes, J.~E., \& Hernquist, L.\ 1992, \araa, 30, 705

\bibitem[Bergvall et al.(1986)]{ber86} 
Bergvall, N., Johansson, L., \& Olofsson, K.\ 1986, \aap, 166, 92 

\bibitem[Berta et al.(2007)]{ber07}
Berta, S., et al.\ 2007, \aap, 467, 565

\bibitem[Bianchi et al.(2008)]{bia08} 
Bianchi, S., Chiaberge, M., Piconcelli, E., Guainazzi, M., \& Matt, G.\ 2008, \mnras, 386, 105 

\bibitem[Blain et al.(2004)]{bla04}
Blain, A.~W., Chapman, S.~C., Smail, I., \& Ivison, R.\ 2004, \apj, 611, 725

\bibitem[Borys et al.(2003)]{bor}
Borys, C., Chapman, S., Halpern, M., \& Scott, D.\ 2003, \mnras, 344, 385

\bibitem[Borys et al.(2006)]{bor06}
Borys, C., et al.\ 2006, \apj, 636, 134

\bibitem[Braito et al.(2009)]{brai09} 
Braito, V., Reeves, J.~N., Della Ceca, R., Ptak, A., Risaliti, G., 
\& Yaqoob, T.\ 2009, \aap accepted, arXiv:0905.1041 

\bibitem[Brandes(2001)]{bra01}
Brandes, U., 2001, Journal of Mathematical Sociology, 25, 163

\bibitem[Bridge et al.(2007)]{bri07}
Bridge, C.~R., et al.\ 2007, \apj, 659, 931

\bibitem[Bushouse et al(2002)]{bus02}
Bushouse H. A., et al, 2002, ApJS, 138, 1

\bibitem[Chapman et al.(2003)]{cha03}
Chapman, S.~C., Windhorst, R., Odewahn, S., Yan, H., \& Conselice, C.\ 2003, \apj, 599, 92

\bibitem[Connolly et~al.(2006)]{connolly}
Connolly, B.~M. {\it et al.}, 2006, Phys.Rev. D74, 043001

\bibitem[Corcoran \& Wainwright(1995)]{libga}
Corcoran, A.~L., Wainwright, R.~L., 1995, {\it Using LibGA to Develop Genetic Algorithms for Solving Combinatorial Optimization Problems}, in {\it The Application Handbook of Genetic Algorithms,}, Volume I, Lance Chambers, editor, pages 143-172, CRC Press.

\bibitem[Cui et al.(2001)]{cui01}
Cui, J., Xia, X.-Y., Deng, Z.-G., Mao, S., \& Zou, Z.-L.\ 2001, \aj, 122, 63

\bibitem[Darling \& Giovanelli(2006)]{dar06} 
Darling, J., \& Giovanelli, R.\ 2006, \aj, 132, 2596 

\bibitem[Dasyra et al.(2006a)]{das06a} 
Dasyra, K.~M., et al.\ 2006a, \apj, 638, 745 

\bibitem[Dasyra et al.(2006b)]{das06} 
Dasyra, K.~M., et al.\ 2006b, \apj, 651, 835 

\bibitem[Desai et al.(2007)]{des07}
Desai, V., et al.\ 2007, \apj, 669, 810

\bibitem[Dole et al.(2001)]{dol}
Dole, H., et al. 2001, \aap, 372, 364

\bibitem[Dubinski et al.(1999)]{dub99}
Dubinski, J., Mihos, J.~C., \& Hernquist, L.\ 1999, \apj, 526, 607

\bibitem[Duc et al.(1997)]{duc97} 
Duc, P.-A., Mirabel, I.~F., \& Maza, J.\ 1997, \aaps, 124, 533 

\bibitem[Eales et al.(2000)]{eal}
Eales, S., Lilly, S., Webb, T., Dunne, L., Gear, W., Clements, D., \& Yun, M.\ 2000, \aj, 120,
2244

\bibitem[Erdos \& Renyi(1959)]{erd} 
Erdos, P., Renyi, A., 1959, Publ. Math. Debrecen, 6, 290

\bibitem[Farrah et al.(2001)]{far01}
Farrah, D., et al.\ 2001, \mnras, 326, 1333

\bibitem[Farrah et al.(2002)]{far02b}
Farrah D., Verma A., Oliver S., Rowan-Robinson M., McMahon R.,
2002, MNRAS, 329, 605

\bibitem[Farrah et al.(2003)]{far03}
Farrah, D., Afonso, J., Efstathiou, A., Rowan-Robinson, M., Fox, M.,
\& Clements, D.\ 2003, \mnras, 343, 585

\bibitem[Farrah et al.(2005)]{far05}
Farrah, D., Surace, J.~A., Veilleux, S., Sanders, D.~B., \& Vacca, W.~D.\ 2005, \apj, 626, 70

\bibitem[Farrah et al.(2006)]{far06a}
Farrah, D., et al.\ 2006, \apjl, 641, L17

\bibitem[Farrah et al.(2007)]{far07b}
Farrah, D., et al.\ 2007, \apj, 667, 149

\bibitem[Farrah et al.(2008)]{far08}
Farrah, D., et al.\ 2008, \apj, 677, 957

\bibitem[Franceschini et al.(2003)]{fra03}
Franceschini, A., et al.\ 2003, \mnras, 343, 1181

\bibitem[Freeman(1979)]{fre79}
Freeman, L. C., 1979, Social Networks, 1, 215

\bibitem[Fruchterman \& Reingold(1991)]{fru91}
Fruchterman, T. M. J., Reingold, E. M., 1991, Software: Practice \& Experience, 21, 11:1129-1164

\bibitem[Genzel et al(1998)]{gen98}
Genzel R., et al, 1998, ApJ, 498, 579

\bibitem[Genzel et al.(2001)]{gen01} 
Genzel, R., Tacconi, L.~J., Rigopoulou, D., Lutz, D., \& Tecza, M.\ 2001, \apj, 563, 527 

\bibitem[Gradshteyn \& Rhyzik(2000)]{Gradshteyn:2000}
Gradshteyn, I.~S., \& Rhyzik, I.~M., 2000, {\it Table of Integrals, Series and Products},
Edited by A. Jeffrey and D. Zwillinger, Academic Press, New York, 6th edition.

\bibitem[Hao et al.(2005)]{hao05} 
Hao, L., et al.\ 2005, \apjl, 625, L75 

\bibitem[Higdon et al.(2006)]{hig06}
Higdon, S.~J.~U., Armus, L., Higdon, J.~L., Soifer, B.~T., \& Spoon, H.~W.~W.\ 2006, \apj, 648, 323

\bibitem[Houck et al(2004)]{hou04}
Houck, J.~R., et al.\ 2004, \apjs, 154, 18

\bibitem[Huang et al.(2009)]{hua09} 
Huang, J.~-., et al.\  2009, ApJ accepted, arXiv:0904.4479 

\bibitem[Hughes et al.(1998)]{hug}
Hughes, D.~H., et al.\ 1998, \nat, 394, 241

\bibitem[Imanishi et al.(2007)]{ima07}
Imanishi, M., Dudley, C.~C., Maiolino, R., Maloney, P.~R., Nakagawa, T.,
\& Risaliti, G.\ 2007, \apjs, 171, 72

\bibitem[Iono et al.(2004)]{ion04}
Iono, D., Yun, M.~S., \& Mihos, J.~C.\ 2004, \apj, 616, 199

\bibitem[Jaynes \& Bretthorst(2005)]{jaynes}
Jaynes, E.T., Bretthorst, G. L., (Ed.), 2005, `{\it Probability Theory:
The Logic of Science}', Cambridge University Press

\bibitem[Jeffreys(1961)]{jeffreys}
Jeffreys, H., 1961, {\it Theory of Probability}, Third Edition, Oxford University Press

\bibitem[Kamada \& Kawai(1989)]{kam89}
Kamada, T., Kawai, S., Information Processing Letters, 31, 7-15.

\bibitem[Kawakatu et al.(2006)]{kaw06}
Kawakatu, N., Anabuki, N., Nagao, T., Umemura, M., \& Nakagawa, T.\ 2006, \apj, 637, 104

\bibitem[Kawakatu et al.(2007)]{kaw07}
Kawakatu, N., Imanishi, M., \& Nagao, T.\ 2007, \apj, 661, 660

\bibitem[Kewley et al.(2001)]{kew01} 
Kewley, L.~J., Heisler, C.~A., Dopita, M.~A., \& Lumsden, S.\ 2001, \apjs, 132, 37 

\bibitem[Kim \& Sanders(1998)]{kim98}
Kim, D.-C., \& Sanders, D.~B.\ 1998, \apjs, 119, 41

\bibitem[Klaas et al.(2001)]{kla01}
Klaas, U., et al.\ 2001, \aap, 379, 823

\bibitem[Lahuis et al.(2007)]{lah07}
Lahuis, F., et al.\ 2007, \apj, 659, 296

\bibitem[Lawrence et al.(1999)]{law99} 
Lawrence, A., et al.\ 1999, \mnras, 308, 897 

\bibitem[Le Floc'h et al.(2005)]{lef05}
Le Floc'h, E., et al.\ 2005, \apj, 632, 169

\bibitem[Leech et al.(1994)]{lee94} 
Leech, K.~J., Rowan-Robinson, M., Lawrence, A., \& Hughes, J.~D.\ 1994, \mnras, 267, 253 

\bibitem[L{\'{\i}}pari et al.(2005)]{lip05}
L{\'{\i}}pari, S., Terlevich, R., Zheng, W., Garcia-Lorenzo, B., Sanchez, S.~F.,
\& Bergmann, M.\ 2005, \mnras, 360, 416

\bibitem[Lonsdale et al.(2006)]{lfs06}
Lonsdale, C.~J., Farrah, D., \& Smith, H.~E.\ 2006, Astrophysics Update 2, 285

\bibitem[Lonsdale et al.(2009)]{lon09} 
Lonsdale, C.~J., et al.\ 2009, \apj, 692, 422 

\bibitem[Lusseau \& Newman(2004)]{lus04}
Lusseau, D., Newman, M. E. J., 2004, Proceedings of the Royal Society B, 271, S477

\bibitem[Lutz et al.(1998)]{lut98}
Lutz, D., Kunze, D., Spoon, H.~W.~W., \& Thornley, M.~D.\ 1998, \aap, 333, L75

\bibitem[Magliocchetti et al.(2007)]{mag07}
Magliocchetti, M., Silva, L., Lapi, A., de Zotti, G., Granato, G.~L., Fadda, D.,
\& Danese, L.\ 2007, \mnras, 375, 1121

\bibitem[Melbourne et al.(2008)]{mel08}
Melbourne, J., et al.\ 2008, \aj, 135, 1207

\bibitem[Meusinger et al.(2001)]{meu01}
Meusinger, H., Stecklum, B., Theis, C., \& Brunzendorf, J.\ 2001, \aap, 379, 845

\bibitem[Mirabel et al.(1991)]{mir91} 
Mirabel, I.~F., Lutz, D., \& Maza, J.\ 1991, \aap, 243, 367 

\bibitem[Mortier et al.(2005)]{mor}
Mortier, A.~M.~J., et al.\ 2005, \mnras, 363, 563

\bibitem[Pastor-Satorras et al.(2001)]{pas}
Pastor-Satorras, R., Vazquez, A., Vespignani, A., 2001, Phys. Rev. Lett., 87, 258701

\bibitem[Press et al.(1992)]{numerical}
Press, W.~H., Teukolsky, S.~A., Vetterling, W.~T., and Flannery, B.~P., 1992,
{\it Numerical Recipes in C}, Cambridge University Press

\bibitem[Ptak et al(2003)]{pta03}
Ptak A., Heckman T., Levenson N. A., Weaver K., Strickland D., 2003, ApJ, 592, 782

\bibitem[Rieke \& Low(1972)]{rie72}
Rieke G. H., Low F. J., 1972, ApJ, 176, L95

\bibitem[Rigopoulou et al.(1999)]{rig99}
Rigopoulou, D., Spoon, H.~W.~W., Genzel, R., Lutz, D., Moorwood, A.~F.~M.,
\& Tran, Q.~D.\ 1999, \aj, 118, 2625

\bibitem[Rowan-Robinson et al.(1997)]{rr97}
Rowan-Robinson, M., et al. 1997, \mnras, 289, 490

\bibitem[Rupke et al.(2005)]{rup05} 
Rupke, D.~S., Veilleux, S., \& Sanders, D.~B.\ 2005, \apjs, 160, 87 

\bibitem[Sanders et al.(1988)]{san88}
Sanders, D.~B., Soifer, B.~T., Elias, J.~H., Madore, B.~F., Matthews, K., Neugebauer, G., \&
Scoville, N.~Z.\ 1988, \apj, 325, 74

\bibitem[Sanders \& Mirabel(1996)]{san96}
Sanders, D.~B., \& Mirabel, I.~F.\ 1996, \araa, 34, 749

\bibitem[Sekiguchi \& Wolstencroft(1993)]{sek93} 
Sekiguchi, K., \& Wolstencroft, R.~D.\ 1993, \mnras, 263, 349

\bibitem[Sellke et al(2001)]{sel01}
Sellke, T., Bayarri, M.~J., Berger, J.~O., 2001, The American Statistician, Vol 55, p62

\bibitem[Siganos et al(2003)]{sig03}
Siganos G., Faloutsos M., Faloutsos P., Faloutsos C., 2003, ACM/IEEE Transactions on Networking, vol. 11, no. 4, pp. 514-524

\bibitem[Sirocky et al.(2008)]{sir08} 
Sirocky, M.~M., Levenson, N.~A., Elitzur, M., Spoon, H.~W.~W., \& Armus, L.\ 2008, \apj, 678, 729 

\bibitem[Sivia(1996)]{sivia}
Sivia, D. S., 1996, `Data Analysis: A Bayesian Tutorial', Oxford University Press

\bibitem[Smail et al.(2003)]{sma03}
Smail, I., Chapman, S.~C., Ivison, R.~J., Blain, A.~W., Takata, T., Heckman, T.~M., Dunlop,
J.~S., \& Sekiguchi, K.\ 2003, \mnras, 342, 1185

\bibitem[Smail et al.(2004)]{sma04}
Smail, I., Chapman, S.~C., Blain, A.~W., \& Ivison, R.~J.\ 2004, \apj, 616, 71

\bibitem[Soifer et al.(1984)]{soi84}
Soifer, B.~T., et al.\ 1984, \apjl, 278, L71

\bibitem[Soifer et al.(2008)]{soi08} 
Soifer, B.~T., Helou, G., \& Werner, M.\ 2008, \araa, 46, 201 

\bibitem[Spoon et al.(2004)]{spo04}
Spoon, H.~W.~W., et al.\ 2004, \apjs, 154, 184

\bibitem[Spoon et al.(2006)]{spo06}
Spoon, H.~W.~W., et al.\ 2006, \apj, 638, 759

\bibitem[Spoon et al.(2007)]{spo07}
Spoon, H.~W.~W., Marshall, J.~A., Houck, J.~R., Elitzur, M., Hao, L., Armus, L., Brandl,
B.~R., \& Charmandaris, V.\ 2007, \apjl, 654, L49

\bibitem[Spoon et al.(2009)]{spo08} 
Spoon, H.~W.~W., Armus, L., Marshall, J.~A., Bernard-Salas, J., Farrah, D., Charmandaris, V., 
\& Kent, B.~R.\ 2009, \apj, 693, 1223 

\bibitem[Stanford et al.(2000)]{sta00}
Stanford, S.~A., Stern, D., van Breugel, W., \& De Breuck, C.\ 2000, \apjs, 131, 185

\bibitem[Strauss et al.(1990)]{str90}
Strauss, M.~A., Davis, M., Yahil, A., \& Huchra, J.~P.\ 1990, \apj, 361, 49

\bibitem[Surace et al.(2000)]{sur00}
Surace, J.~A., Sanders, D.~B., \& Evans, A.~S.\ 2000, \apj, 529, 170

\bibitem[Tacconi et al.(2002)]{tac02}
Tacconi, L.~J., Genzel, R., Lutz, D., Rigopoulou, D., Baker, A.~J., Iserlohe, C.,
\& Tecza, M.\ 2002, \apj, 580, 73

\bibitem[Takata et al.(2006)]{tak06}
Takata, T., Sekiguchi, K., Smail, I., Chapman, S.~C., Geach, J.~E., Swinbank, A.~M., Blain, A., \&
Ivison, R.~J.\ 2006, \apj, 651, 713

\bibitem[Taniguchi et al.(1999)]{tan99}
Taniguchi, Y., Ikeuchi, S., \& Shioya, Y.\ 1999, \apjl, 514, L9

\bibitem[Tran et al.(2001)]{tra01}
Tran, Q.~D., et al.\ 2001, \apj, 552, 527

\bibitem[Tremaine et al.(2002)]{tre02} 
Tremaine, S., et al.\ 2002, \apj, 574, 740 

\bibitem[V{\"a}is{\"a}nen et al.(2008)]{vai08}
V{\"a}is{\"a}nen, P., et al.\ 2008, \mnras, 384, 886

\bibitem[Valiante et al.(2007)]{val07}
Valiante, E., Lutz, D., Sturm, E., Genzel, R., Tacconi, L.~J., Lehnert, M.~D., \& Baker, A.~J.\
2007, \apj, 660, 1060

\bibitem[Vega et al.(2008)]{veg08}
Vega, O., Clemens, M.~S., Bressan, A., Granato, G.~L., Silva, L., \& Panuzzo, P.\ 2008, \aap, 484, 631

\bibitem[Veilleux et al.(1995)]{vei95} 
Veilleux, S., Kim, D.-C., Sanders, D.~B., Mazzarella, J.~M., 
\& Soifer, B.~T.\ 1995, \apjs, 98, 171 

\bibitem[Veilleux et al.(1999)]{vei99}
Veilleux, S., Kim, D.-C., \& Sanders, D.~B.\ 1999, \apj, 522, 113

\bibitem[Veilleux et al.(1999)]{vei99b} 
Veilleux, S., Sanders, D.~B., \& Kim, D.-C.\ 1999, \apj, 522, 139 

\bibitem[Veilleux et al.(2002)]{vei02}
Veilleux, S., Kim, D.-C., \& Sanders, D.~B.\ 2002, \apjs, 143, 315

\bibitem[Veilleux et al.(2006)]{vei06}
Veilleux, S., et al.\ 2006, \apj, 643, 707

\bibitem[V{\'e}ron-Cetty \& V{\'e}ron(2001)]{ver01} 
V{\'e}ron-Cetty, M.-P., \& V{\'e}ron, P.\ 2001, \aap, 374, 92 

\bibitem[Weingartner \& Draine(2001)]{wein}
Weingartner, J.~C., \& Draine, B.~T., 2001, ApJ., 548, 296-309

\bibitem[Werner et al.(2004)]{wer04}
Werner, M.~W., et al.\ 2004, \apjs, 154, 1

\bibitem[Zauderer et al.(2007)]{zau07}
Zauderer, B.~A., Veilleux, S., \& Yee, H.~K.~C.\ 2007, \apj, 659, 1096

\end{thebibliography}
\end{document}